\pdfoutput=1
\documentclass[acmsmall]{acmart}

\def\BibTeX{{\rm B\kern-.05em{\sc i\kern-.025em b}\kern-.08emT\kern-.1667em\lower.7ex\hbox{E}\kern-.125emX}}

\usepackage{microtype}
\usepackage[export]{adjustbox}
\usepackage{graphicx}
\usepackage{wrapfig}
\usepackage{subcaption}
\usepackage[normalem]{ulem}
\useunder{\uline}{\ul}{}
\usepackage{tikz}
\usepackage{cleveref}
\usepackage{xr}
\usepackage{booktabs} 

\usepackage[ruled]{algorithm2e} 

\usepackage{enumerate}
\usepackage{ifthen,xspace}

\newboolean{draft}
\setboolean{draft}{false}
\ifthenelse{\boolean{draft}}{%
  \newcounter{comments}
  \newcommand{\adb}[1]{\addtocounter{comments}{1}{\color{Blue}[Ted \thecomments: #1]}}
  
   \newcommand{\ted}[1]{\addtocounter{comments}{1}{\color{Blue}[Ted \thecomments: #1]}}
 \newcommand{\todoted}[1]{\addtocounter{comments}{1}{\color{DarkRed}[For Ted: \thecomments: #1]}} \newcommand{\tmm}[1]{\addtocounter{comments}{1}{\color{DarkGreen}[TMM \thecomments: #1]}}

}
{

\newcommand{\adb}[1]{}
\newcommand{\ted}[1]{}
\newcommand{\todoted}[1]{}
\newcommand{\tmm}[1]{}
\newcommand{\kurt}[1]{}
}

\newcommand{\Digcola}{\emph{Dig-Cola}\xspace}
\newcommand{\Ipsep}{\emph{IPSEP-Cola}\xspace}
\newcommand{\Spring}{\emph{Spring Electrical}\xspace}
\newcommand{\SA}{\emph{Simulated Annealing}\xspace}
\newcommand{\Crowd}{\emph{Crowd}\xspace}
\newcommand{\CrowdRandom}{\emph{Crowd-Random}\xspace}

\newcommand{\etal}{\textit{et al.}\xspace}

\begin{document}

%
\title[Flud]{Flud: a hybrid crowd--algorithm approach for visualizing biological networks}

%

\author{Aditya Bharadwaj}
\orcid{1234-5678-9012}
\affiliation{%
  \institution{Virginia Tech}
  \city{Blacksburg}
  \state{Virginia}
  \country{USA}
  \postcode{24060}
}
\email{adb@vt.edu}

\author{David Gwizdala}
\affiliation{%
  \institution{Bridgewater Associates}
  \city{Stamford}
  \state{Connecticut}
  \country{USA}
}
\email{gwizdala@vt.edu}

\author{Yoonjin Kim}
\affiliation{%
  \institution{Virginia Tech}
  \city{Blacksburg}
  \state{Virginia}
  \country{USA}
  \postcode{24060}
}
\email{ykim05@vt.edu}

\author{Kurt Luther}
\affiliation{%
  \institution{Virginia Tech}
  \city{Arlington}
  \state{Virginia}
  \country{USA}
  \postcode{22203}
}
\email{kluther@vt.edu}

\author{T. M. Murali}
\affiliation{%
  \institution{Virginia Tech}
  \city{Blacksburg}
  \state{Virginia}
  \country{USA}
  \postcode{24060}
}
\email{murali@vt.edu}

\renewcommand{\shortauthors}{Bharadwaj et al.}

%
\begin{abstract}
Modern experiments in many disciplines generate large quantities of network (graph) data. Researchers require aesthetic layouts of these networks that clearly convey the domain knowledge and meaning. However, the problem remains challenging due to multiple conflicting aesthetic criteria and complex domain-specific constraints. In this paper, we present a strategy for generating visualizations that can help network biologists understand the protein interactions that underlie processes that take place in the cell. Specifically, we have developed Flud, an online game with a purpose (GWAP) that allows humans with no expertise to design biologically meaningful graph layouts with the help of algorithmically generated suggestions. Further, we propose a novel hybrid approach for graph layout wherein crowdworkers and a simulated annealing algorithm build on each other's progress. To showcase the effectiveness of Flud, we recruited crowd workers on Amazon Mechanical Turk to lay out complex  networks that represent signaling pathways. Our results show that the proposed hybrid approach outperforms state-of-the-art techniques for graphs with a large number of feedback loops.  We also found that the algorithmically generated suggestions guided the players when they are stuck and helped them improve their score. Finally, we discuss broader implications for mixed-initiative interactions in human computation games.
\end{abstract}

%
%


%

\vbadness=99999

\maketitle

\section{Introduction}
\label{sec:introduction}
Many fields of science require meaningful and visually appealing layouts of networks (also known as graphs). A prominent example is the discipline of network biology, where scientists use networks to understand the chemical reactions and protein interactions that underlie processes that take place in the cell~\cite{bo-nbucf-2004}. 
In order to present and analyze these networks, researchers require aesthetic layouts of these networks that clearly convey the relevant biological information. A \emph{layout} of a network assigns $x$ and $y$ coordinates to each node and routes each edge using a straight line or a curve in order to create a meaningful visual representation.

There are two major approaches for creating network layouts. The first approach views humans as the deciding agent and primary creator, with computers seen as support tools. An example is a graph (or network) drawing interface in Cytoscape~\cite{smoot2010cytoscape} that provides the layout tools for designers to use while drawing networks. This approach offers creative freedom and allows the user create meaningful layouts by capturing complex, domain-specific constraints. However, it is time-consuming to create layouts manually. 
The second approach uses fully-automated methods and does not require human intervention. In contrast to the first approach, it can generate data visualizations at scale~\cite{tamassia2013handbook,battista-ioannis-algos-drawing-graphs-compgeo-1994,pavlopoulos-schneider-survey-bio-net-vis-2008,gibson-2d-graph-layout-survey-infovis-2013}. However, these methods lack the ability to capture complex visualization constraints and domain-specific needs.  As a result, it is a common for biologists to depend on the time-consuming practice of manually improving automatically generated visualizations. 

Crowdsourcing has emerged as a promising solution to scale up such manual tasks~\cite{van-rogowitz-perceptual-organization-itvcg-2008,
singh-luther-crowdlayout-chi-2018,yuan-xin-intelligent-many-users-itvcg-2012} by tapping into the human intelligence and creativity of crowdworkers.
However, it is unclear how to design interfaces and tasks that will allow novice crowds---who lack expertise in biology and computer science in general---to make domain-specific modifications to network layouts and balance various constraints. Our research seeks to bridge this fundamental gap.


 
\begin{figure}[h]
  \centering
  \includegraphics[width=0.8\linewidth]{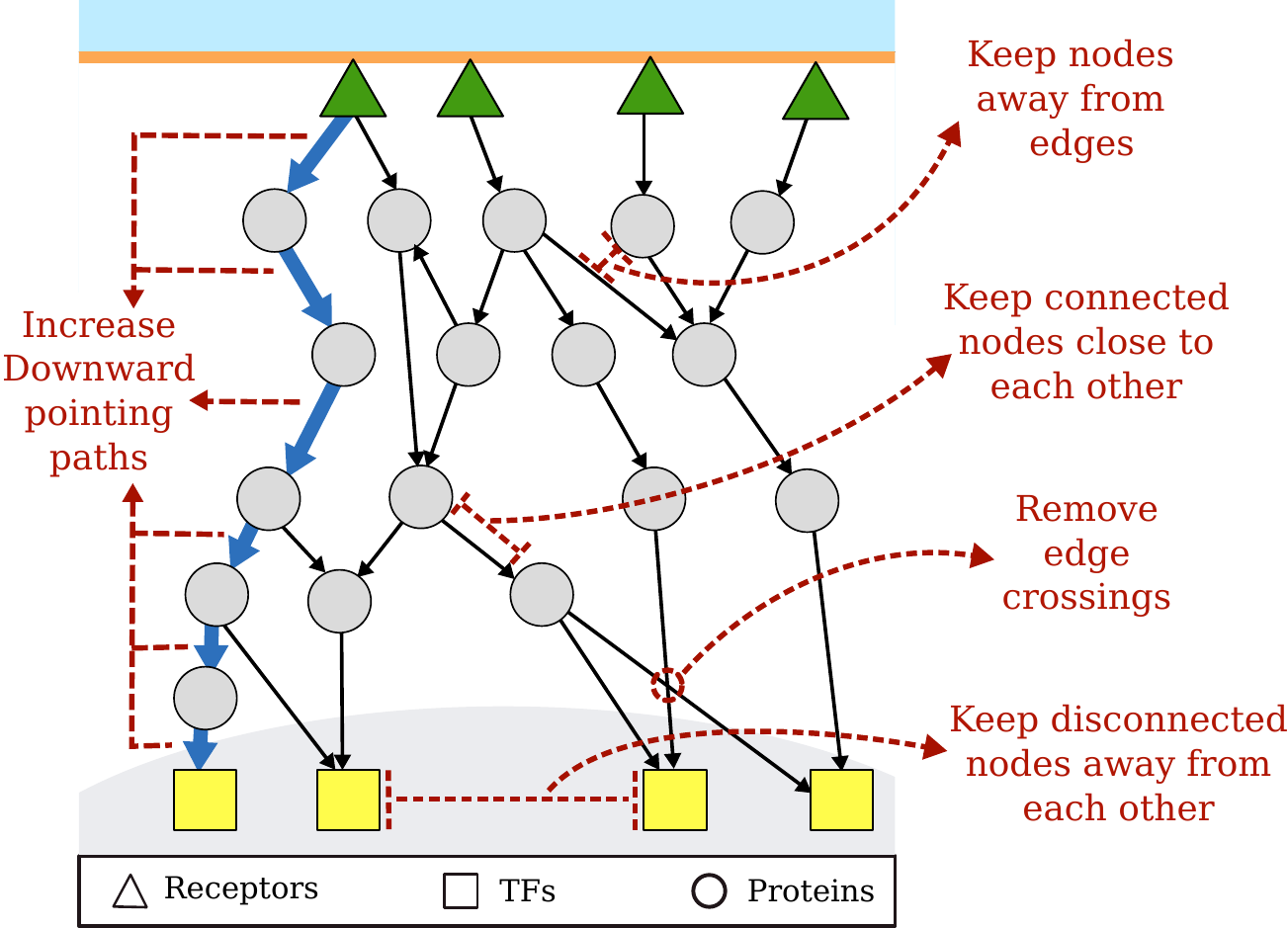}
  \caption{Five layout criteria used in Flud. 1) Minimize the number of edge crossings, 2) Keep nodes connected by an edge close to each other, 3) Disperse disconnected node pairs, 4) Increase the separation between nodes and edges, and 5) Increase the number of downward pointing paths.}
  \label{fig:layout-criteria}
\end{figure}

In this work, we present Flud, an online game with purpose (GWAP) that allows humans with no expertise to design biologically meaningful network layouts with the help of algorithmically-generated suggestions. The goal of the game is to move nodes in a given network so as to create a layout that optimizes a score based on pre-specified design criteria. These criteria include four previously defined aesthetic considerations (Figure~\ref{fig:layout-criteria}): 1) minimizing the number of edge crossings, 2) keeping nodes connected by an edge close to each other, 3) dispersing disconnected node pairs, and 4) increasing the separation between nodes and edges. We also introduce a new, fifth criterion inspired by a biological application to cellular signaling: 5) maximize the number of downward pointing paths in the layout (Figure~\ref{fig:layout-criteria}). These types of paths draw visual attention to sequences of edges that lead from receptor proteins in the cell membrane (green triangles, placed at the top of the layout) through internal nodes to effector molecules in the nucleus (transcription factors, yellow squares, placed at the bottom of the layout).

The role of the game players is to make modifications to the layout so as to balance the criterion-specific scores based on the direction of flow of information in the network, edge crossings, and relative distances between nodes and edges. Since, these criteria may conflict with each other, Flud allows the players to track the changes in layout score and the corresponding criterion-specific scores with every move.
Another important feature of Flud is the availability of a ``clue'' for each criterion. If the player is stuck, perhaps because the network is complex or it is unclear what the next move should be, the game highlights a small subset of nodes and edges in the network as clues. These nodes and edges are selected by Flud such that changing the positions of these elements is likely to improve the score for the corresponding criterion. With the help of criterion-specific scores, clues, and layout tools in Flud interface, we expect that game players will succeed in generating aesthetic and meaningful layouts. 

Flud also facilitates a novel mixed-initiative network layout strategy that utilizes a sequential and collaborative process wherein human players and a simulated annealing based layout algorithm \cite{davidson-harel-simulated-annealing-drawing-1996} build on each other's progress. The simulated annealing algorithm seeks to iteratively reach a layout with higher score by randomly searching the neighborhood of an initial guess. Its strength is that it avoids getting stuck at local optima --- a solution that is better than the others nearby but is not the very best --- by probabilistically accepting an inferior solution. In contrast, human players use their intelligence and creativity to optimize the score by making modifications that may be global in nature and challenging to achieve using an automated method. 
To showcase the effectiveness of Flud and the novel mixed-initiative approach, we recruited nearly 2,000 
novice crowd workers on Amazon Mechanical Turk to lay out and visualize complex protein networks that represent signaling pathways in human cells. We presented three different versions of the hybrid crowd--algorithm approach and evaluated them against a crowd-only strategy and algorithmic baselines. Our results show that such a collaboration between humans and algorithms leads to higher scoring layouts than either from humans or from algorithms alone. We found that the game elements such as criterion-specific modes and clues supported crowd workers in the visualization tasks. The results also show that the crowd workers who moved one of the nodes suggested by Flud contributed more to the overall layout score in comparison to rest of the nodes in the network. 

In summary, our contributions include: 
\begin{itemize}
    \item A novel mixed-initiative approach that combines crowdsourcing with computational engines to create high-quality visualizations of biological networks.
    \item A game with a purpose, Flud, that facilitates this approach with two components: a) an interface that gamifies the network visualization task to make it more accessible to crowd workers with no biological or computer science expertise, and b) an implementation that allows crowd workers and algorithms to build on each others progress.
    
    \item Experiments that provide empirical evidence of the benefits of mixed-initiative layout methods compared to algorithmic baselines. 
\end{itemize}

\section{Related Work}
\label{sec:related-work}
\subsection{Algorithms for network layout}
Several algorithms exist to automatically compute network layouts. Popular methods include circular~\cite{baur-ulrik-crossings-circular-layouts2004,gansner-improved-circular-layout-2006}, hierarchical~\cite{eades-xuemin-straight-hierarchical-1996,sugiyama-mitsuhiko-methods-hierarchical-1981,gansner-drawing-directed-trans-soft-eng-1993}, orthogonal \cite{eiglsperger-gunnar-orthogonal-graph-drawing-2001}, force directed~\cite{fruchterman-edward-force-directed-graph-1991,kamada-kawai-algorithm-undirected-1989}, and spectra~\cite{koren-spectral-iccc-2003,koren-drawing-eigenvectors-2005} layouts. Several of these methods can generate layouts in a few seconds for moderately-sized networks. Implementations are available in various tools or libraries, e.g., Cytoscape \cite{shannon-ideker-cytoscape-genome-research-2003}, Gephi \cite{jacomy-bastian-forceatlas2-gephi-2014}, Graphviz~\cite{ellson2002graphviz}, NetworkX \cite{hagberg-chult-networkx-2008}, NetBioV \cite{tripathi-netbiov-2014}, and Pajek \cite{batagelj-Mrvar-pajek-2004}. 

Despite their wide availability, these methods have several drawbacks. They may produce very specific types of layouts, e.g., arrange nodes in circles~\cite{baur-ulrik-crossings-circular-layouts2004,gansner-improved-circular-layout-2006}, or be appropriate for restricted classes of networks, e.g., directed acyclic networks~\cite{eades-xuemin-straight-hierarchical-1996,sugiyama-mitsuhiko-methods-hierarchical-1981}. They may compute layouts that try to optimize specific aesthetic criteria, e.g., prevent edges from being stretched and disconnected nodes from coming too close~\cite{fruchterman-edward-force-directed-graph-1991,kamada-kawai-algorithm-undirected-1989} or hierarchically arrange nodes and remove edge crossings \cite{sugiyama-mitsuhiko-methods-hierarchical-1981,gansner-drawing-directed-trans-soft-eng-1993}. One such popular method is Dig-Cola \cite{dwyer-yehuda-dig-cola-infovis-2005}, which arranges the nodes in layers while preserving edge length and symmetries. Since highlighting information flow is an important criterion for experiments in this paper, we use Dig-Cola as a baseline to compare performance against Flud. In general, these methods do not have the flexibility to accommodate a diverse variety of layout criteria and domain-specific constraints. 

\looseness=-1
As discussed earlier, a general purpose approach to accommodate multiple layout criteria is to formulate network layout as an optimization problem and solve it using simulated annealing~\cite{davidson-harel-simulated-annealing-drawing-1996}. 
Taking inspiration from these ideas, we propose to examine how a mixed-initiative approach that combines human intelligence with simulated annealing can yield better layouts than either of these methods alone.


\subsection{Leveraging human abilities through crowdsourcing to lay out networks}

Some research has examined how users lay out networks, including in comparison to automated layouts. Two studies~\cite{dwyer-north-comparison-user-auto-graph-layouts-2009,van-rogowitz-perceptual-organization-itvcg-2008} asked participants to draw an aesthetic layout of a given network. Dwyer~\etal~\cite{dwyer-north-comparison-user-auto-graph-layouts-2009} concluded that the best user-generated layouts performed as well as or better than automated layouts based on physical models. Although these studies do not use crowd workers, they provide evidence that non-expert humans can create network layouts comparable to algorithms.


Researchers have explored using non-expert volunteers (e.g., citizen scientists) to perform other visualization tasks. For example, systems such as ManyEyes~\cite{viegas-manyeyes-itvcg-2007} and Sense.us~\cite{heer-voyagers-senseus-chi-2007} have used novice crowds to create, collaborate on, and analyze data visualizations. Connect the Dots~\cite{li_crowdsourcing_2018} and CRICTO~\cite{chung_cricto:_2017} ask novice crowds to build social networks for intelligence analysis, but not to create layouts for them. Inspired by these projects, we explore the use of non-expert workers on Amazon Mechanical Turk to create biological network layouts.



Most related to this paper, our prior work on CrowdLayout leverages crowd workers to design layouts of biological networks that satisfy domain-inspired guidelines specified in natural language~\cite{singh-luther-crowdlayout-chi-2018}. CrowdLayout also uses crowd workers to evaluate how well these layouts satisfy the guidelines. Our work in CrowdLayout shows that translating domain knowledge into representational guidelines can allow non-expert crowds to create effective layouts.


Our work differs from and complements these efforts in several important ways. First, we focus on combining multiple aesthetic criteria with a specific domain knowledge guideline, all of which can be translated into mathematical formulae. Second, we score layouts to provide real-time feedback to the user who is modifying the layout rather than (as in the case of CrowdLayout) wait for other crowd workers to evaluate the layout.  Third, we formulate layout creation as a collaboration among multiple crowd workers or between crowd workers and computational engines, in contrast to single-user design tasks.

\subsection{Designing network layouts by combining algorithms with crowdsourcing}


Prior work has explored using mixed-initiative systems to perform a variety of tasks like data wrangling~\cite{kandel-wrangler-chi-2011}, exploratory analysis~\cite{wongsuphasawat-voyager-tvcg-2016}, and natural language translation~\cite{green-ptm-uist-2014}. In a recent work, Heer~\cite{heer-agency-automation-pnas-2019} discussed how these systems use \textit{shared representations} like  text-editing interfaces and domain-specific languages through which humans and algorithms can work together to accomplish a goal. In contrast to these systems, the shared representation in Flud involves three components: (i) a network layout interface, (ii) 2D coordinates on the screen as potential actions, and (iii) a scoring scheme as a shared objective. More importantly, we use these shared representations to support mixed-initiative interaction in a novel network layout task.

Some mixed-initiative systems also leverage the complementary strengths of human intelligence and automated techniques to help users understand network data. These include a visualization tool that allows users to analyze legal citation networks with degree of interest functions~\cite{van_ham_show_2009}, and a system that helps users explore clusters of similar research papers via machine learning and visualization~\cite{chau_apolo:_2011}. These systems can be highly effective for individual expert users, but this paper considers how non-expert crowds can work with algorithms to pre-process network layouts for expert users.




Researchers have also begun exploring mixed-initiative systems that use crowdsourcing. 
For example, Flock~\cite{cheng2015flock} helps build accurate classifiers by supporting a mixed-initiative interaction where crowd nominates features and machine learning weighs them. Jellybean~\cite{sarma2015surpassing} was able to accurately count objects in images by combining results from crowds and computer vision algorithms. Ideahound~\cite{ideahound-siangliulue-uist-2016} combines human judgment with machine learning techniques to create a computational semantic model of the emerging solution space. Mobi~\cite{zhang_human_2012} demonstrated how a mixed-initiative crowdsourcing system could support trip planning by allowing a traveler to specify both quantitative global constraints (e.g., allotted time, number of activities) enforced by the software, and qualitative constraints (e.g., visiting the beach vs. the downtown) satisfied by crowd workers. Subsequent research explored these ideas for other tasks and domains, such as conference planning~\cite{kim_cobi:_2013}. Moreover, while Mobi guided worker effort through todo items, systems like Flock, Jellybean, and Ideahound used simple instructions to guide them. In contrast, we consider how gameplay mechanisms can motivate desired behavior in crowd workers. Also, Flud differs in that we adapt these ideas for a novel task type (i.e., network layout) and task domain (i.e., biology).

Yuan et al. \cite{yuan-xin-intelligent-many-users-itvcg-2012} proposed a strategy that utilized crowd workers to lay out subnetworks of a complex network. Subsequently, the authors used a constrained distance-embedding algorithm to compose the layouts of the subnetworks into one for the entire network. In contrast, our work seeks to facilitate the design process for crowd workers in order to improve the overall quality of layouts they generate. 

\subsection{Designing a game with a purpose to lay out networks}

\looseness=-1
The idea of using games to solve real-world scientific problems is not new~\cite{schrier2018designing, cooper2014framework, ferrara2013games} and has been used to solve a wide range of complex problems, such as protein folding~\cite{cooper-foldit-nature-2010}, RNA sequence design~\cite{lee-das-rna-design-rules-pnas-2014}, DNA sequence alignment~\cite{kawrykow2012phylo}, molecular design~\cite{lee-eterna-2016}, and neuron reconstruction~\cite{kim-seung-wiring-retina-nature-2014}. Inspired by the success of these games, we developed Flud, a game with a purpose that leverages the creativity and cognitive abilities of the players to create biological network visualizations, a challenge that has not previously been tackled using games. We also explore a game mechanic that differs from most prior citizen science games. While the players in most of these games compete against each other, Flud players sequentially collaborate on a layout by building on each others' progress.

\looseness=-1
Some computer games have been developed for network drawing purposes~\cite{duncan2012graph}, including the popular one-player game  Planarity~\cite{eppstein2013drawing}. Planarity (also known as UntangleManiak or The Plateau) asks the player to move nodes in order to untangle (remove) the edge crossings in the given network. CycleXing~\cite{cyclexings-bauer2012} is a two-player game with an adversarial game mechanic where one player tries to maximize the number of crossings, and the other tries to minimize them. Similar to Planarity and CycleXing, Flud asks the players to move the nodes around to manipulate layouts. In contrast, Hashi~\cite{hashi} is a single-player network drawing based game where the players draw an edge between nodes. The goal of a Hashi game is to create a single connected component while following specific rules like the edges cannot cross and should be orthogonal. While minimizing the number of edge crossings is one of our criteria, Flud players need to consider four other layout criteria as well. More importantly, none of these games asks the players to optimize the number of downward pointing paths. 



In a recent study, Hamari~\etal found that leaderboards are the most commonly studied gamification technique for motivating the players~\cite{hamari2014does}. They are generally used to help players judge their success by comparing their performance in against other players (e.g., Foldit~\cite{cooper-foldit-nature-2010}, Peekaboom~\cite{von2006peekaboom}) and even themselves (e.g., Planarity~\cite{eppstein2013drawing}). In contrast to these games, since Flud players sequentially collaborate by building on the previous best layout, we only show them the best score so far instead of showing a leaderboard. However, it might be interesting to explore the use of leaderboard to motivate the players to play multiple games by aggregating the points. 

The scoring system is another essential gaming element and is used to reward players for their performance. The scoring system generally depends on more than one component. While games like Phylo~\cite{kawrykow2012phylo} provide detailed scoring information by showcasing values for all of the components that make up the overall score, games like Foldit~\cite{cooper-foldit-nature-2010} and EyeWire~\cite{eyewire-kim2014space} only show the overall score. In Flud, our players need to consider multiple layout criteria while playing the game, and therefore, we show all of the component scores along with the overall scores. Our scoring information differs from these systems in two important ways. First, to help players understand the scoring method, we transparently show the weights of all the component scores. Second, we use visual cues like up/down arrows and green/red colors to allow players to track progress after each move.

Game modes is another gaming element used in Flud. They are commonly used in games to offer players different gameplay settings, difficulty levels, tools, rules, and even graphics. A typical example of the game mode is the choice between single-player versus multi-player setting. Expectedly, game modes are common in serious games as well. For example, Phylo has three different games modes (Story, Ribo, Phylo) with different gameplay settings including one for RNA molecules. Foldit has fives game modes: Modeless, Pull, Structure, Note, Design. Each of these modes enable the players with unique abilities, which are otherwise not available in other modes. In Flud, we adapt these ideas and use five modes where we enable players with a game visualization and clues unique to a given mode. However, our use of modes differs from other serious games in that we allocate a mode to players instead of giving them an option to choose it. Moreover, in this paper, we empirically study the best strategy to assign modes to Flud players. 

\section{Description of the Flud System}
\label{sec:system-description}
Flud is a web-based game with a purpose that allows a \emph{requester} --- a biologist seeking to visualize his or her network data --- to crowdsource the layout design task for biological network visualizations to novice game \emph{players}. Flud, as a system, has two main components: the requester interface and the game interface. We now describe how Flud assists novice \emph{players} in visualizing and laying out a network in the context of layout criteria specified by a \emph{requester}.

\subsection{Flud requester interface}
\label{sec:requester-interface}
This interface allows a requester to send a network to Flud and crowdsource the layout design task. The requester can use the interface to control parameters such as the number of players per game, layout criteria, and the number of minutes a player can play the game. We detail some of the most important parameters below. 

\subsubsection{Flud layout criteria} \label{sec:layout-criteria}
A requester can ask Flud players to optimize the network layout for five types of criteria (see Figure~\ref{fig:layout-criteria}). The first criterion is domain-inspired, while the other four are aesthetic and have been previously used as design guidelines in network drawing (also called graph drawing).

\begin{enumerate}
\item \emph{Downward pointing paths}: This guideline asks players to maximize the number of downward pointing paths. This domain-specific constraint is especially useful for analyzing the flow of information in biological networks that represent cellular signaling. 

\item \emph{Non-crossing edge pairs}~\cite{purchase-carrington-graph-aesthetics-empirical-2001,taylor-applying-2005,tamassia-batini-automatic-drawings-1988}: The goal of this guideline is to maximize the number of edge pairs that do not intersect, i.e., minimize the number of edge crossings.

\item \emph{Edge length}~\cite{taylor-applying-2005,tamassia-batini-automatic-drawings-1988}: This guideline asks players to minimize the length of the edges in the network. 

\item \emph{Node distribution}~\cite{bennett-aesthetics-2007,davidson-harel-simulated-annealing-drawing-1996,taylor-applying-2005,tamassia-batini-automatic-drawings-1988}: In this guideline, unconnected nodes should be far apart from each other. 

\item \emph{Node edge separation}~\cite{davidson-harel-simulated-annealing-drawing-1996}: This guideline seeks to create layouts where nodes are positioned away from edges. 
\end{enumerate}

Requesters may also assign priorities or weights to each layout criterion to convey their relative importance to Flud players. These priorities help  players to prioritize layout criteria in case of conflicts. A requester can also exclude a layout criterion by assigning it a priority of zero.

\subsubsection{Crowdsourcing approach} \label{sec:crowd-hybrid-approaches}
One of the challenges in creating network layouts is that different criteria may conflict with each other. Heuristics used by automated methods may compute non-optimal solutions or get stuck in local optima. 
For instance, the correct orientation of several edges in a path may be required to make it point downwards.
Therefore, it is common practice for experts (biologists in our case) to manually improve automatically generated visualizations. In Flud, we use crowdsourcing to leverage the visual and cognitive abilities of humans to observe patterns and identify solutions that escape local optima. Flud allows requesters to specify the total number of game players for a layout design task. They can also select one of two available crowdsourcing approaches: 

\begin{enumerate}
    \item \emph{Crowd}: In this approach, Flud asks a fixed number 
    of players (specified by the requester) to play the game in a sequence. In each game session, a player starts with the highest-scoring layout created so far, i.e., across all earlier sessions. During a game session, a player may create multiple layouts. Flud stores the highest-scoring layout of these. If this layout scores better than the current leader, Flud updates the best overall layout. In this fashion, players can iteratively improve upon one another's results. 
    \item \emph{Hybrid}: Here, Flud alternates sessions of gameplay between players and simulated annealing~\cite{davidson-harel-simulated-annealing-drawing-1996}. In each session, either a player or simulated annealing starts with the highest-scoring layout created so far, with the this layout updated as in the crowd-only approach. In this fashion, human players and the algorithm can iteratively improve upon one another's results. Flud allows requesters to specify the initial temperature (default = 100) and number of iterations (default = 500) in a given simulated annealing session. 

\end{enumerate}

\subsection{Flud game interface}
\label{subsec:game_interface}

\begin{figure}
\includegraphics[width=\textwidth, keepaspectratio, clip, trim=0.6in 0in 0 1mm]{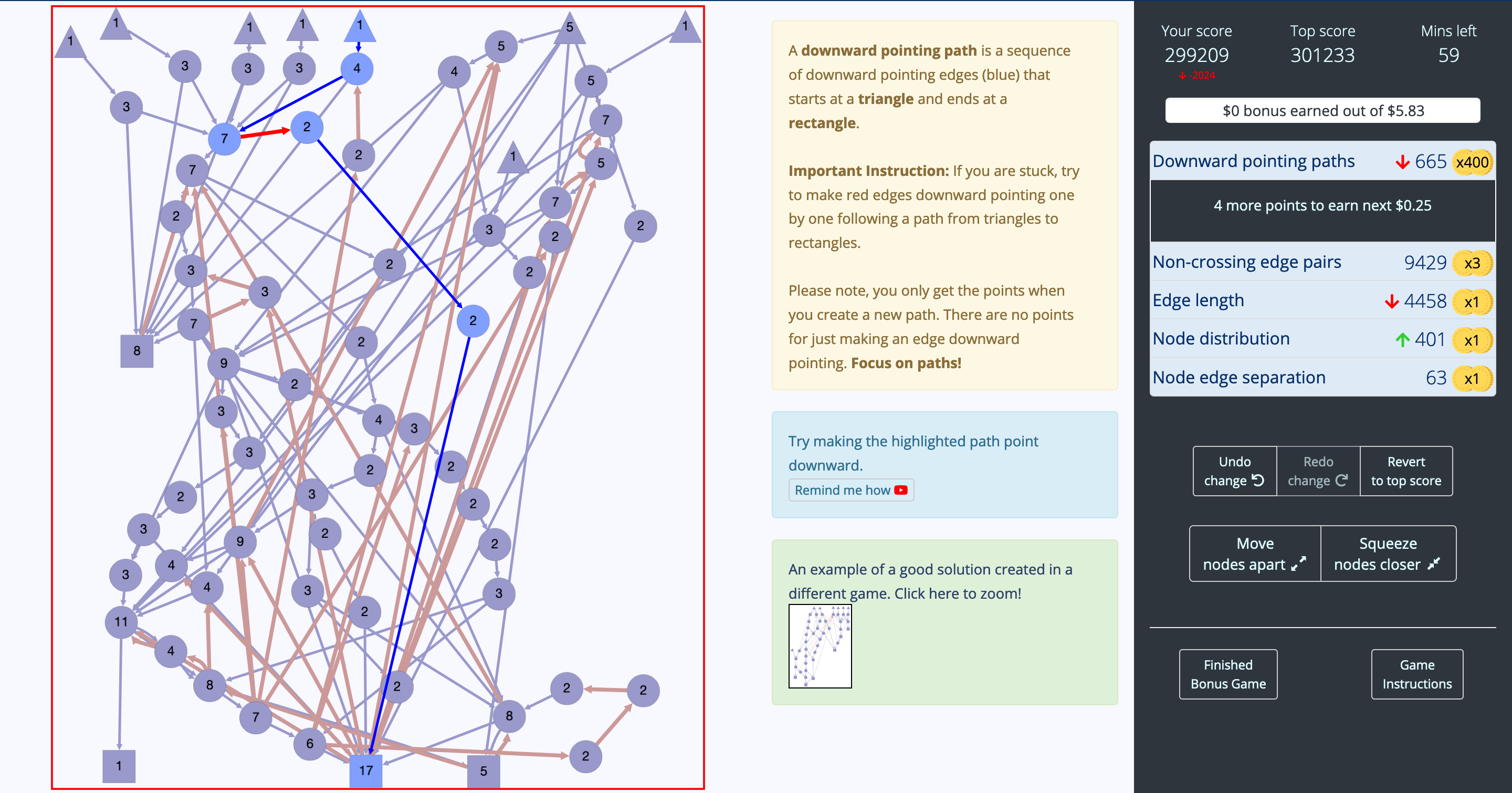}
  \caption{Screenshot of Flud's game interface in the ``Downward Pointing Paths'' mode. The interface has three major parts: a visualization of the network being laid out (left),  text panels with game-related tips (middle), and a sidebar with game controls 
  and score information (right). The upward pointing edges are shown in red to draw the player's attention. The path highlighted in saturated colors is an automatically generated hint. Reorienting the red edge to make it point downward will increase the number of downward pointing paths. }
\label{fig:flud-screenshot-interface}
\end{figure}

The game interface has three major parts: a visualization of the network being laid out on the left, text panels with game-related tips in the middle, and a sidebar with game controls and score information on the right (Figure~\ref{fig:flud-screenshot-interface}). 

\subsubsection{Network visualization} Flud provides an interactive network visualization that supports both touch and mouse gestures to select and drag one or more nodes. Since Flud players do not need biological expertise, they do not see any node or any edge labels describing their biological meaning. Instead, each node displays the number of edges incident on it as a signal of that node's importance. 

\subsubsection{Scores}
The top portion of the sidebar displays the highest score achieved by a layout for the network, the score of the current layout, and a criterion-specific score board. 
The score board displays the scores and priorities (in coin-shaped badges) for each individual criterion. In the score board, we sort the criteria in decreasing order of the priorities assigned by the requester.
We scale the per-criterion scores between 0 and 10,000 to avoid displaying floating point values. After the player makes a move, Flud recalculates and displays, in real time, each per-criterion score as well as the total score. A green up arrow ($\uparrow$) or a red down arrow ($\downarrow$) next to each per-criterion score 
allows players to track the impact of their last move on the score. In addition, we display the change in the overall score with a similar color scheme.


\subsubsection{Criterion-specific modes and clues} 
\begin{figure}[htbp]
  \subcaptionbox{Mode for downward pointing paths. \label{fig:flud-screenshots-criteria-downward-mode}}[.40\linewidth][c]{%
    \includegraphics[width=0.40\textwidth,keepaspectratio,clip,trim=0.71in 0.1in 10.73in 2.12in,frame]{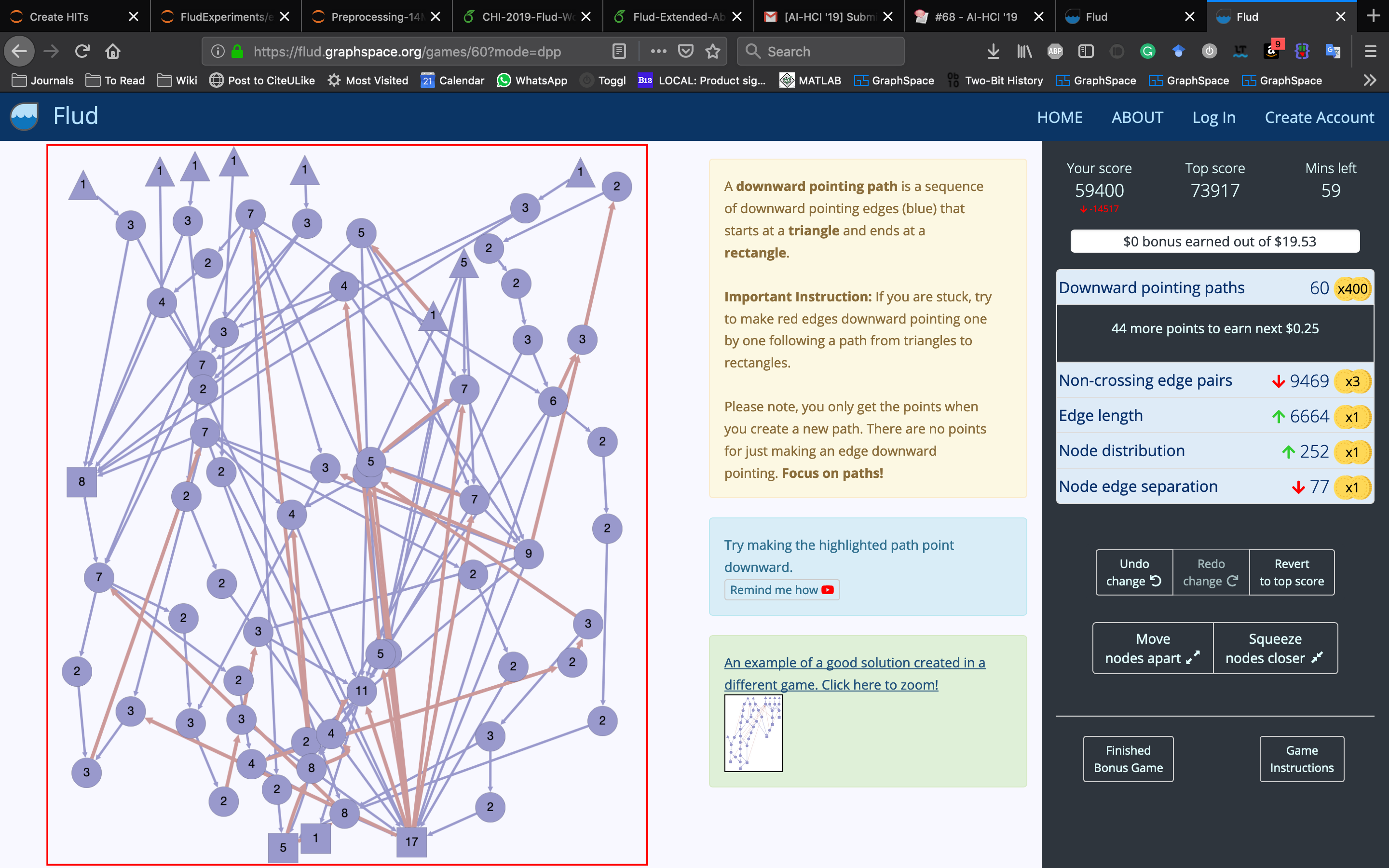}}
  \quad
  \subcaptionbox{Clue for downward pointing paths. \label{fig:flud-screenshots-criteria-downward-clue}}[.40\linewidth][c]{%
    \includegraphics[width=0.40\textwidth,keepaspectratio,clip,trim=0.7in 0.1in 10.73in 2.12in,frame]{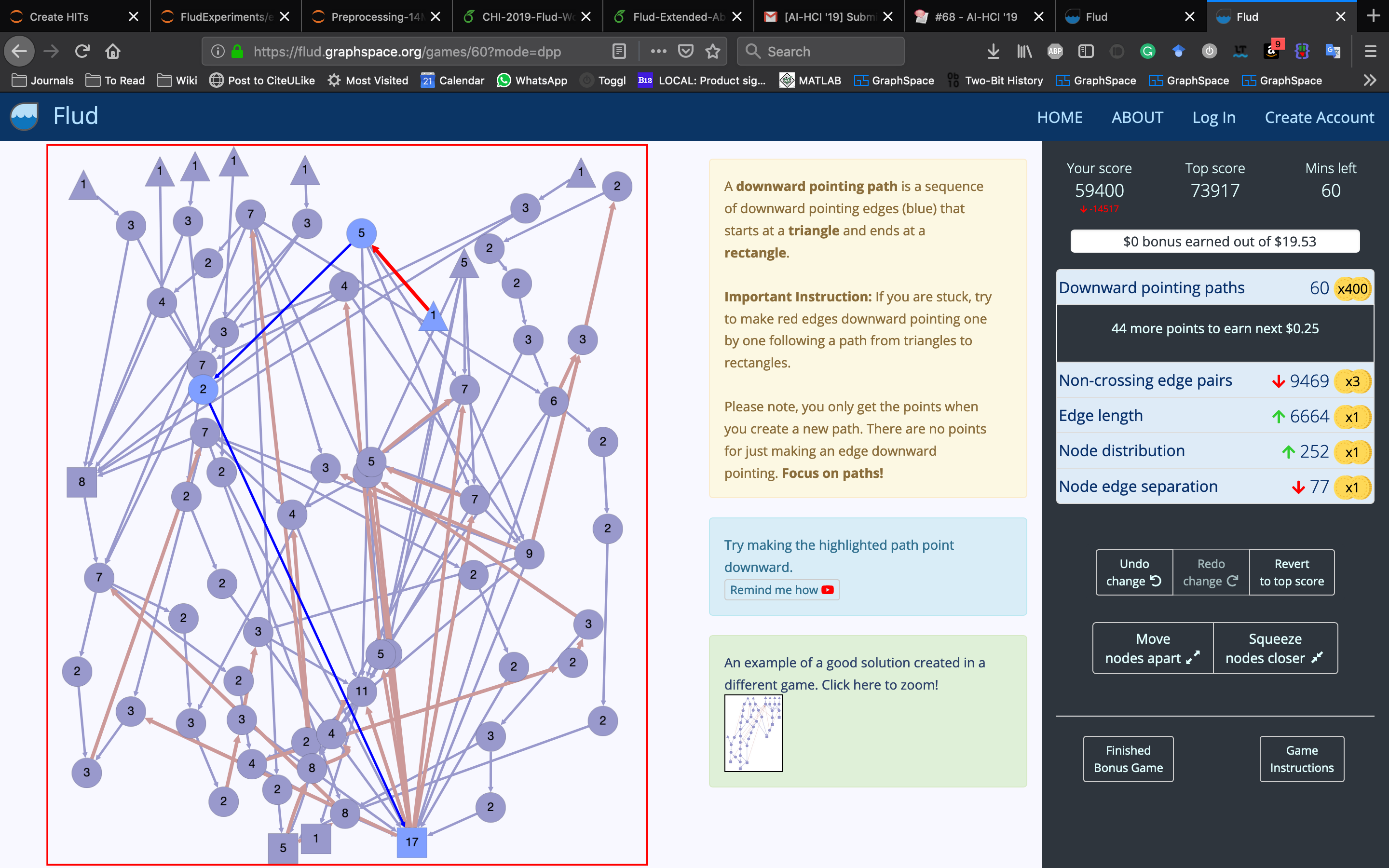}}

  \subcaptionbox{Mode for edge crossings.  \label{fig:flud-screenshots-criteria-crossings-mode}}[.40\linewidth][c]{%
    \includegraphics[width=0.40\textwidth,keepaspectratio,clip,trim=0.71in 0.1in 10.73in 2.12in,frame]{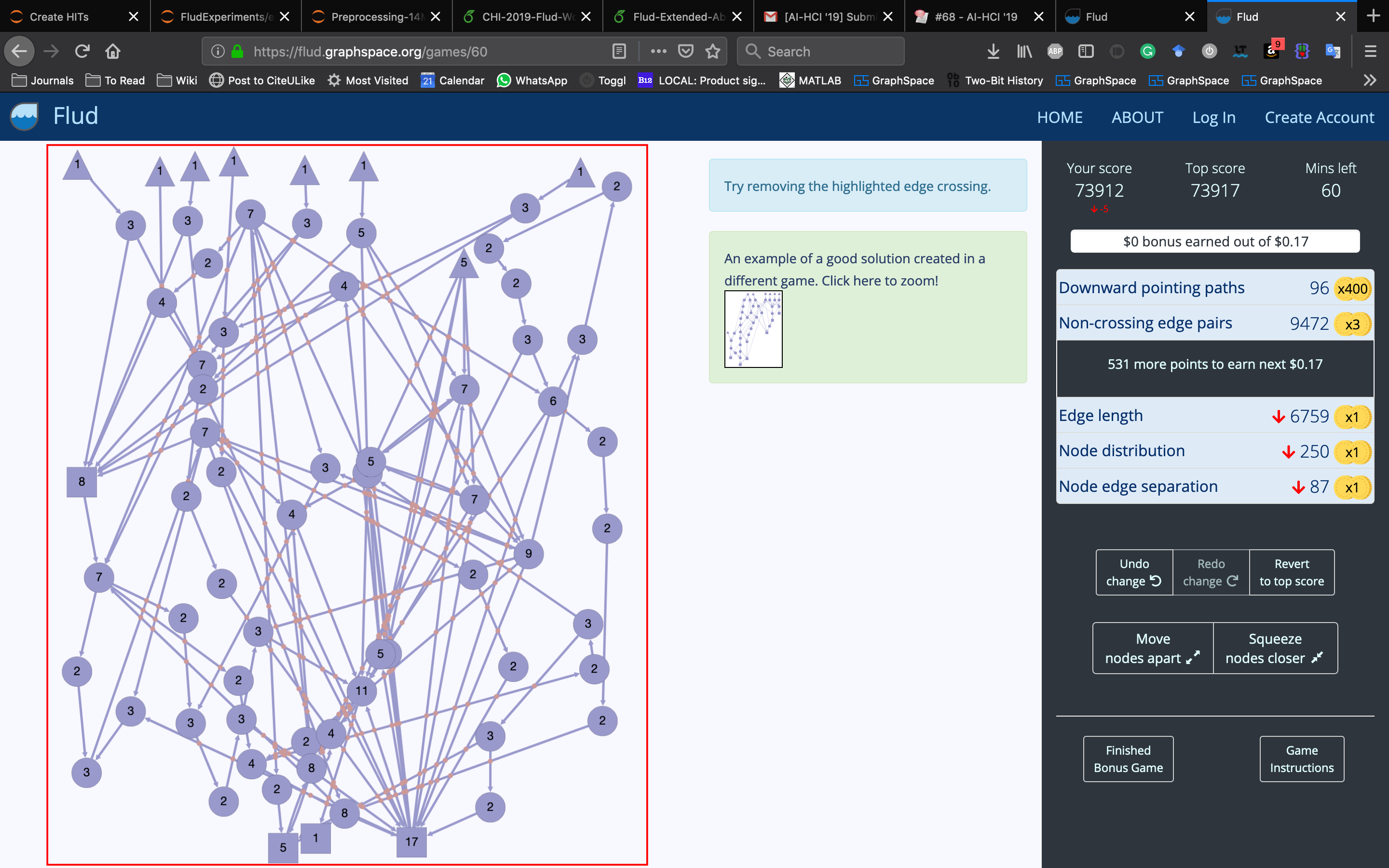}}
    \quad
  \subcaptionbox{Clue for edge crossings. \label{fig:flud-screenshots-criteria-crossings-clue}}[.40\linewidth][c]{%
    \includegraphics[width=0.40\textwidth,keepaspectratio,clip,trim=0.71in 0.1in 10.73in 2.12in,frame]{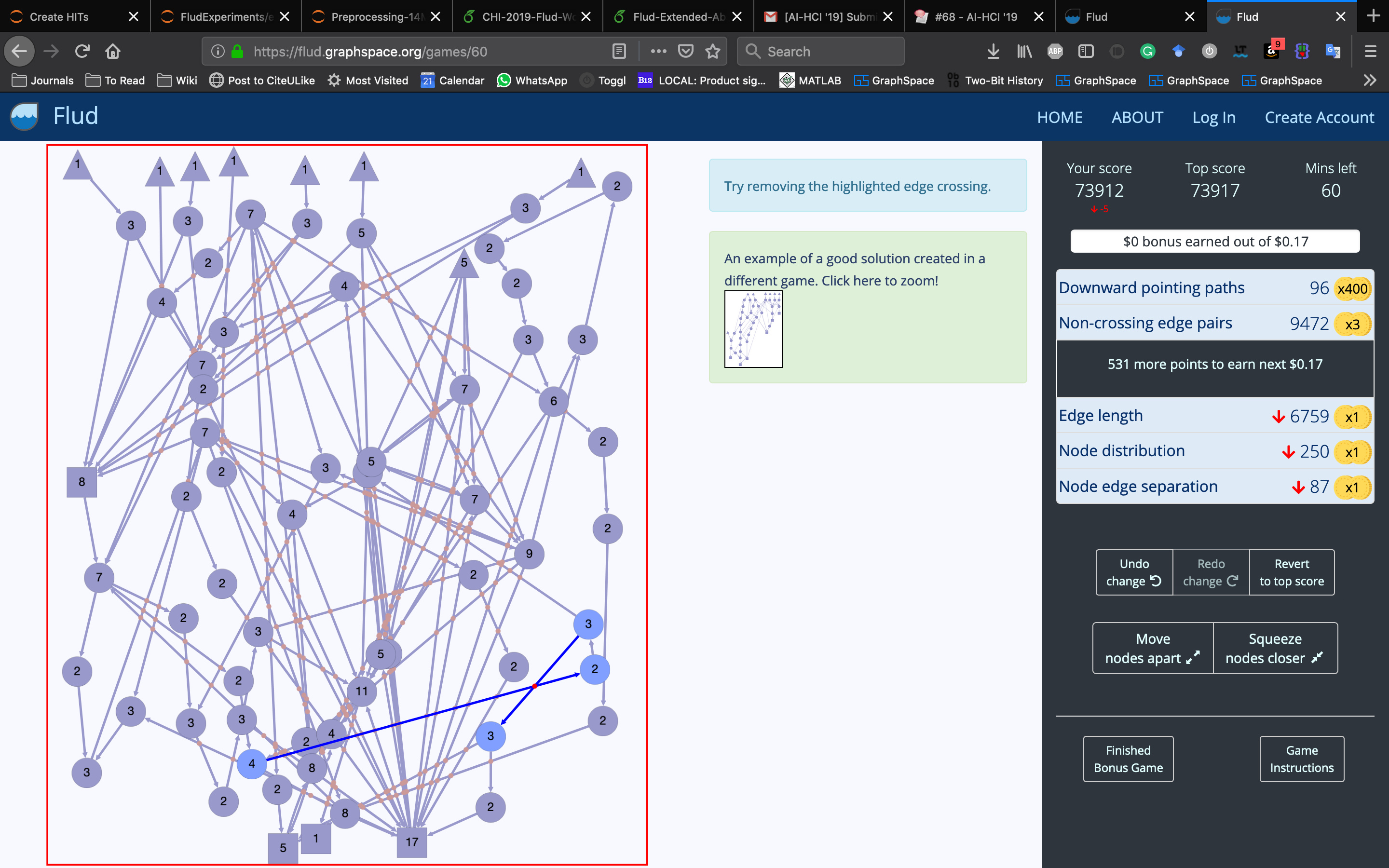}}
    
  \caption{Screenshots of the network visualization and the clue in the ``Downward pointing paths'' and ``Non-crossing edge pairs'' modes. }
  \label{fig:flud-screenshots-criteria-dpp-ec}
\end{figure}

\begin{figure}[htbp]
  \subcaptionbox{Clue for edge length. \label{fig:flud-screenshots-criteria-connected-node-pairs-clue}}[.31\linewidth][c]{%
    \includegraphics[width=0.31\textwidth,keepaspectratio,clip,trim=0.71in 0.1in 10.73in 2.12in, frame]{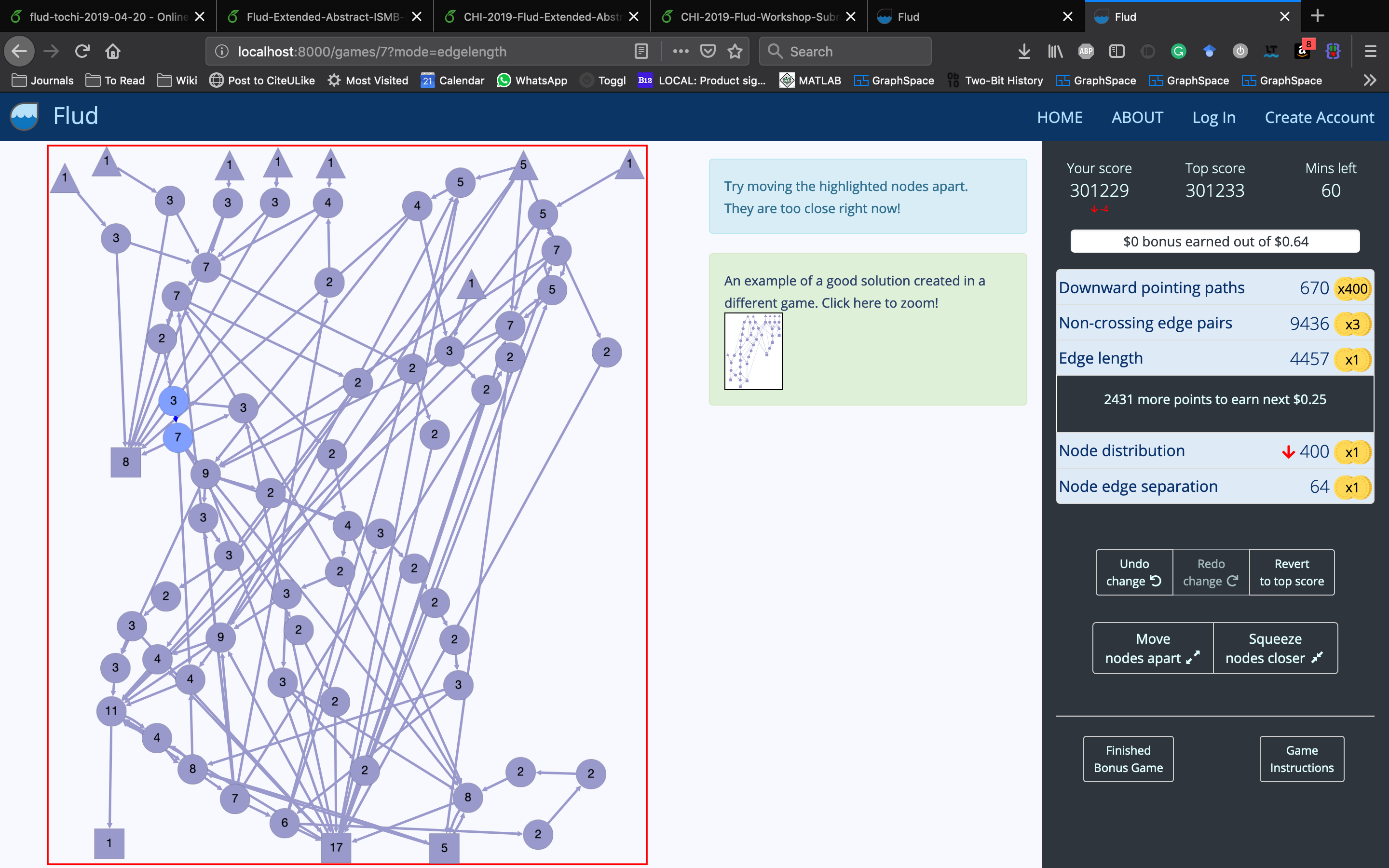}}
    \quad
   \subcaptionbox{Clue for node distribution. \label{fig:flud-screenshots-criteria-node-distribution-clue}}[.31\linewidth][c]{%
    \includegraphics[width=0.31\textwidth,keepaspectratio,clip,trim=0.72in 0.1in 10.73in 2.13in,frame]{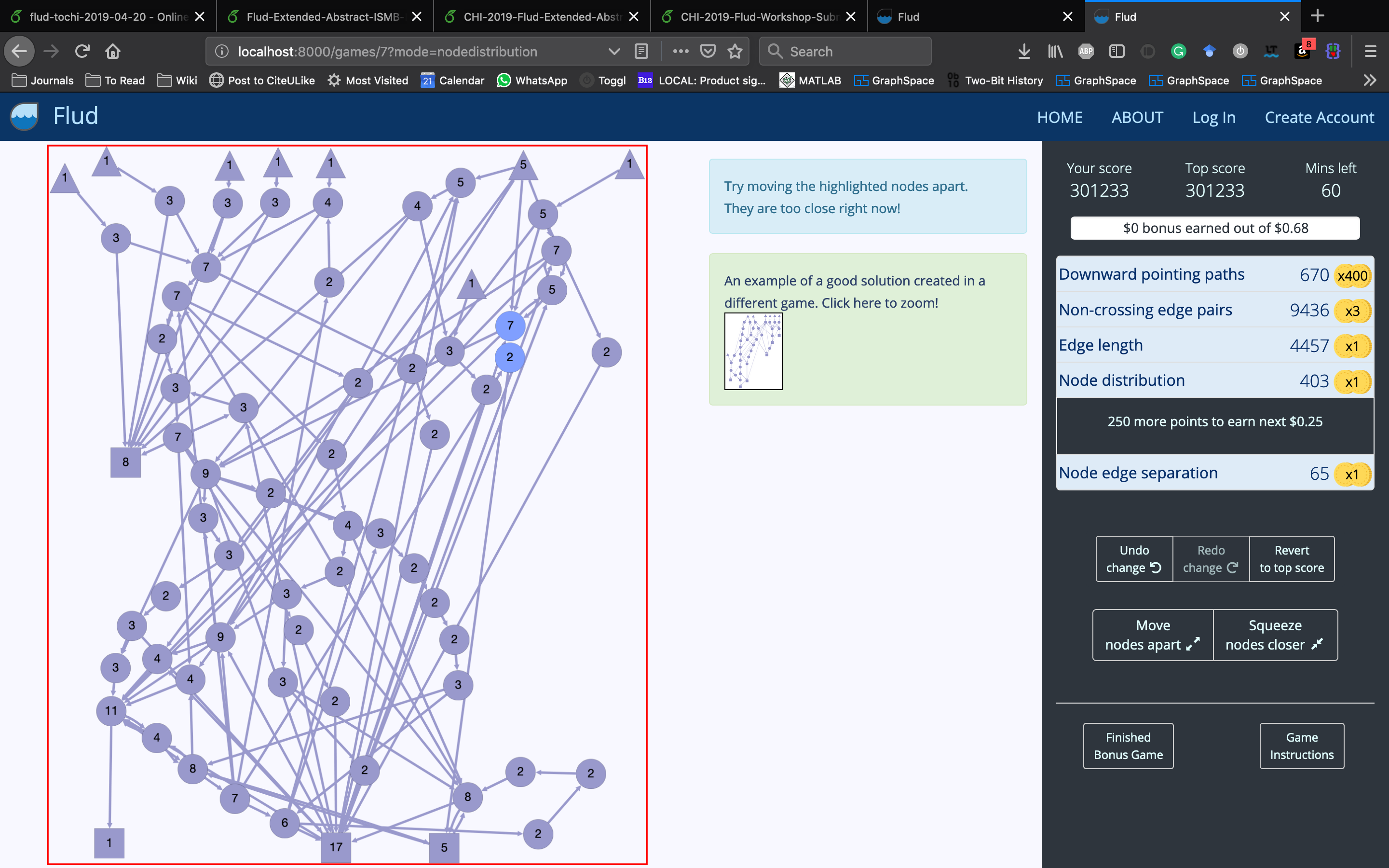}}
    \quad
   \subcaptionbox{Clue for node edge separation. \label{fig:flud-screenshots-criteria-node-edge-separation-clue}}[.31\linewidth][c]{%
    \includegraphics[width=0.31\textwidth,keepaspectratio,clip,trim=0.72in 0.1in 10.73in 2.13in,frame]{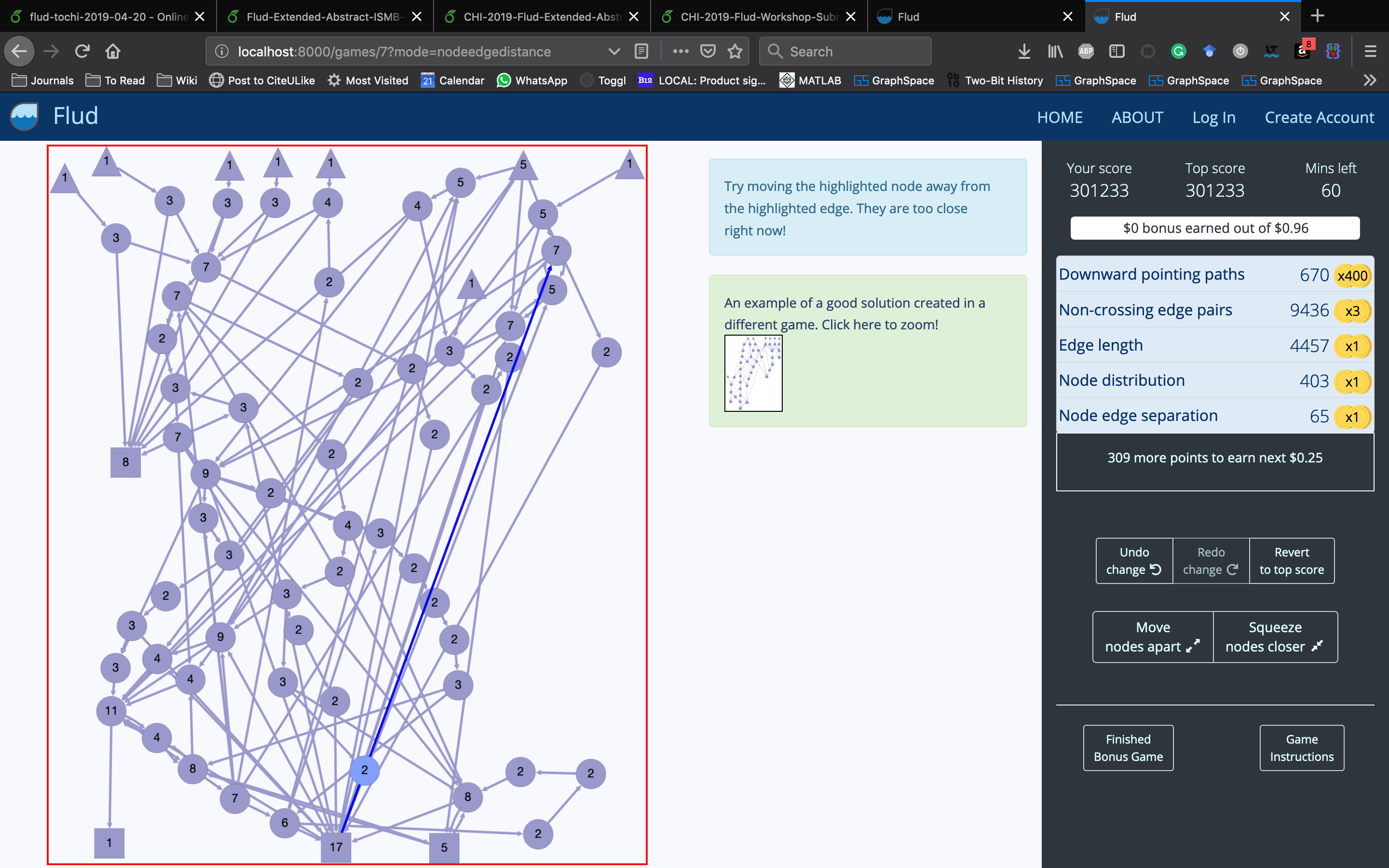}}
   
  \caption{Screenshots of the network visualization in Edge length, Node distribution, and Node edge separation mode. }
  \label{fig:flud-screenshots-criteria-el-nd-ned}
\end{figure}

Csikszentmihalyi~\cite{csikszentmihalyi1990flow} argues that a user can achieve \textit{flow}, a state of being ``in the zone,'' if they attain multiple component states.
These states include challenge-skill level balance, immediate and unambiguous feedback from the system, concentration on the task at hand, and clarity of goals. Inspired by flow theory, which has been influential in game design~\cite{Cowley-Darry-Flow-Theory-Games-2008}, we implemented two important gameplay features --- criterion-specific modes and clues (Figures~\ref{fig:flud-screenshots-criteria-dpp-ec} and~\ref{fig:flud-screenshots-criteria-el-nd-ned}) --- to help players focus on their goals and make progress without becoming bored or frustrated. At the start of a game session, the player is assigned a single criterion-specific mode. In this mode, to delineate the task, the visual representation of the network highlights (in red) the elements that are relevant to the criterion-specific task. Flud has five such modes: Downward pointing paths, Non-crossing edge pairs, Edge length, Node distribution, and Node edge separation. 

Moreover, if the player is stuck while playing, Flud reduces the challenge level by presenting them with an algorithmically-generated, mode-specific ``clue'' that highlights a small subset of nodes and edges in the network and further narrows the focus of a player to a very specific task \cite{von-dabbish-gwap-2008}. Flud selects specific elements in the clue such that changing their positions is likely to improve the score for the corresponding criterion. We now describe each criterion-specific mode and the corresponding clue.

\begin{figure}[htbp]
 \centering
  \includegraphics[width=0.5\textwidth]{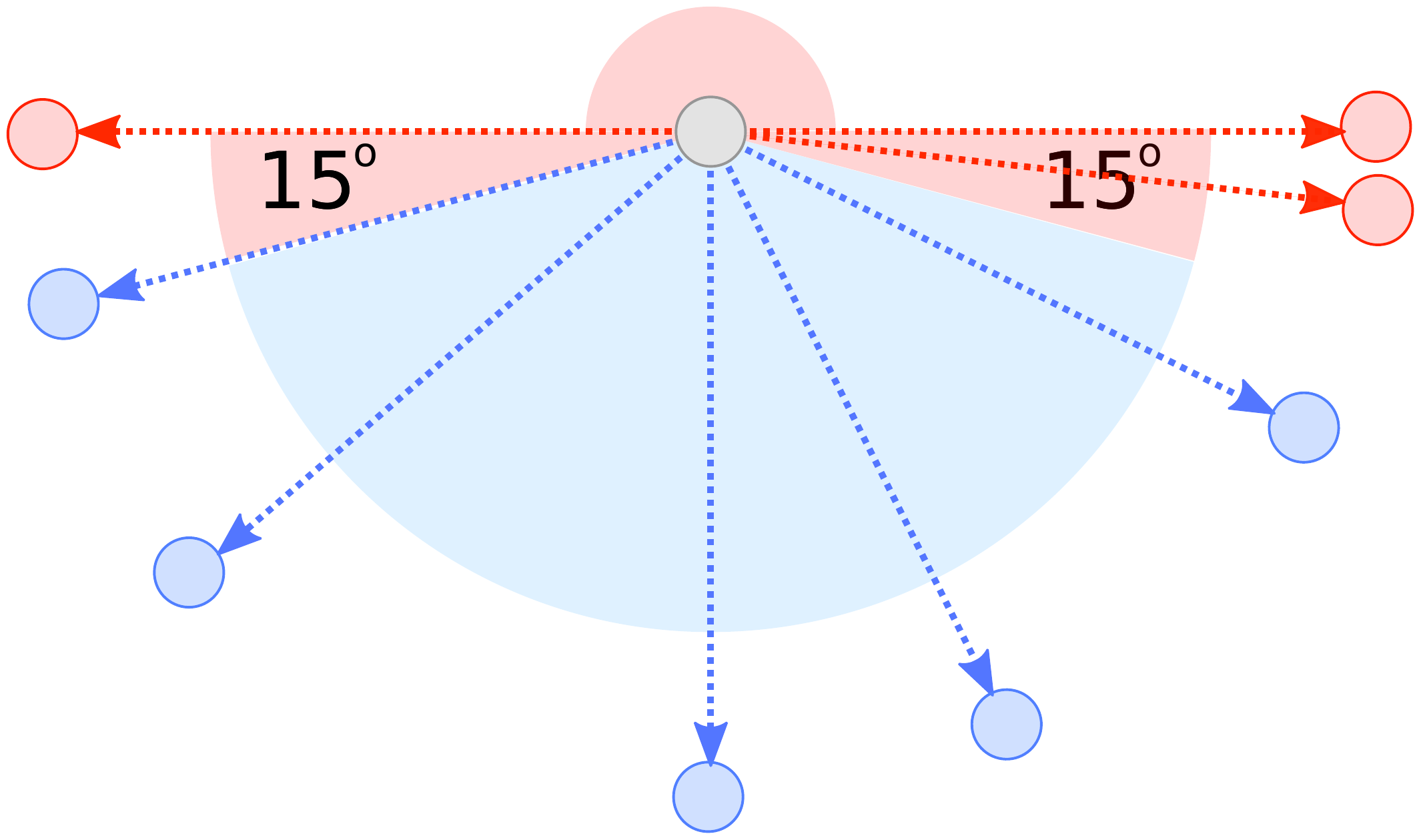}
  \caption[Illustration of downward and upward pointing edges.]{Illustration of downward and upward pointing edges. A directed edge is \emph{downward pointing} if the $y$-coordinate of its head is below the $y$-coordinate of the tail and angle between the edge and the $x$-axis is greater than equal to 15 degrees. The blue and red regions show the angles at which a directed edge will be considered as downward pointing and upward pointing, respectively. }
  \label{fig:downward-edge}
 \end{figure}

\begin{figure}[htbp]
 \centering
  \includegraphics[width=\textwidth]{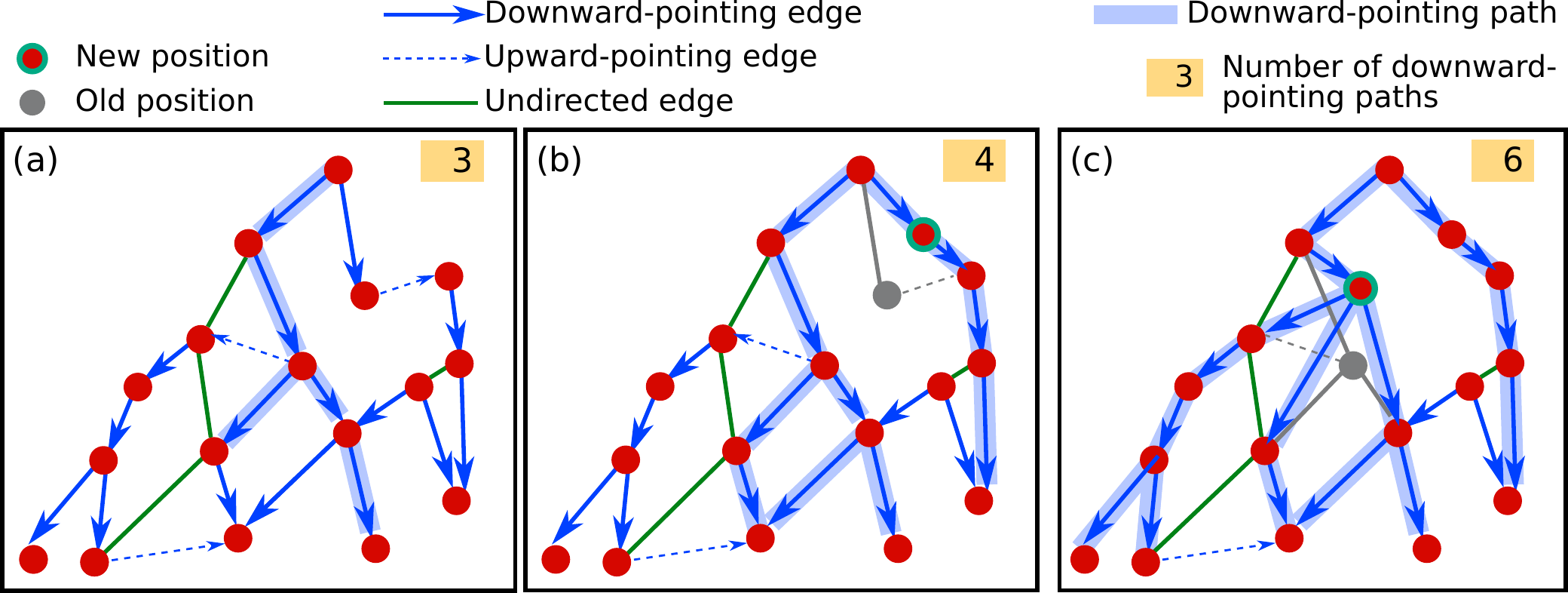}
  \caption[Illustration of downward-pointing paths.]{Downward-pointing paths. (a) Three paths. (b)  Moving the node with a green border increases the number of paths to four. (c) Another move increases the number to six. }
  \label{fig:downward-paths}
 \end{figure}

\paragraph{(i) Downward pointing paths:} We say that an individual directed edge is \emph{downward pointing} if the $y$-coordinate of its head is below the $y$-coordinate of the tail and angle between the edge and the $x$-axis is greater than equal to a fixed degree (15 degrees, in our implementation; Figure~\ref{fig:downward-edge}). A path is \emph{downward pointing} if all the edges in it have this property (Figure~\ref{fig:downward-paths}). 

In the downward pointing paths mode, Flud highlights every upward pointing edge in red (Figure~\ref{fig:flud-screenshots-criteria-downward-mode}). A player can reorient such an edge in an attempt to increase the number of downward pointing paths. A clue in this mode (Figure~\ref{fig:flud-screenshots-criteria-downward-clue}) highlights a path in the network that (a) contains at least one upward pointing edge, and (b) has the property that reorienting all upward-pointing edges in the path is guaranteed to increase the number of downward pointing paths. If there is no such path, the player does not see any clue. 

\paragraph{(ii) Non-crossing edge pairs:} This mode displays every edge crossing as a red point (Figure~\ref{fig:flud-screenshots-criteria-crossings-mode}). When a player hovers on an edge crossing, the mode highlights the intersecting edges. Decreasing the number of edge crossings is perhaps one of the most challenging layout criteria since the player has to manipulate nodes without introducing new intersections. Intuitively, the higher the degree of a node, the harder it may be to remove intersections involving edges incident on that node. Therefore, the clue in this mode highlights a pair of crossing edges such that the total degree of the four nodes involved is the smallest over all the intersections in the layout (Figure~\ref{fig:flud-screenshots-criteria-crossings-clue}). Note that we cannot guarantee that the player will indeed be able to decrease the number of crossings by moving one or more of these nodes. 

    

\paragraph{(iii) Edge length:} For this criterion, players need to consider the length of all the edges in the network. Therefore, the mode presents all the nodes and edges in blue. 
 The clue highlights an edge that is either very long or very short (Figure~\ref{fig:flud-screenshots-criteria-connected-node-pairs-clue}). Correcting the length of this edge should lead to the improvement of the edge length score. 

\paragraph{(iv) Node distribution:} 
In this mode, the clue highlights the unconnected pair of nodes that are the closest to each other (Figure~\ref{fig:flud-screenshots-criteria-node-distribution-clue}).  Thus, the clue provides a greedy way for players to consider node pairs that need to be moved further apart. 

    

\paragraph{(v) Node edge separation:} 
Here, the clue highlights the node-edge pair that is the closest to each other (Figure~\ref{fig:flud-screenshots-criteria-node-edge-separation-clue}).

    

\subsubsection{Layout controls} The bottom part of the sidebar contains multiple  controls that allow players to rapidly perform non-trivial changes to the layout. These changes include expanding and squeezing the spacing between selected nodes, undoing and redoing earlier actions, and reverting to the layout with the best score. 

\subsubsection{Game tips} The middle part of the interface contains text panels with game related tips. These tips include an example of a good layout for a toy 
network, a link to the tutorial on how to use criterion-specific clues, and specific instructions on how to improve the per-criterion score for the assigned mode. 

\subsubsection{Bounding box}  To prevent players from creating pathological layouts that move nodes indiscriminately far apart from each other, we set the scores for each criterion to zero if any node is moved outside a bounding box of fixed size. 


\subsection{Flud scoring system} \label{sec:flud-scoring-system}




In describing the scoring system, we use $G=(V, E)$ to denote the network $G$ being laid out with $V$ denoting its nodes and $E$ denoting its edges. We use $n$ to denote the number of nodes and $m$ for the number of edges in $G$. Given the $x$- and $y$-coordinates for every node in $V$, we use $d(u,v)$ to denote the Euclidean distance between nodes $u$ and $v$ and $l(e)$ to denote the length of an edge $e$ (the Euclidean distance between its endpoints). We use $w$ and $h$ to represent the fixed width and height of the bounding box. We now describe how we computed the score corresponding to each criterion (per-criterion score).


\begin{enumerate}
\item \emph{Downward pointing paths}: We compute a normalized \emph{downward pointing} score as follows: 
$$\mathit{DP}(G) = \frac{\pi(G)}{\rho(G)}.$$

where $\pi(v)$ is the number of downward-pointing paths and $\rho(v)$ is the approximated upper bound of the number of downward-pointing paths. The closer $DP(G)$ is to one, the larger the number of downward-pointing paths in the layout.

\item \emph{Non-crossing edge pairs}:  Since the maximum number of non-crossing edge pairs possible in $G$ is $m(m-1)/2$, we compute the \emph{non-crossing edge pairs score} as  
$$ \mathit{EC}(G) = \frac{2\chi(G)}{m(m-1)}.$$
The closer $EC(G)$ is to one, the smaller is the number of edge crossings in the layout. We have intentionally defined $\mathit{EC}(G)$ as above so that high values reflect good performance and players focus on increasing their scores. 

\item \emph{Edge length}:  We define the cost $c(e)$ of an edge to be equal to its length $l(e)$ if $l(e) \geq $ size of a node or equal to a large penalty, otherwise.
We normalize the cost of each edge by the largest possible edge length and therefore, penalty should be greater the diagonal of the screen. We compute the \emph{edge length score} of a layout as
 $$\mathit{EL}(G) = \max \Big\{0, 1 - \frac{1}{m}\sum_{e \in E} \frac{c(e)}{\sqrt{w^2 + h^2}} \Big\}$$ 
Thus, the closer $EL(G)$ is to one, the closer the nodes connected by an edge are, on average. 

\item \emph{Node distribution}: Here, we want to maximize the average distance in the layout between a node and its closest unconnected node. Unlike the previous criterion, it applies only to pairs of nodes not connected by an edge. We compute the \emph{node distribution score} as
$$\mathit{ND}(G) = \frac{1}{n}  \mathlarger{\sum}\limits_{u \in V} \frac{\min\limits_{\substack{v \in V, (u,v) \notin E}} d(u,v)}{\sqrt{w^2 + h^2}},$$

\item \emph{Node edge separation}: In this criterion, we want to maximize the average distance between a node and the closest edge to it in the layout. Therefore, we define the \emph{node edge separation score} as 
$$\mathit{NED}(G) = \frac{1}{n}  \mathlarger{\sum}\limits_{u \in V} \frac{\min\limits_{\substack{e=\{t,v\} \in E \\ u \neq t \neq v}} d(u,e)}{\sqrt{w^2 + h^2}}$$
where $d(v,e)$ denotes the distance between node $v$ and edge $e$. 
\end{enumerate}

We define the overall score $\mathit{OS}(G)$ for a layout as a weighted sum of the five per criterion scores 
$$\mathit{OS}(G) = w_{\mathit{DP}} \times \mathit{DP}(G) + w_{\mathit{EC}} \times \mathit{EC}(G) + w_{\mathit{EL}} \times \mathit{EL}(G) + w_{\mathit{ND}} \times \mathit{ND}(G) + w_{\mathit{NED}} \times \mathit{NED}(G),$$ 
where the coefficients are the priorities assigned to each criterion by the requester. We describe the algorithm and parameters used to compute the per criterion scores in Appendix.


\subsection{Simulated Annealing}
We use simulated annealing~\cite{davidson-harel-simulated-annealing-drawing-1996} as the primary baseline for comparison and as the algorithm in the hybrid approach. We selected this method primarily  due to its flexibility to accommodate all of our criteria (Section \ref{sec:layout-criteria}).  We run the algorithm for a fixed number of iterations (e.g., 500) with the temperature decreasing by a factor of 0.995 after every iteration. Each iteration involves $10n$ steps, where $n$ is the number of nodes in the network. In every step, we move a node to a random position $(x_{\mathit{new}}, y_{\mathit{new}})$ computed as follows from the current position $(x_{\mathit{current}}, y_{\mathit{current}})$: 
$$x_{new} = x_{current} + \rho_x \frac{1}{4}\bigg(\frac{T}{T_\mathit{max}}\bigg)^2 w$$
$$y_{new} = y_{current} + \rho_y\frac{1}{4}\bigg(\frac{T}{T_\mathit{max}}\bigg)^2 h$$
where $T_\mathit{max}=100$ and $T$ represent the maximum and current temperatures, respectively, and we draw $\rho_x$ and $\rho_y$ uniformly at random from the interval $[0, 1]$. Thus, we select the new position uniformly at random from a rectangle centered at the current location of the node. The size of this rectangle is proportional to the fixed page size and decreases quadratically with the temperature.  
We compute the score of the layout with this new position. We accept the move if it improves the score. Otherwise, we accept this move, even though it worsens the score, with a probability  $\Pr(\Delta s) = e^{-\frac{\Delta s}{T}}$, where $\Delta s$ is the change in score. Therefore, as the temperature decreases, so does the probability of accepting a move that worsens the score. Once the algorithm stops, we use the best layout in the entire run as the final layout. Overall, the annealing schedule took approximately 1 hour to cool down from high initial temperature ($T=T_0=100$) to low temperature ($T\approx1$). 

\subsection{Implementation details}
\label{sec:flud-implementation-details}
We implemented Flud in Python using the Django web framework, with network visualization supported by Cytoscape.js \cite{franz-bader-cytoscapejs-bioinfo-2016}. Flud interfaces with GraphSpace~\cite{bharadwaj-murali-graphspace-bioinfo-2017} for storing and sharing networks and their layouts. To create a game of Flud, a requester interface shares the networks posted by the requester to a public group called `Flud' on GraphSpace. The Flud system pings GraphSpace for new networks in `Flud' group and makes it available on the Flud website for a player to lay out. When a player selects a graph, Flud presents the highest-scoring layout so far for that graph. During a game session, a player may create multiple layouts to maximize their score, and Flud stores the highest-scoring layout. If this layout scores better than the current leader, Flud updates GraphSpace with this layout. In this fashion, players can iteratively improve upon one another's results. Finally, the requesters can access all of the crowd-generated layouts, including the best overall layout on GraphSpace.




\section{Evaluation}
\label{sec:evaluation}



Having implemented the Flud system as described above, we aimed to showcase its effectiveness to crowdsource the layout design task for biological network visualizations. To this end, we posed the following research questions:

\begin{itemize}
    
    \item[\emph{RQ1:}] \emph{How should Flud assign criterion-specific modes to players to optimize the scores they achieve?} 
    \item[\emph{RQ2:}] \emph{How do the crowd and hybrid approaches perform in comparison to automated methods?} 
    
    \item[\emph{RQ3:}] \emph{What are the dynamics of the mixed-initiative collaboration in the hybrid approach?}  
    
\end{itemize}
In the rest of this section, we describe how we selected crowd workers, the networks we laid out, the task each worker had to perform, and how we compensated them before presenting our experiment design. 

\subsection{Networks} \label{subsec:networks}
We selected three different complex protein networks of similar size (Table~\ref{table:graph-structure}) that represent signaling pathways in cells. These networks contained both directed and undirected edges. These networks contained a small (G1), medium (G2), and high (G3) number of cycles. The presence of a large number of cycles makes it difficult to create layouts with many downward-pointing paths. Hence, these networks should present different levels of challenges to crowd workers and to automated algorithms.  

\begin{table}[h]
\resizebox{\textwidth}{!}{%
\begin{tabular}{|c|l|c|c|c|c|}
\hline
\textbf{Network} & \multicolumn{1}{c|}{\textbf{Description}}    & \textbf{\#Nodes} & \textbf{\#Edges} & \textbf{\#Simple Cycles} & \textbf{\begin{tabular}[c]{@{}c@{}}\#Simple Paths from \\ sources to targets\end{tabular}} \\ \hline
G1             & \begin{tabular}[c]{@{}l@{}}Crosstalk from Estrogen signaling\\ pathway to HIF-1 signaling pathway\end{tabular}                                                                                                                                                          & 71              & 112             & 3                       & 1952                                                                                      \\  \hline
G2             & \begin{tabular}[c]{@{}l@{}}Network representing Epithelial-\\to-mesenchymal transition (EMT)\end{tabular} & 69              & 131             & 8463                    & 10066                                                                                     \\ 
\hline
G3             & \begin{tabular}[c]{@{}l@{}}Signaling from SCH9 to TPK2\\ in budding yeast 
\end{tabular}                                                                                                                                                                       & 58              & 136             & 269437                  & 2361                                                                                      \\ \hline
\end{tabular}%
}
\caption{Networks used for evaluation.} 
\label{table:graph-structure}
\end{table}


\subsection{Crowd workers} \label{subsec:crowd-workers}
To showcase that Flud can support broad player bases, we did not require the participants to have network or biology expertise. Therefore, we recruited novice crowd workers from the Amazon Mechanical Turk (MTurk) platform as our game players. We used MTurk's built-in qualification types to only recruit workers from the US with a Human Intelligence Task (HIT) approval rate of at least 97\% and at least 100 completed HITs. 

In a pilot study where we asked crowd workers to play the game, we noticed that some players (a) did not carefully go through the tutorial and (b) moved the nodes aimlessly to use up the fixed number of moves required to be paid for completing the game. In order to prevent such unproductive crowd activity, we only invite crowd workers to play the real game who correctly solved two small puzzles towards the end of the tutorial. The first puzzle asks the crowd worker to increase one of the per-criterion scores in a toy network by at least one point with the help of a clue. The second puzzle again asks the worker to increase the per criterion score for the same network except that there is no clue available. To avoid learning effects, we recruited the crowd workers such that no individual repeated the same network or criterion-specific task.


\subsection{Crowd worker task} \label{subsec:crowd-worker-task}
We start the task for a crowd worker by randomly selecting a game corresponding to one of the three networks. Next, we assign them one of the layout criteria (using one of the strategies described in Section~\ref{subsec:experiment-1}) to restrict their gameplay to a criterion-specific mode. We ask the crowd worker to go through a two-part interactive tutorial that introduces the visual elements, game rules, and use of the criterion specific clues. 
In the first part of the tutorial, we train the worker to improve the criterion-specific subscore both with and without the help of a clue. At the end of part one of the tutorial, we give the worker an option to submit the HIT or to continue with part two of the tutorial and play the game to earn a bonus (Refer to Section~\ref{subsec:compensation}). The second part of the tutorial introduces the worker to the remaining the subscores and the complete rules of the game. During the game session, each worker had one hour to generate an improved layout. A player can quit the game at any point.
Irrespective of the assigned mode, the crowd worker's goal is to create a layout that optimizes the total score based on the layout criteria (see Section~\ref{sec:layout-criteria}).


\subsection{Compensation} \label{subsec:compensation}
For the first half of the tutorial, we compensate the workers with at least the minimum hourly wage rate (\$7.25 per hour) in our region. 
If a worker elects to continue playing Flud in order to obtain bonus compensation, we pay this amount according to how they increase the score of the layout. We use this strategy to motivate a worker to fully utilize the time available to them.

We assign budgets to each criterion in proportion to its priorities. These budgets determine the amount of bonus a player can earn while playing the game for the assigned criterion-specific mode. The bonus amount earned when a player improves the score from $s_i$ to $s_j$ is $$(b+1)^{\frac{\min\{s_\mathit{j},s_{\mathit{target}}\}}{s_{\mathit{target}}}} - (b+1)^{\frac{\min\{s_\mathit{i},s_{\mathit{target}}\}}{s_{\mathit{target}}}},$$
where $s_j > s_i$, $b$ is the budget assigned to the given criterion and $s_{\mathit{target}}$ is the target score we want the players to achieve. 
The exponential nature of the bonus computation ensures that the workers earn more money per point as they approach the target score. The goal is to motivate the workers to continue playing the game, even though the task of improving the score gets harder as the player gets closer to the target score.
Ideally, we should aim to set $s_{\mathit{target}}$ to an achievable value in order to ensure that the bonus is fair. For example, it is not possible for a player to achieve a score 10,000 (maximum possible score) if the network has many cycles. Therefore, we suggest using an achievable target (say, 2,000). Once the workers achieve the target score, the requester has the option to increase the budget and the desired score.

\subsection{Experiment design} \label{subsec:experiment-design}

\begin{figure}[htbp]
 \centering
  \includegraphics[width=\textwidth]{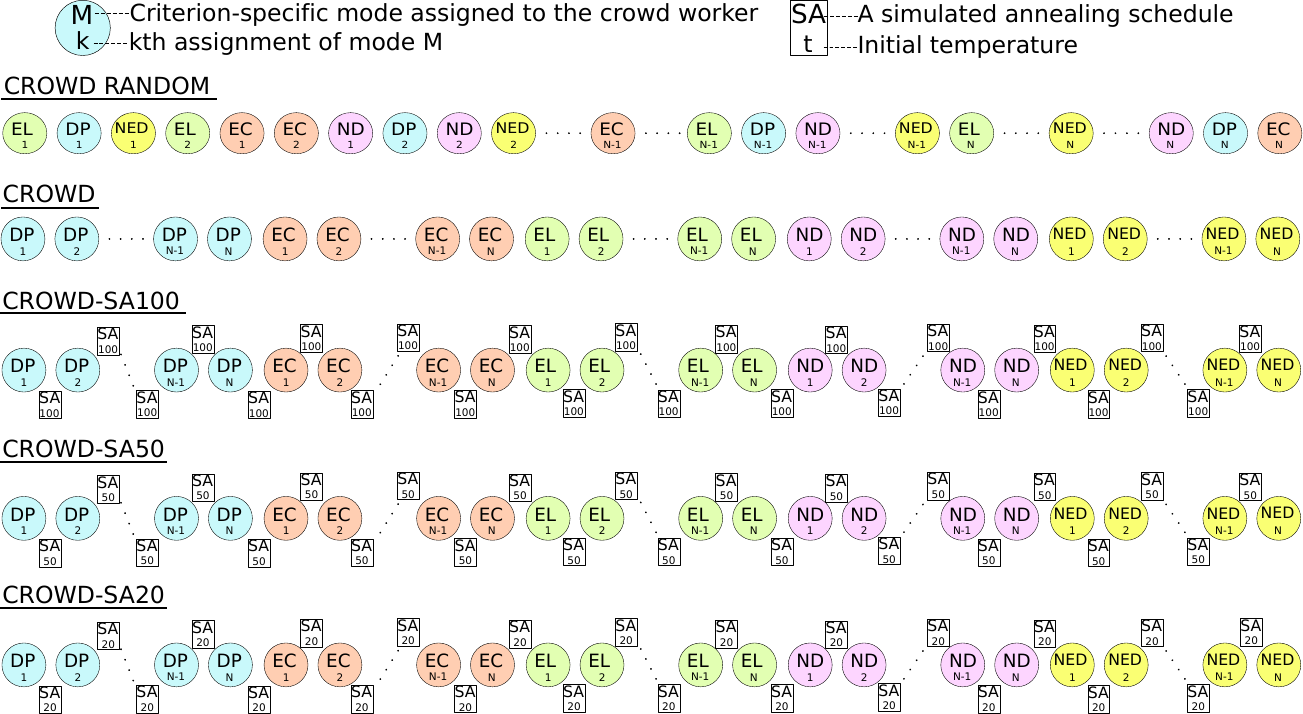}
  \caption{Figure illustrating the \Crowd, \CrowdRandom, and \emph{Hybrid} (Crowd-SA) approaches evaluated in this paper. Each circle represents a crowd worker in a game sequence where a crowd worker is assigned a criterion-specific mode $M$. Each criterion-specific mode is assigned N times in a game sequence. A rectangle represent a simulated annealing schedule with initial temperature $t$.}
  \label{fig:game-sequence-illustration}
\end{figure}

Given the central importance of domain-specific downward pointing path criterion, we assigned it a high priority of 400. We assigned the non-crossing edge-pairs criterion a priority of 3 and rest of the criteria a priority of 1. We conducted two experiments to address our research questions.

\subsubsection{Experiment 1} \label{subsec:experiment-1}
We designed the first experiment to answer \emph{RQ1}, where we want to find a good strategy to assign a criterion-specific mode to a game player in a sequence. To this end, we evaluated the performance of two different approaches: \Crowd 
and \CrowdRandom. In the \Crowd approach, we assigned criterion-specific modes to workers in the order of their priorities, i.e., downward pointing paths (DP), non-crossing edge pairs (EC), edge length (EL), node distribution (ND), and node-edge separation (NED), as shown in Figure~\ref{fig:game-sequence-illustration}.

We recruited $20$ crowd workers for each game sequence where four ($N=4$) workers focused on each of the five layout criteria used in Flud. 
In contrast, for the \CrowdRandom approach, we randomly assigned each criterion four ($N=4$) times in a sequence of $20$ crowd workers (refer to Figure~\ref{fig:game-sequence-illustration}). We recruited crowd workers for three game sequences per network for each approach. Overall, we recruited $360$ crowd workers for this experiment.

\subsubsection{Experiment 2}
\begin{figure}[H]
 \centering
  \includegraphics[width=\textwidth]{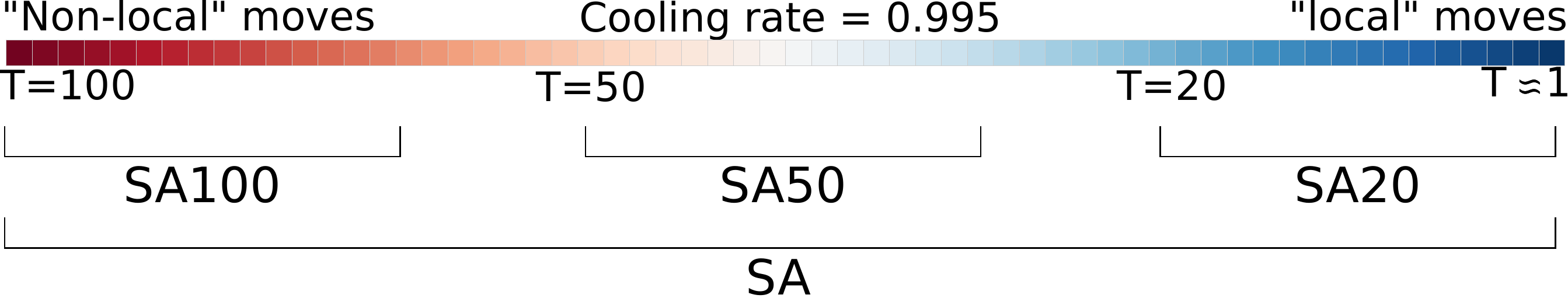}
  \caption{Figure illustrating the three simulated annealing schedules used in our hybrid approaches.} 
  \label{fig:illustration-simulated-annealing-schedules}
\end{figure}

 In our second experiment, we evaluated the performance of the \Crowd  approach described in Experiment 1 and the \emph{Hybrid} approach 
 against automated baseline methods with the goal of answering RQ2 and RQ3. Similar to the \Crowd approach, for \emph{Hybrid} approach, we recruited $5N$ crowd workers for each game sequence where $N$ workers focused on each of the five criteria; we provide the values of $N$ below. However, in \emph{Hybrid}, we alternate sessions of gameplay between crowd worker and simulated annealing (refer to Figure~\ref{fig:game-sequence-illustration}). 
 Davidson and Harel's simulated annealing based layout algorithm~\cite{davidson-harel-simulated-annealing-drawing-1996} starts with ``non-local'' moves ($T=100$), where a node's next location is not restricted to nearby positions, and ends with ``local'' moves ($T\approx1$) where a node can move only to a nearby position.  In this experiment, we tried three different types of hybrid approaches (\emph{Crowd-SA100}, \emph{Crowd-SA50}, and \emph{Crowd-SA20}), where we alternate the \Crowd approach and simulated annealing schedules with three different initial temperatures (refer to Figure~\ref{fig:illustration-simulated-annealing-schedules}). 

\begin{itemize}
    \item \emph{SA100 (High temperature).} In this schedule, we started with a high temperature ($T_0=100$) and only ran the initial part of the schedule (approximately 15 minutes). In this schedule, we only made random moves that are ``non-local'' in nature.
    \item \emph{SA20 (Low temperature).} Here, we started with a low temperature ($T_0=20$) and only ran the later part of the schedule (approximately 15 minutes). Due to the low initial temperature, we only made random moves that are ``local'' in nature. Such moves further optimized the score while preserving the overall structure of the layout.
    \item \emph{SA50 (Medium temperature).} In this variant, we started with an intermediate temperature ($T_0=50$) and ran the middle part of a SA schedule (approximately 15 minutes).
\end{itemize}

Overall, we aimed to recruit $N = 10$, $N = 15$, and $N = 20$ crowd workers per criterion for networks G1, G2, and G3, respectively, for both \emph{Hybrid} and \Crowd approach. We stopped recruiting crowd workers once the total gameplay time taken by the workers in the sequence exceeded 24 hours. 
At the end of the sequence, we fine-tuned the best layout so far to generate the final output layout by using the fine-tuning procedure described in Harel and Sardas's work on incremental improvement of layouts of planar networks.

In order to highlight the quality of layouts generated by the Flud system, we compared our results against four automatic layout algorithms.

\begin{itemize}
    \item \emph{Simulated annealing (SA)}~\cite{davidson-harel-simulated-annealing-drawing-1996}. We selected this algorithm as our primary baseline due to its flexibility to accommodate all of our criteria (Section \ref{sec:layout-criteria}). We ran the algorithm for a fixed number of iterations (e.g., 500) where the temperature cools down from high initial temperature ($T=T_0=100$) to low temperature ($T\approx1$). To compare performance of SA against \Crowd and \emph{Hybrid}, we ran the simulated annealing schedule multiple times for 24 hours such that each schedule builds upon the best layout created so far. We ran the baseline simulated annealing algorithm on a machine with an Intel Xeon (16~cores at 2.40~GHz) CPU. 
    
    \item \emph{Dig-Cola}~\cite{dwyer-yehuda-dig-cola-infovis-2005}. Our second baseline was Dig-Cola, which may indirectly optimize the number of downward pointing paths criterion since it lays out nodes in a hierarchical manner. Dig-cola also provides a way to conserve aesthetic criteria such as edge lengths and symmetries. We used the implementation of Dig-Cola in the Neato program in the Graphviz package \cite{ellson2002graphviz}. We ran Dig-Cola with nine different values of the minimum gap between levels $\{0.1, 0.2, 0.5, 1, 2, 5, -0.5, -1, -2\}$, four different values of the preferred edge length parameter ${1,3.125,5}$, and all possible values of the overlap parameter. Finally, we used the layout that produced the largest total score.
    
    \item \emph{IPSEP-Cola}~\cite{dwyer2006ipsep}. Our third baseline was IPSEP-Cola, which tries to ensure that node $s$ is placed above node $t$ if there is a directed edge from $s$ to $t$. If the network contains directed cycles, IPSEP-Cola computes the largest acyclic subgraph and relaxes the constraints on the excluded edges. 
    This approach may optimize the number of downward pointing paths.
    We used the implementation of IPSEP-Cola in the Neato program in the Graphviz package \cite{dwyer2006ipsep}. We ran a parameter search similar to Dig-Cola and used the layout that produced the largest total score in our comparisons.
    
    \item \emph{Spring-electrical model}~\cite{hu-efficient-force-directed-sfdp-mathematica-2005}. This force-directed algorithm uses spring elasticity to keep connected nodes closer and electric charge repulsion to keep disconnected nodes away from each other. The method is available as the `sfdp' program in the Graphviz package \cite{ellson2002graphviz}. We ran the program with six different values of the spring constant $K = \{0.1, 0.2, 0.5, 1, 2, 5, 10\}$ and of the power of repulsive force $R = \{0.1, 0.2, 0.5, 1, 2, 5, 10\}$. We used the layout from the parameter pair with the highest score in our comparisons.
\end{itemize}


Finally, we also instrumented Flud to record each action taken by the crowd workers during their gameplay as well as the corresponding scores and layout. We used the collected data to analyze how the crowd workers played the game, and understand the impact of features such as criterion-specific modes and clues on the scores.   

\section{Results}
\label{sec:results}

\subsection{RQ1: How should Flud assign criterion-specific modes to players to optimize the scores they achieve?}
\label{subsec:RQ1-mode-seq-results}

\begin{figure}[H]
    \begin{tikzpicture}
    \node[inner sep=0pt] at (0,-6.2)
        {\includegraphics[width=\textwidth]{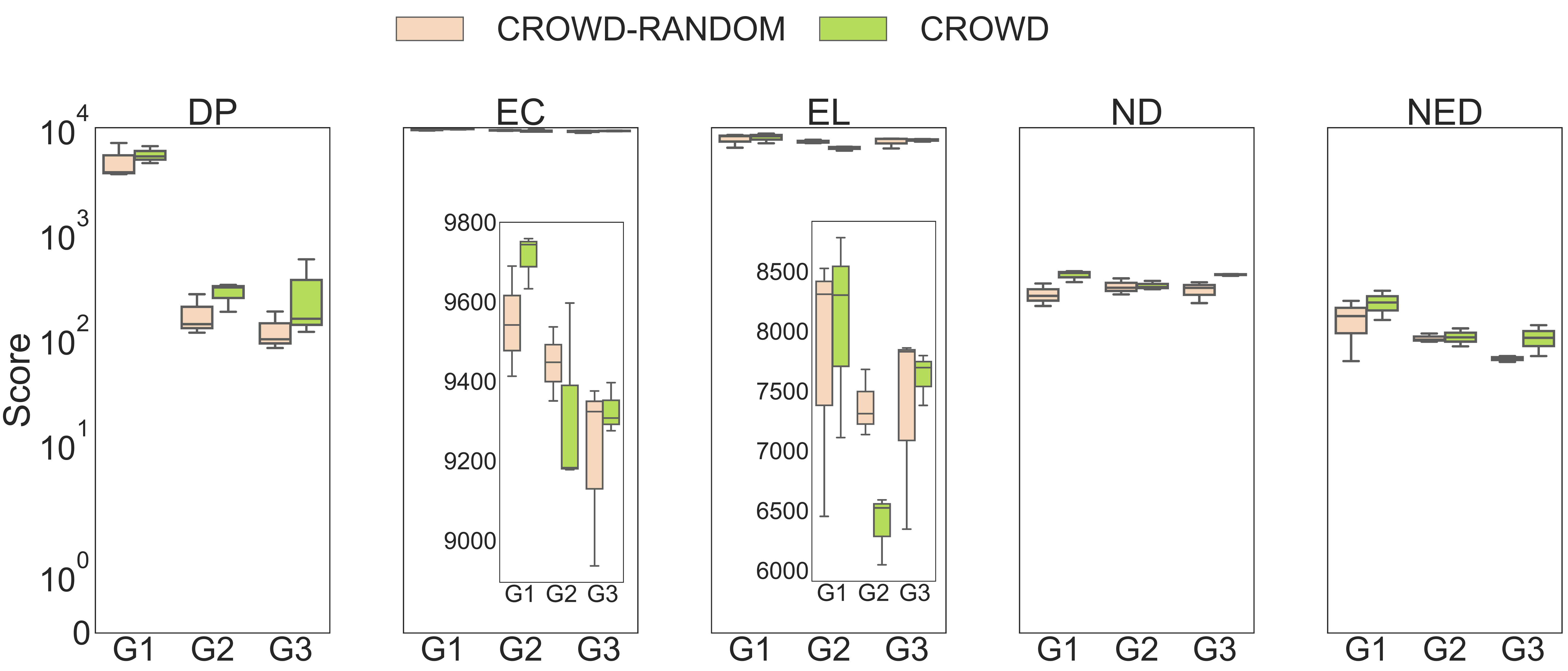}};
    \node[inner sep=0pt] at (0,0)
        {\includegraphics[width=\textwidth]{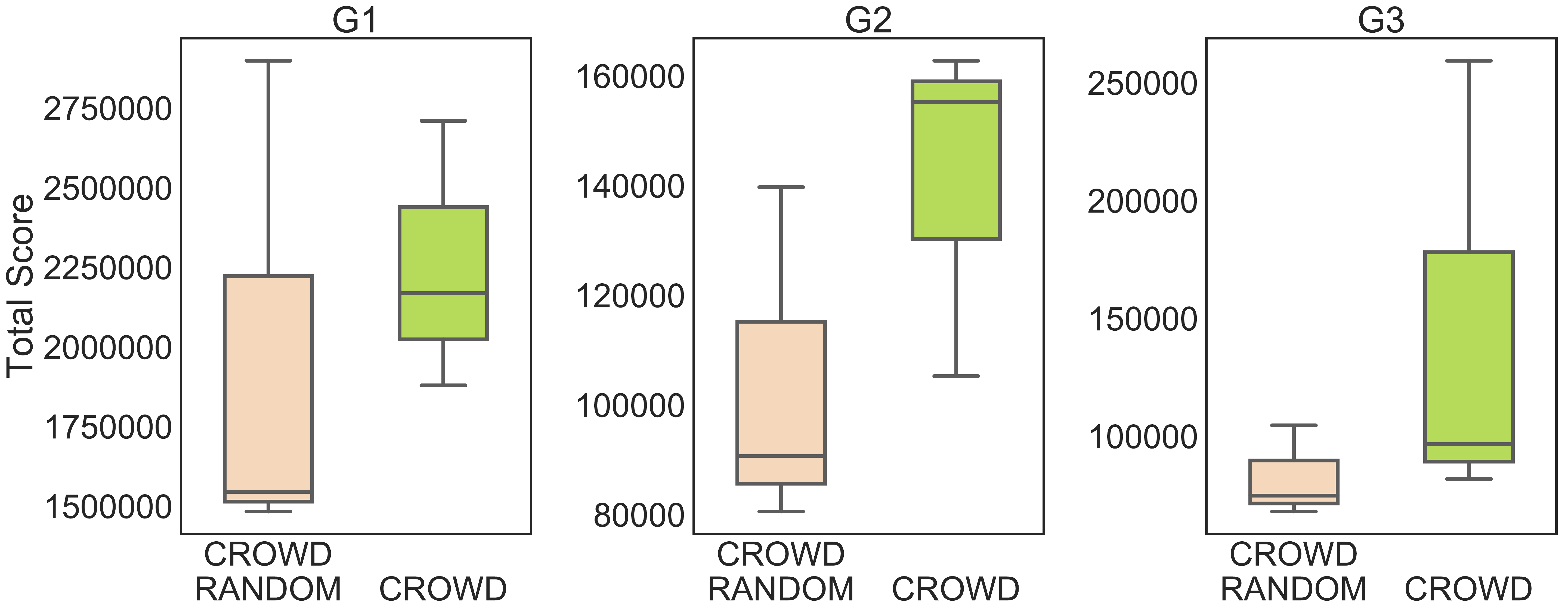}};
        
    \node[inner sep=0pt] at (-6.5, 2.3) {\textbf{(A)}};
    
    \node[inner sep=0pt] at (-6.5, -3.8) {\textbf{(B)}};
    \end{tikzpicture}
    \caption{ Comparison of scores achieved by crowd workers when modes are assigned randomly (\emph{Crowd-Random}) and in decreasing order of priorities (\emph{Crowd}). (A) Distributions of total scores. (B) Distributions of per-criterion scores.  }
    \label{fig:random-order-vs-weighted-order-scores-box-plot}
\end{figure}

We used Experiment 1 to compare the performance of the crowdsourcing approach when the criterion-specific modes are assigned in decreasing order of the priorities (i.e., \emph{Crowd}) versus a random assignment of modes (i.e., \emph{Crowd-Random}). Figure~\ref{fig:random-order-vs-weighted-order-scores-box-plot}A shows that for each of the three networks, the median total score achieved by the \emph{Crowd} approach was greater than the median score achieved by \emph{Crowd-Random} approach. On further investigation, we found out that the \emph{Crowd} approach achieved a better median per-criterion score for downward pointing paths (DP), node distribution (ND), and node edge separation (NED) criterion for all three networks (Figure~\ref{fig:random-order-vs-weighted-order-scores-box-plot}B). On the other hand, the \emph{Crowd-Random} approach achieved a higher median per-criterion score for non-crossing edge pairs (EC) and edge length (EL) criterion for networks G2 and G3. Since the \emph{Crowd} approach achieves a higher score for three out of five layout criteria, including the important domain-specific DP criterion (reflected in the total layout score), we decided to assign criterion-specific modes in order of priorities in Experiment 2.

In the \emph{Crowd} approach, we also observed that while four crowd workers were able to lay out as many as 1,303 downward pointing paths in network G1, the same number of crowd workers could only lay out 321 and 131 downward pointing paths in networks G2 and G3, respectively. We attributed this difference in performance to a large number of cycles in G2 and G3 compared to G1 (Table~\ref{table:graph-structure}). We noticed that in the presence of large numbers of cycles, crowd workers had to make more moves in network G2 (mean=899) and G3 (mean=850) compared to network G1 (mean=403) where the number of cycles is very low (Table \ref{table:graph-structure}). Therefore, 
we decided to recruit crowd workers in proportion to the number of simple paths from sources to targets (Table \ref{table:graph-structure}) in Experiment 2 to balance out the crowd workers' low throughput for downward pointing paths criterion in networks with large number of cycles.

\subsection{RQ2: How do the crowdsourcing approaches perform in comparison to automated methods?}
\label{subsec:RQ2-performance-analysis-results}

We used the data collected from Experiment 2 to answer this research question. We considered several aspects to answer RQ2: overall scores, number of downward-pointing paths, scores for aesthetic criteria, rate of improvement, and final network layouts.



\begin{figure}[H]
 \centering
  \includegraphics[width=\textwidth]{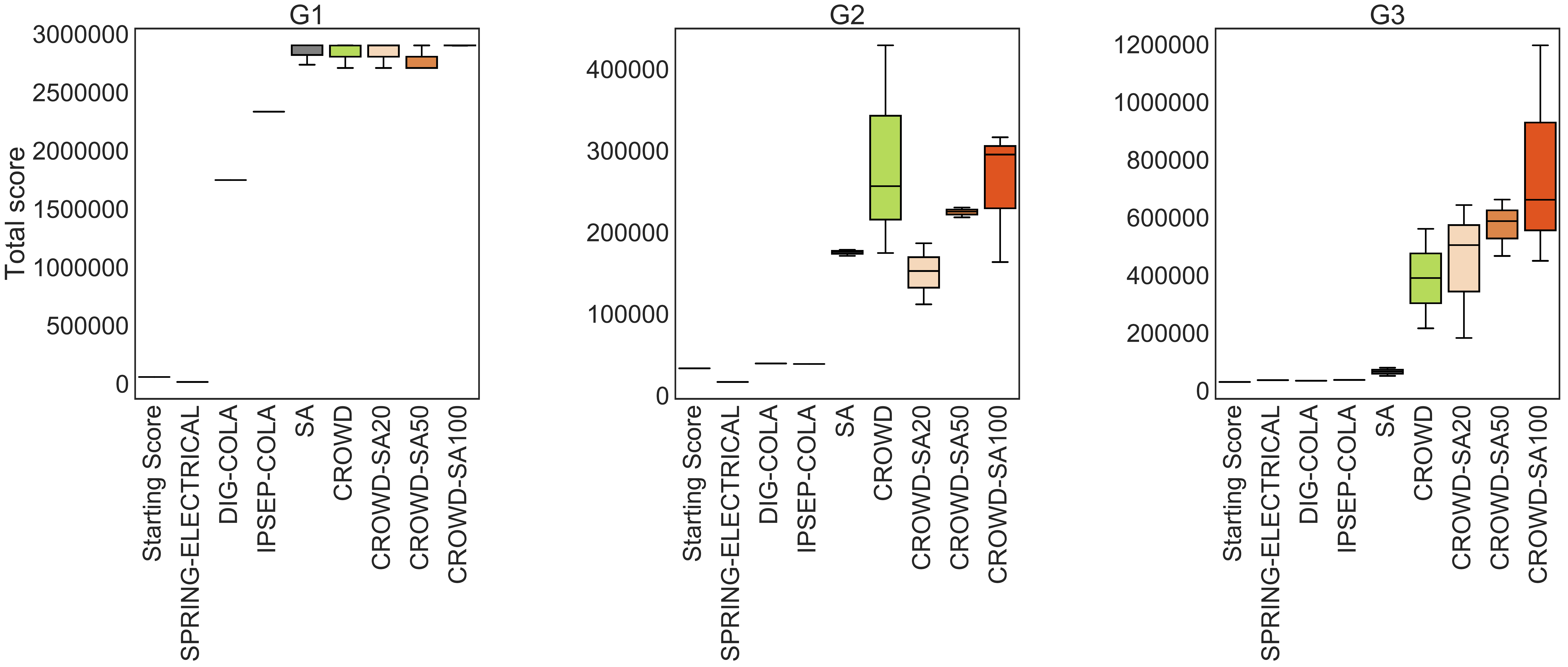}
  \caption{Distributions of the total scores of layouts created by different approaches. 
  }
  \label{fig:total-scores-performance-analysis-box-plot}
\end{figure}

\noindent
\emph{Overall score}. First, we compared the performance of all the approaches on the overall score (Figure~\ref{fig:total-scores-performance-analysis-box-plot}). We observed that the crowd and hybrid approaches clearly outperformed all the automated techniques for networks G2 and G3. In contrast, for network G1, the automated methods, especially SA, had comparable performance to the crowd and hybrid approaches. We also noticed that \Digcola, \Ipsep, and \Spring algorithms hardly improved the starting overall score for networks G2 and G3. However, \Digcola and \Ipsep were able to considerably improve the overall score for network G1. We attributed this difference to the much larger number of cycles in G2 and G3 in comparison to G1 (Table~\ref{table:graph-structure}). To further understand these results, we examined the downward-pointing paths results in detail.

\begin{figure}[H]
 \centering
  \includegraphics[width=\textwidth]{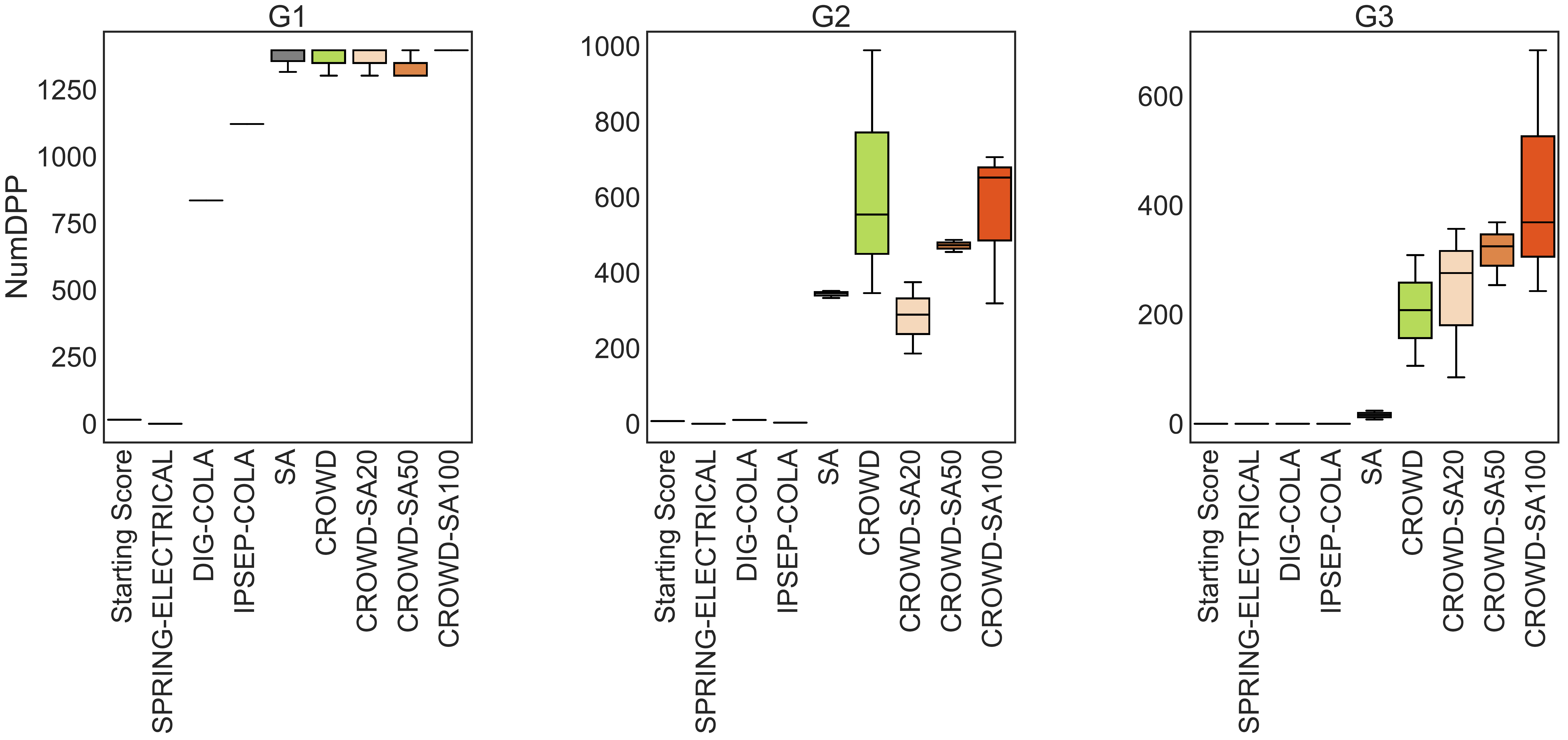}
  \caption{Distributions of the number of downward pointing paths in the layouts created by different approaches. 
  }
  \label{fig:num-dpp-performance-analysis-box-plot}
\end{figure}

\noindent \emph{Downward-pointing paths criterion}. Here, we compared the performance of all the approaches on the number of downward pointing pathways, our sole domain-specific layout criterion (Figure~\ref{fig:num-dpp-performance-analysis-box-plot}). In Figure~\ref{fig:num-dpp-performance-analysis-box-plot}, we saw a trend similar to overall scores shown in Figure~\ref{fig:total-scores-performance-analysis-box-plot}. We attribute this similarity to the high priority 
we assigned to the DP criterion. This explains that DP is the main reason why our baseline methods did not perform well on networks G2 and G3. We believe as the number of cycles increase in a network, it becomes harder for the automated methods to optimize for downward pointing paths. In fact, \Digcola, \Spring, and \Ipsep computed very few downward pointing paths ($\leq 10$) in networks G2 and G3. In contrast, crowd workers seem to be able to observe the direction of the flow along the edges
and develop moves that significantly increase the number of downward pointing paths despite the presence of cycles. Overall, these results indicate that crowd workers are better than automated methods at downward pointing paths task, specially for networks with several cycles. Moreover, \emph{Crowd-SA100} hybrid approach achieves even more number of downward pointing paths than crowd workers alone (\Crowd). We discuss the performance of the hybrid approaches in detail later in the section.

\begin{figure}[H]
 \centering
  \includegraphics[width=\textwidth]{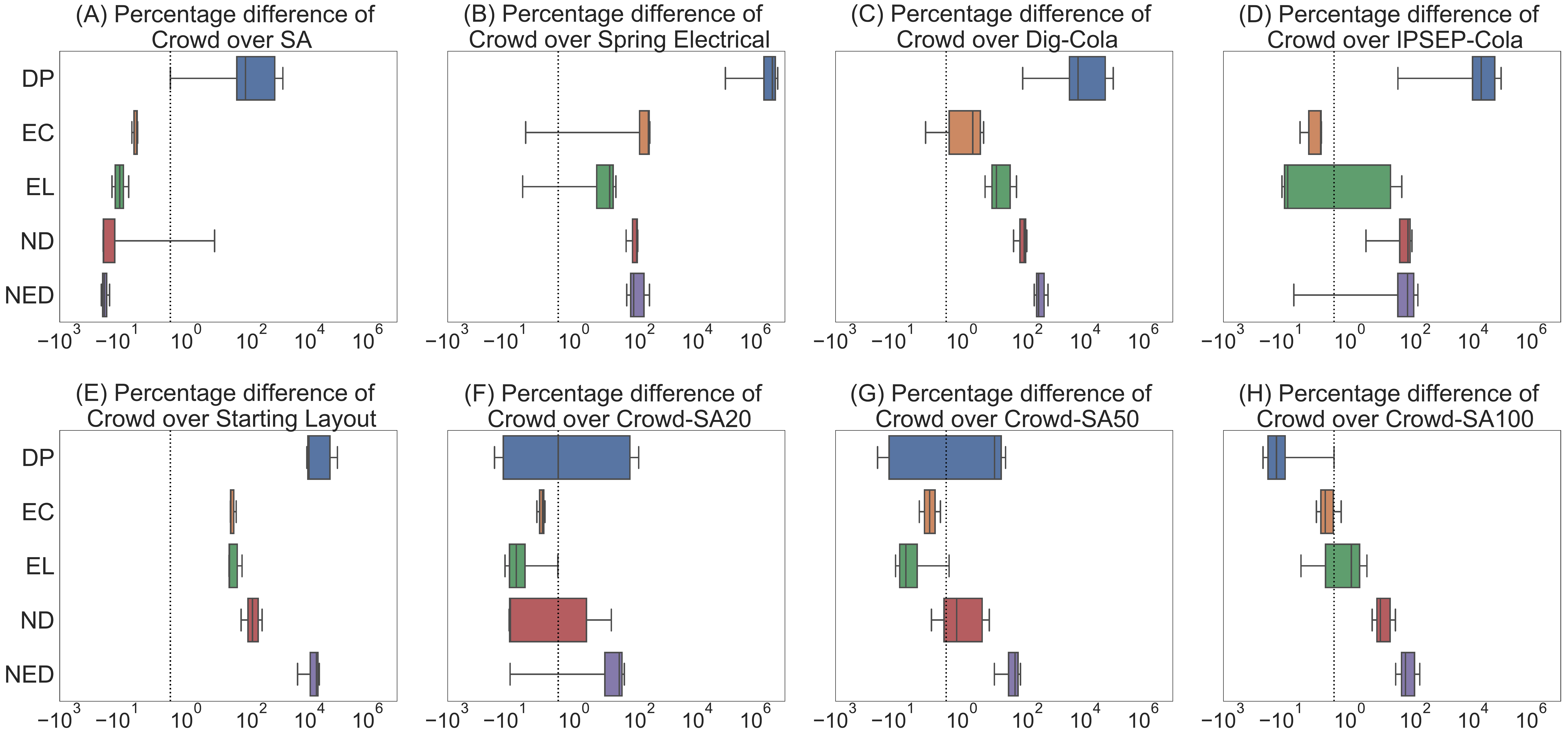}
  \caption{Distributions of the percentage difference between median per-criterion scores achieved by the \emph{Crowd} approach in comparison to other methods. The black dotted line corresponds to a value of zero. Positive values (to the right of this dotted line) indicate that the \emph{Crowd} approach performed better than another method. The negative values (to the left of this dotted line) indicate the opposite.}
  \label{fig:percent-diff-crowd-vs-rest-approaches-box-plot}
\end{figure}

\noindent
\emph{Other aesthetic criteria}. Next, we compared the distributions of per-criterion scores for the remaining layout criteria of all the approaches (Figures~\ref{fig:ec-performance-analysis-box-plot}--\ref{fig:ned-performance-analysis-box-plot} in Appendix). In these figures, we observed that the simulated annealing (\emph{SA}) baseline  outperformed the \emph{Crowd}, \emph{Hybrid}, and other automated approaches on the EC, EL, ND, and NED criteria. 
However, when we analyzed the median scores achieved by the \emph{Crowd} and \emph{SA} (refer to Figure~\ref{fig:percent-diff-crowd-vs-rest-approaches-box-plot}A), we found that the (positive) percentage difference for the DP criterion was one order of magnitude higher than the (negative) percentage difference for each of the remaining criteria. Since, DP is our sole biology inspired criterion, it is important to achieve high DP score to get a biologically meaning drawing. Therefore, we believe there is an advantage (more biologically meaningful drawing) of using \Crowd approach even though it performs slightly worse (less aesthetic) than \SA on the EC, EL, ND, and NED criteria. 
We also observed that the \emph{Crowd} approach outperformed the \emph{Spring Electrical} and \emph{Dig-Cola} methods for all criteria (Figure~\ref{fig:percent-diff-crowd-vs-rest-approaches-box-plot}B--C). \emph{Crowd} also outperformed \emph{IPSEP-Cola} on the DP, ND, and NED criteria (refer to Figure~\ref{fig:percent-diff-crowd-vs-rest-approaches-box-plot}D).

\noindent
\emph{Crowd versus Hybrid}. The \emph{Crowd} approach outperformed the \emph{Crowd-SA100} hybrid approach (Figure~\ref{fig:percent-diff-crowd-vs-rest-approaches-box-plot}H) on distance-based criteria (EL, ND, and NED). On the other hand, \emph{Crowd-SA100} performed better on the DP and EC criteria. We attribute this behavior to assistance from the \emph{SA100} simulated annealing component in the \emph{Crowd-SA100} hybrid approach. 
We believe that since \emph{SA100} starts at a high temperature, it has the ability to make non-local node movements that could result in the re-orientation of an edge to remove a crossing or to make it downward pointing. These non-local moves allow the \emph{Crowd-SA100} approach to use \emph{SA100} to further optimize for the DP and EC criteria. In contrast, the hybrid \emph{Crowd-SA20} approach performed better than the \emph{Crowd} approach for distance-based criteria such as EL and ND. Here, \emph{SA20} can move a node only to a nearby position and, therefore, allows \emph{Crowd-SA20} to explore the local neighborhood of a layout state in a more exhaustive manner compared to crowd workers. Surprisingly, the \emph{Crowd} approach outperformed the hybrid \emph{Crowd-SA20}  approach on the NED criterion, despite assistance in local moves from \emph{SA20} (Figure~\ref{fig:percent-diff-crowd-vs-rest-approaches-box-plot}F). Overall, the results show that the hybrid \emph{Crowd-SA100} approach, where simulated annealing makes non-local jumps, performs as well as or better than the non-hybrid \emph{Crowd} approach.

\begin{figure}[htbp]
 \centering
  \includegraphics[width=\textwidth, keepaspectratio]{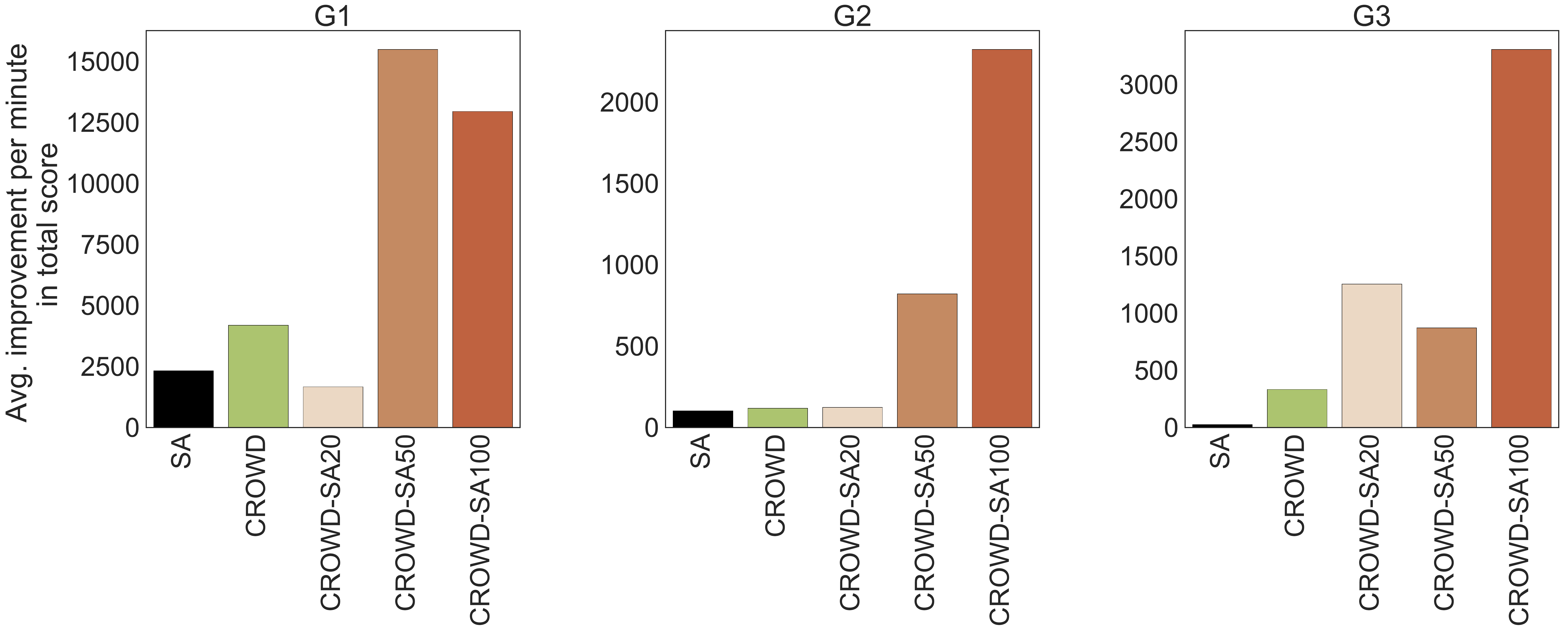}
  \caption{Bar plot of the average improvement in the total score per minute achieved by a crowd worker or a simulated annealing run in the \emph{SA}, \emph{Crowd}, and \emph{Hybrid} approaches.}
  \label{fig:rate-of-improvement}
 \end{figure}

\noindent
\emph{Rate of improvement}. Next, to evaluate the rate of improvement, we computed the average improvement in total score per minute achieved by a crowd worker or a simulated annealing run in the \emph{SA}, \emph{Crowd}, and \emph{Hybrid} approaches (Figure~\ref{fig:rate-of-improvement}). Here, we observed that the \emph{Crowd} and  \emph{Hybrid} approaches improved the scores at a faster rate than \emph{SA} in all three networks, while also considerably outscoring it in networks G2 and G3. We also noted that \emph{Crowd-SA100} had a better average improvement in total score per minute than \emph{Crowd} for all three networks. These results show that while \emph{SA} and \emph{Crowd} generate layouts comparable to \emph{Crowd-SA100} in some networks, \emph{Crowd-SA100} offers a better rate of improvement than these methods. Separately, we also observed that the automated methods --- \Digcola, \Ipsep, and \Spring --- generated the final layout within seconds. The \emph{Crowd} and \emph{Hybrid} approaches were slower, but clearly outperformed the automated methods within a few minutes. This observation is supported by Figure~\ref{fig:all-game-sequences} in Appendix showing the scores achieved by each approach over time for each network for all game sequences. 

\begin{figure}[htbp]

   \subcaptionbox{Crowd approach (Overall Score=2,903,703) \label{fig:flud-best-layout-G1}}[.48\linewidth][c]{%
    \includegraphics[width=0.48\linewidth,keepaspectratio]{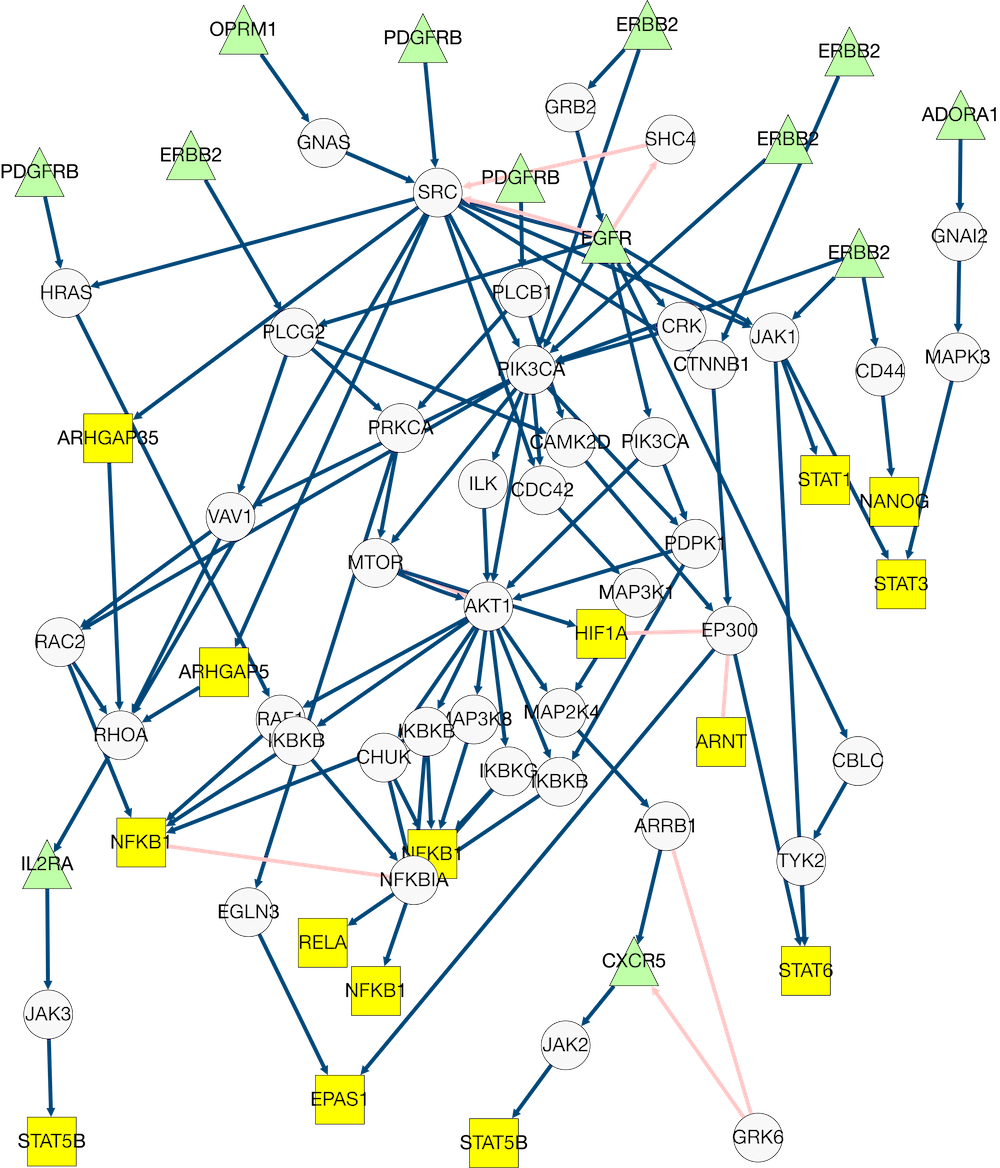}}
    \quad
    \subcaptionbox{Dig-Cola (Overall Score=1,749,105) 
    \label{fig:digcola-best-layout-G1}}[.48\linewidth][c]{%
    \includegraphics[width=0.48\linewidth,keepaspectratio]{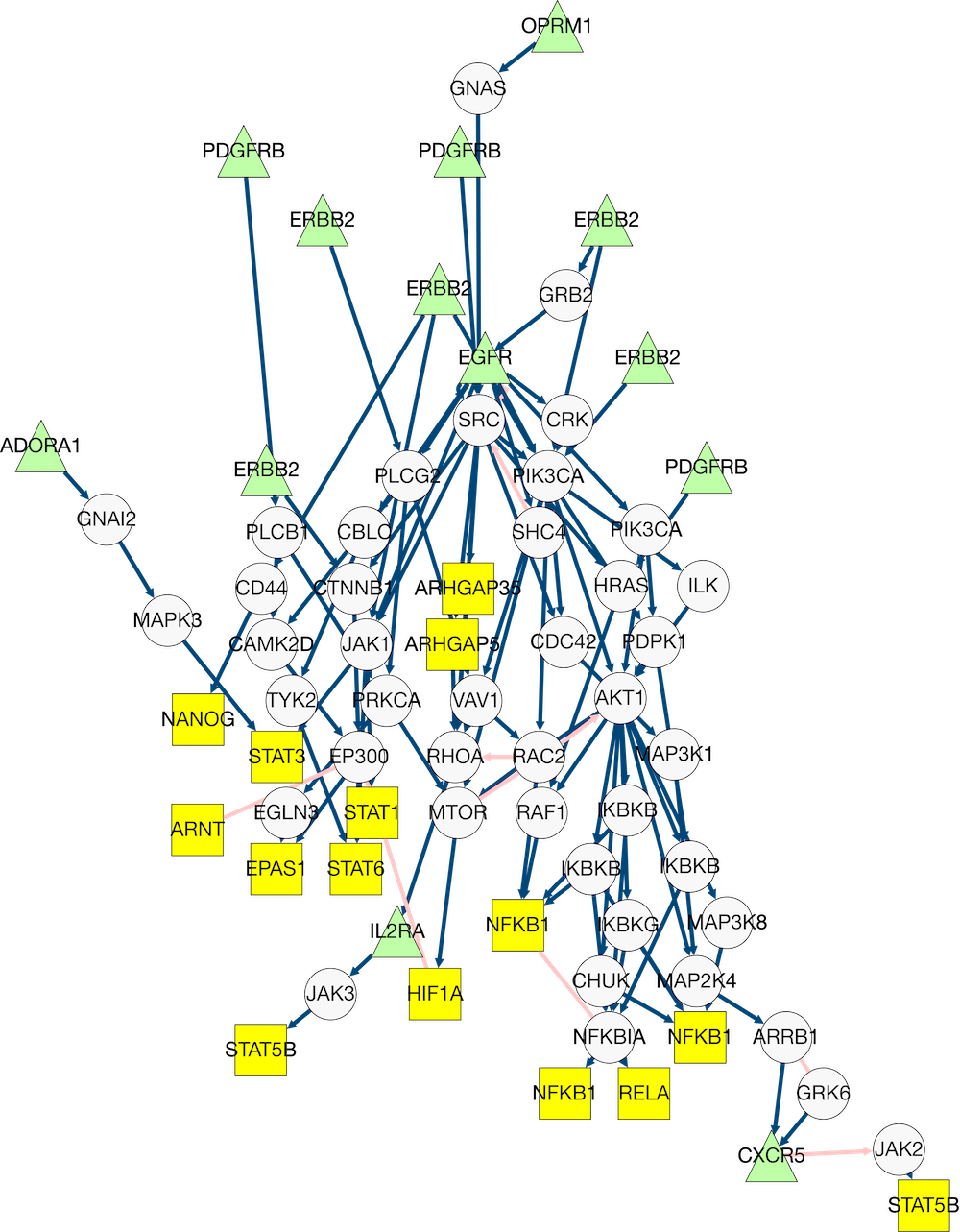}}
      \quad
  \subcaptionbox{SA (Overall Score=2,904,454) 
   \label{fig:sa-best-layout-G1}}[.48\linewidth][c]{%
    \includegraphics[width=0.48\linewidth,keepaspectratio]{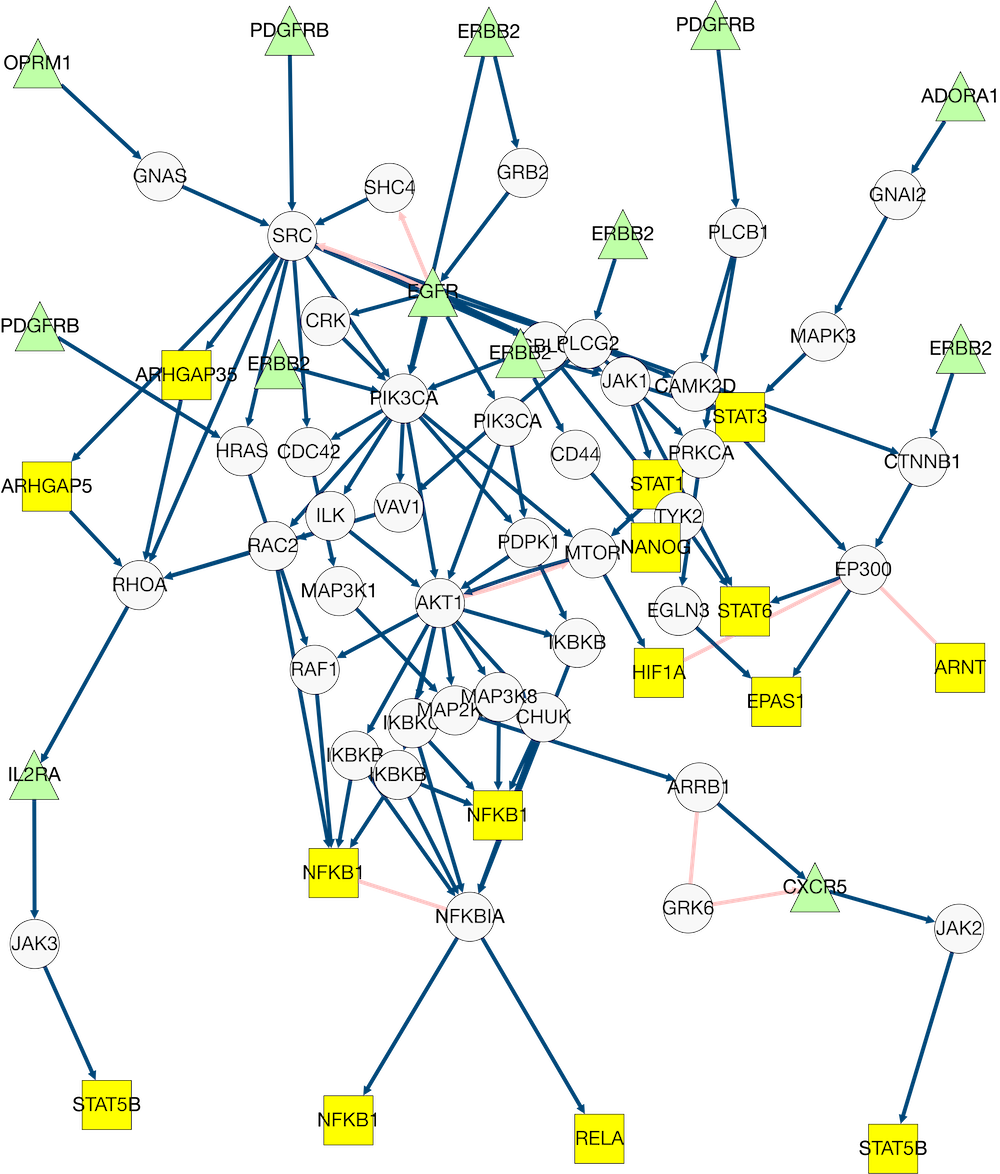}}
    \quad
   \subcaptionbox{Spring Electrical Model (Overall Score=16,941)  \label{fig:spring-best-layout-G1}}[.48\linewidth][c]{%
    \includegraphics[width=0.48\linewidth,keepaspectratio]{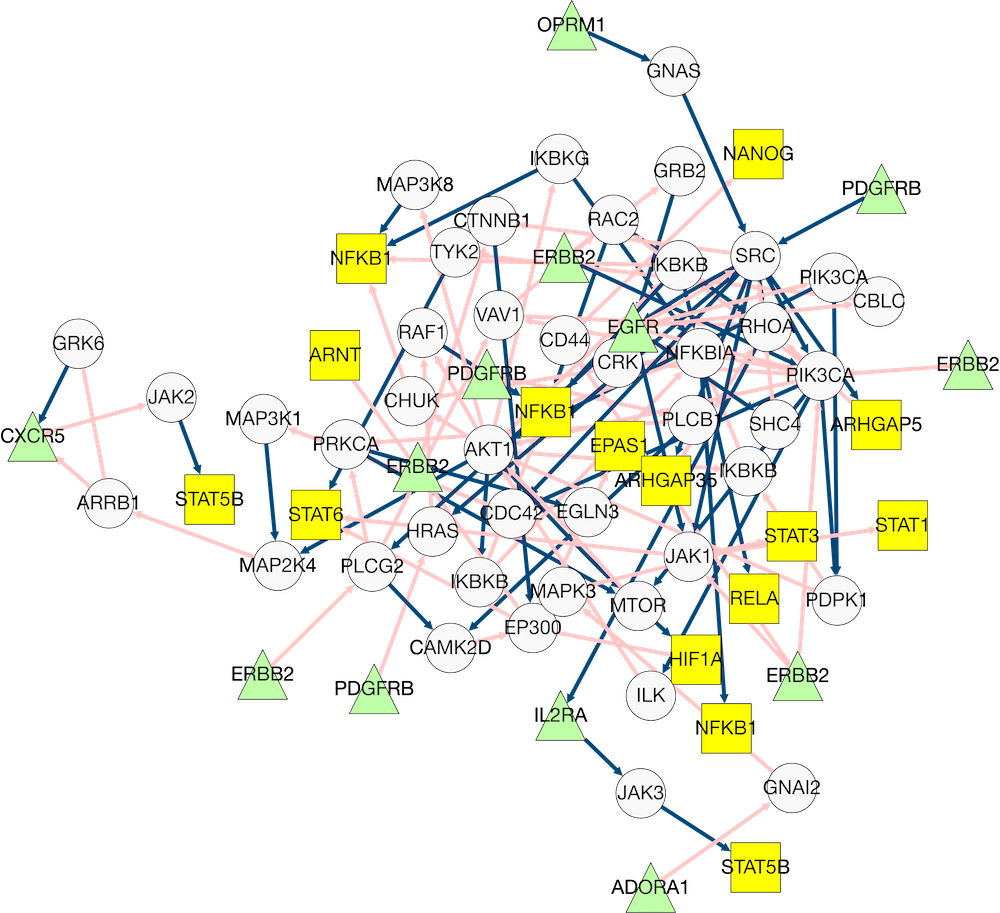}}
  
   
  \caption{Best layouts generated by Crowd and baseline methods for network G1. Green and yellow colored nodes represent the source and target nodes, respectively. The upward pointing edges are shown in red whereas downward pointing edges are shown in blue color.  
  }
  \label{fig:best-layouts-G1}
\end{figure}

\begin{figure}[!htbp]
    \centering
   \subcaptionbox{Crowd approach (Overall Score=429439) \label{fig:flud-best-layout-G2}}[.48\textwidth][c]{%
    \includegraphics[width=0.48\textwidth,keepaspectratio]{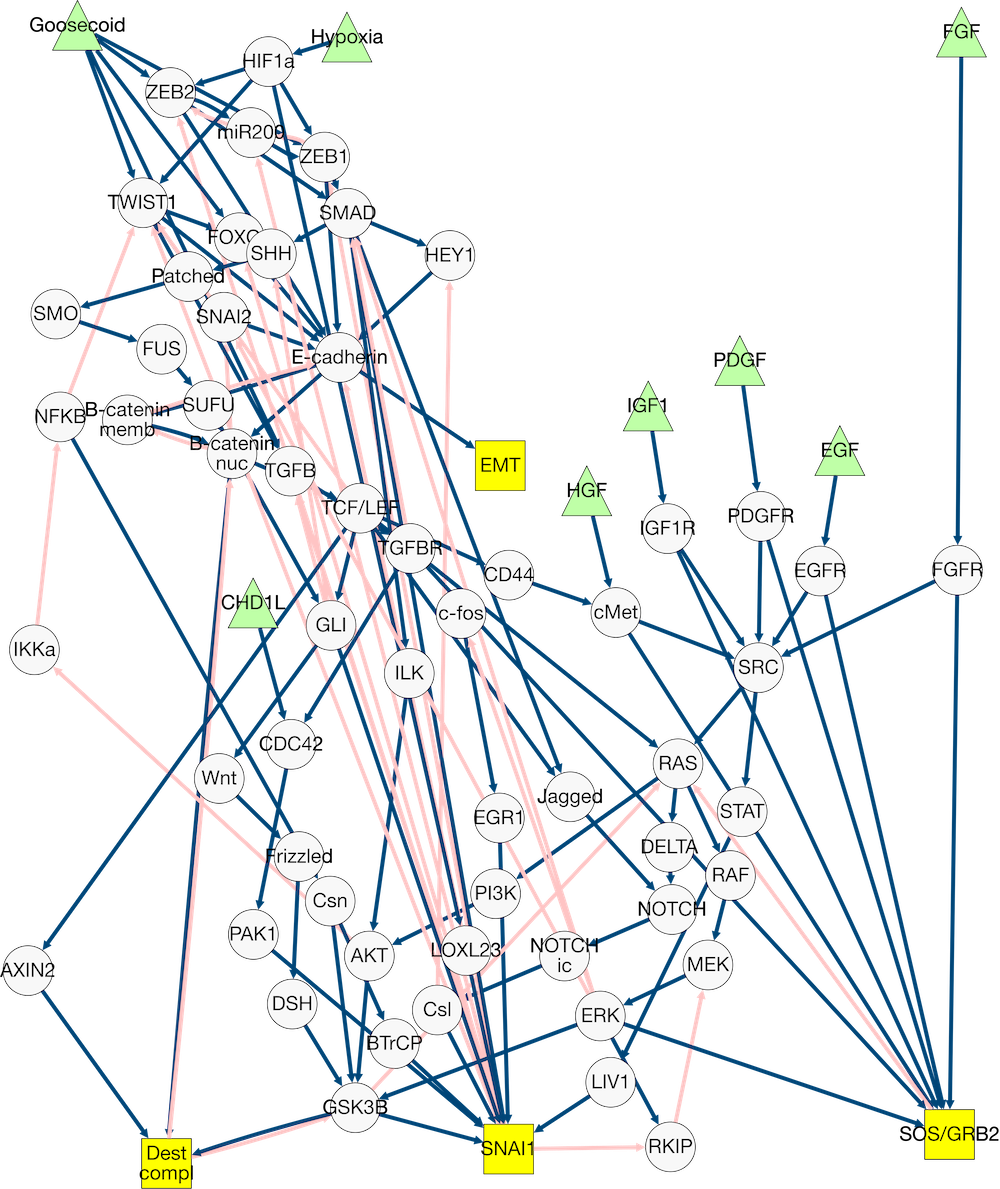}}
    \quad
    \subcaptionbox{Dig-Cola (Overall Score=39349)
    \label{fig:digcola-best-layout-G2}}[.48\textwidth][c]{%
    \includegraphics[width=0.44\textwidth,keepaspectratio]{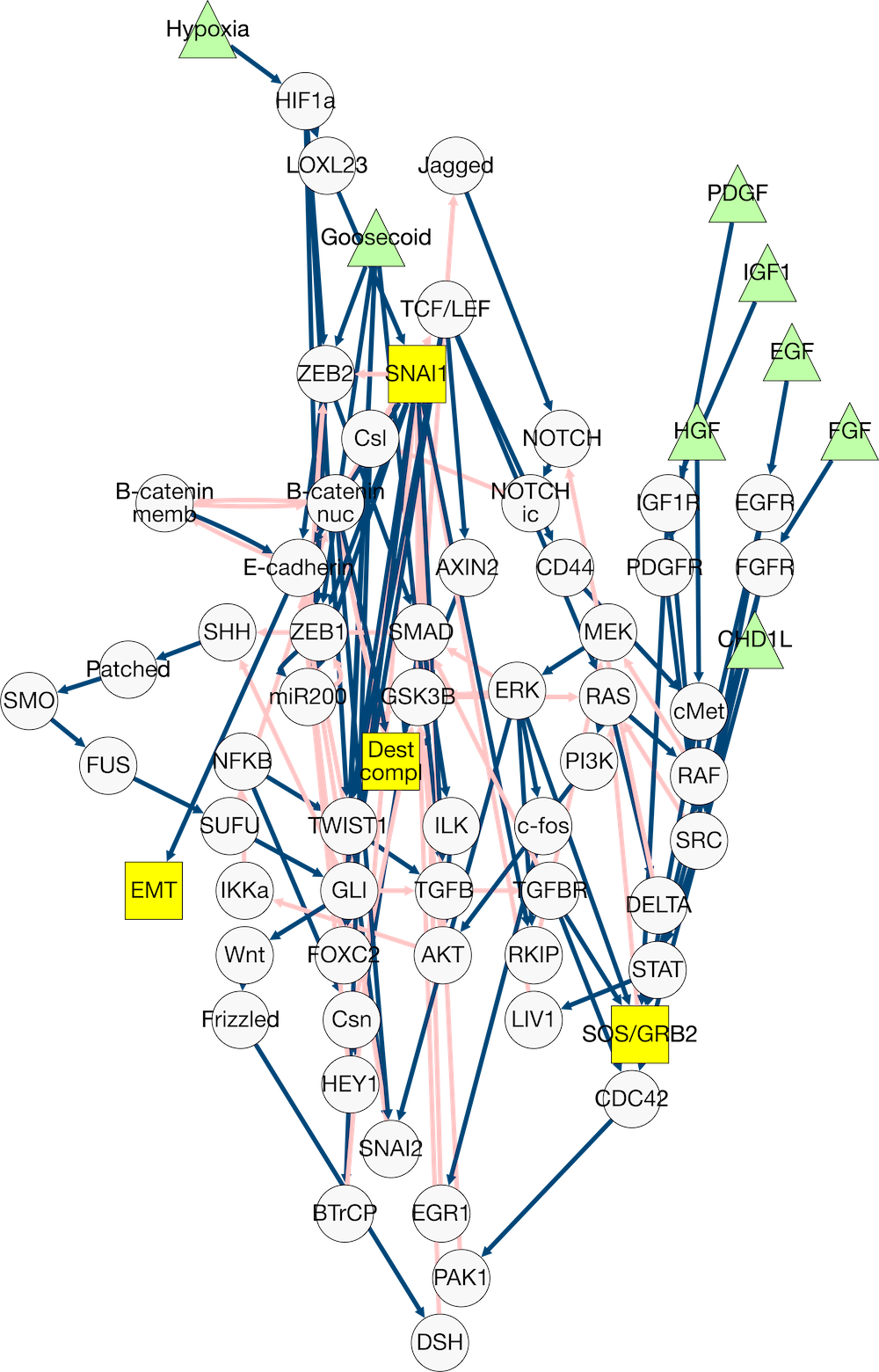}}
    \quad
  \subcaptionbox{SA (Overall Score=178718) 
   \label{fig:sa-best-layout-G2}}[.48\textwidth][c]{%
    \includegraphics[width=0.48\textwidth,keepaspectratio]{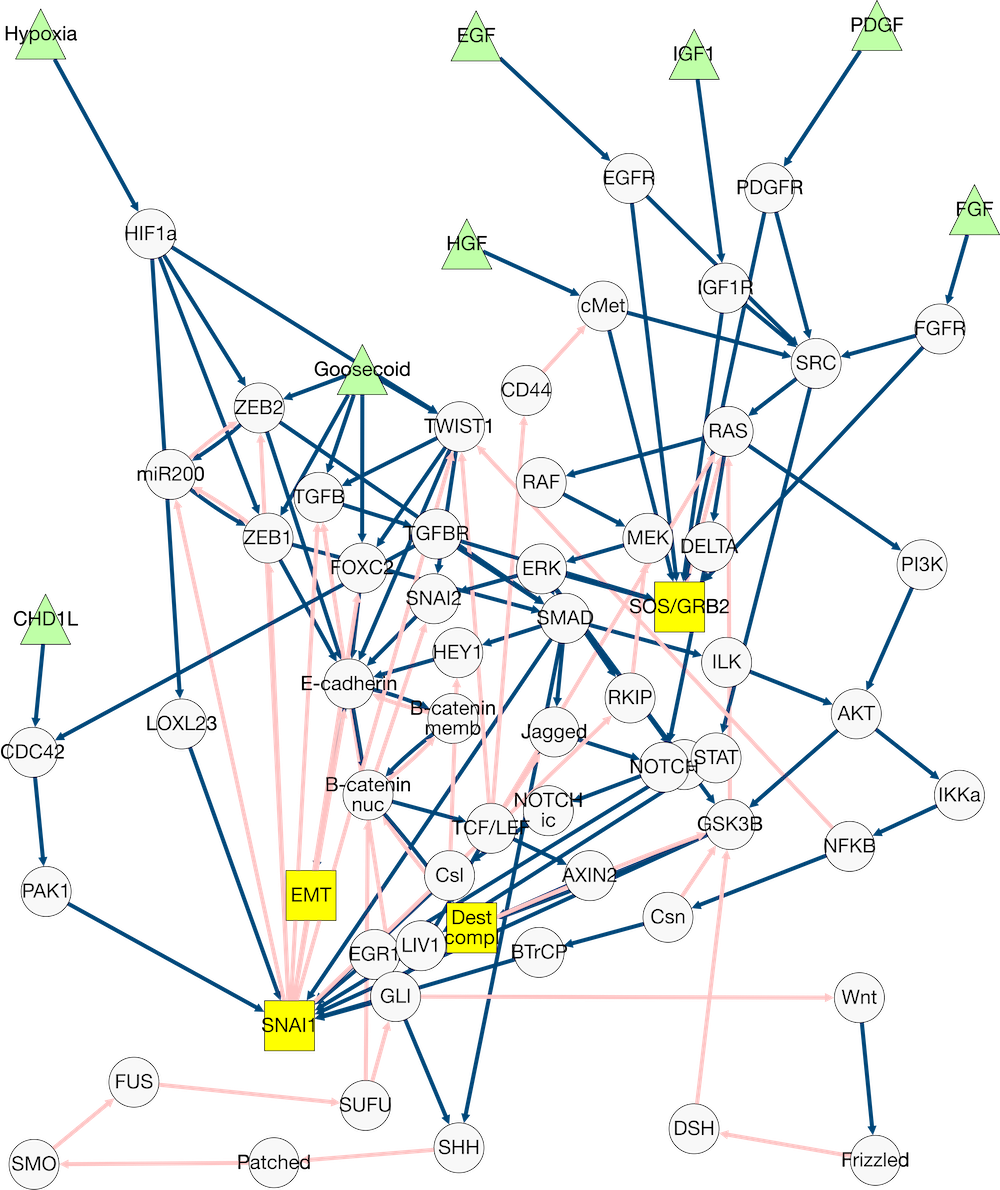}}
    \quad
   \subcaptionbox{Spring Electrical Model (Overall Score=16749) \label{fig:spring-best-layout-G2}}[.48\textwidth][c]{%
    \includegraphics[width=0.44\textwidth,keepaspectratio]{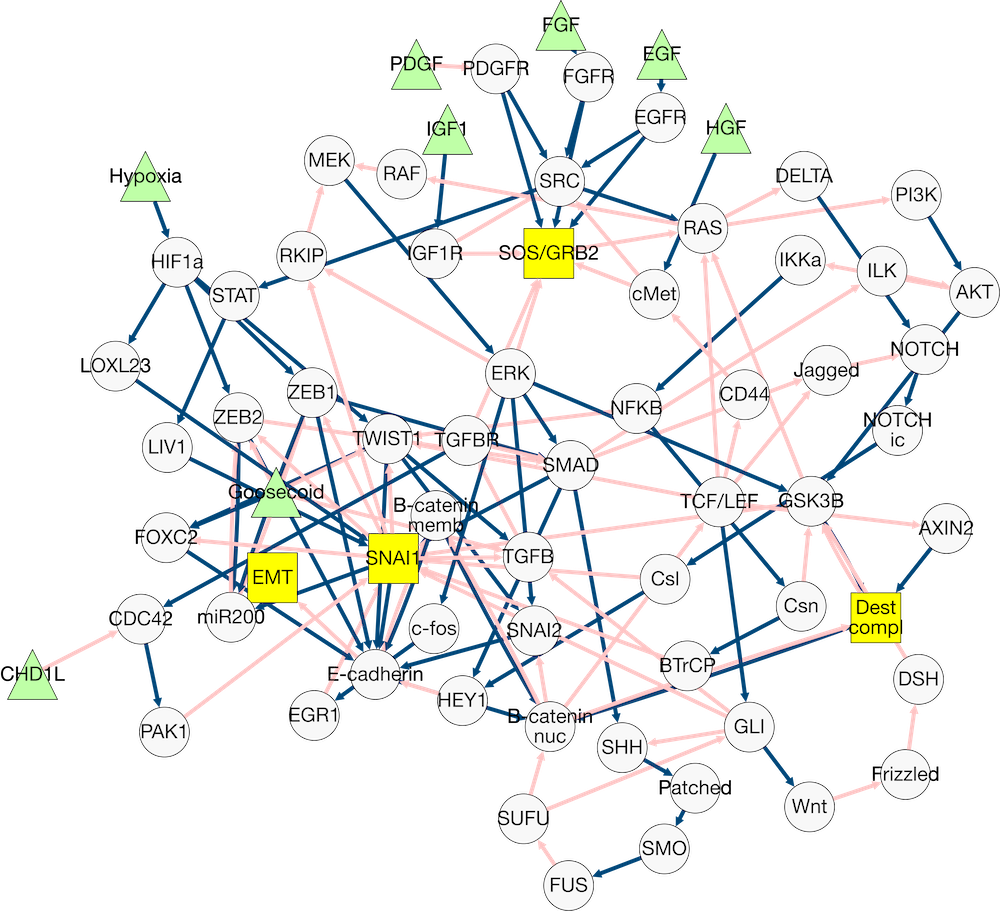}}
    
   
  \caption{Best layouts generated by Crowd and baseline methods for network G2. Green and yellow colored nodes represent the source and target nodes, respectively. The upward pointing edges are shown in red whereas downward pointing edges are shown in blue color.  
  }
  \label{fig:best-layouts-G2}
  \vspace{-3.8pt} 
\end{figure}

\noindent
\emph{Qualitative comparisons of layouts.} 
Finally, we qualitatively compared the best layouts generated using Flud and automated approaches (Figure~\ref{fig:best-layouts-G1} and~\ref{fig:best-layouts-G2}). Figure~\ref{fig:best-layouts-G1} shows the layouts generated for network G1, a network representing crosstalk from the Estrogen signaling pathway to the HIF-1 signaling pathway. While Dig-Cola generated a layout with several downward pointing edges (Figure~\ref{fig:best-layouts-G1}b), it is hard to read the layout because of many edge crossings and node-node and node-edge overlaps. In contrast, the Crowd and SA approaches (Figure~\ref{fig:best-layouts-G1}a and Figure~\ref{fig:best-layouts-G1}c, respectively) created layouts with many downward-pointing paths while achieving better separation among nodes and edges.  Figure~\ref{fig:best-layouts-G2} shows the layouts generated for network G2, a network representing the Epithelial-to-mesenchymal transition (EMT). 
In the crowdsourced layout (Figure~\ref{fig:flud-best-layout-G2}), we can clearly see the downward-pointing path from FGF to SOS/GRB2 on the right, whereas it is hard to clearly see this two-edge path in layouts generated using \Digcola (Figure~\ref{fig:digcola-best-layout-G2}) and \Spring (Figure~\ref{fig:spring-best-layout-G2}) method.  We also note that Figure~\ref{fig:digcola-best-layout-G2} gives the false impression that paths from source nodes to the SNAI1 target node are very short. In contrast, it is clear from the crowdsourced layout (Figure~\ref{fig:flud-best-layout-G2}) that SNAI1 is regulated by multiple upstream source nodes via multiple paths or mechanisms.

\subsection{RQ3: What are the dynamics of the mixed-initiative collaboration in hybrid approach?}
\label{subsec:RQ4-contribution-analysis-results}

\begin{figure}[H]
 \centering
  \includegraphics[width=\textwidth]{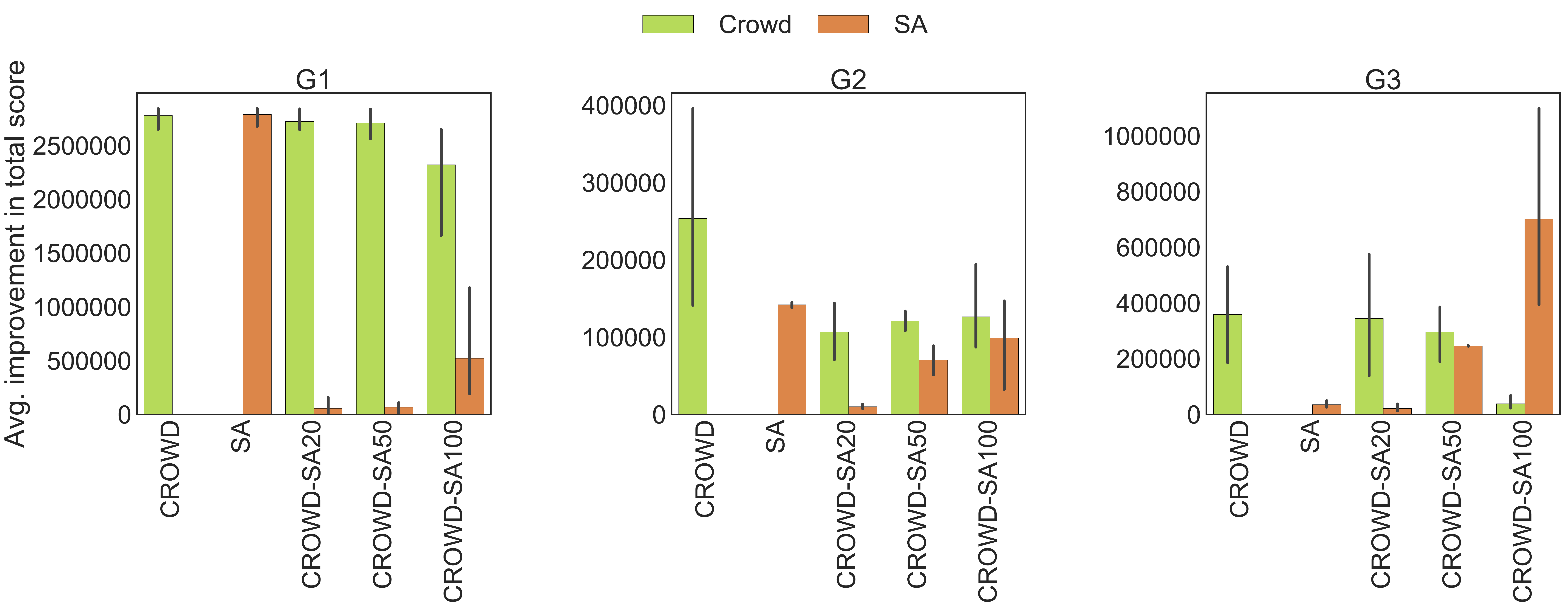}
  \caption{ Bar plot showing average improvement in total score made by simulated annealing and crowd workers in hybrid approach.}
  \label{fig:crowd-sa-contribution-total-score}
\end{figure}

To answer this question, we compared the average improvement to the total score contributed by the crowd workers in different approaches (Figure~\ref{fig:crowd-sa-contribution-total-score}), using the data collected in Experiment 2. We found that the contribution by crowd workers is lower in hybrid approaches than in the \emph{Crowd} approach, where the crowd workers work alone without any help from simulated annealing. We also noticed that on average, the contribution by simulated annealing is lower in hybrid approaches than in the \emph{SA} approach for networks G1 and G2. We believe that this decrease in contribution from crowd and simulated annealing in hybrid approaches is because of the distribution of the total work between these two components.  However, the 
\emph{Crowd} and \emph{SA} collaborate with each other in the \emph{Crowd-SA100} approach, leading to higher scoring layouts than either from \emph{Crowd} or from \emph{SA} alone (Figure~\ref{fig:total-scores-performance-analysis-box-plot}). 

\begin{figure}[H]
  \subcaptionbox{Before crowd started. \label{fig:flud-screenshots-G3-crowd-sa100-initial}}[.24\textwidth][c]{%
    \includegraphics[width=0.24\textwidth,keepaspectratio,frame]{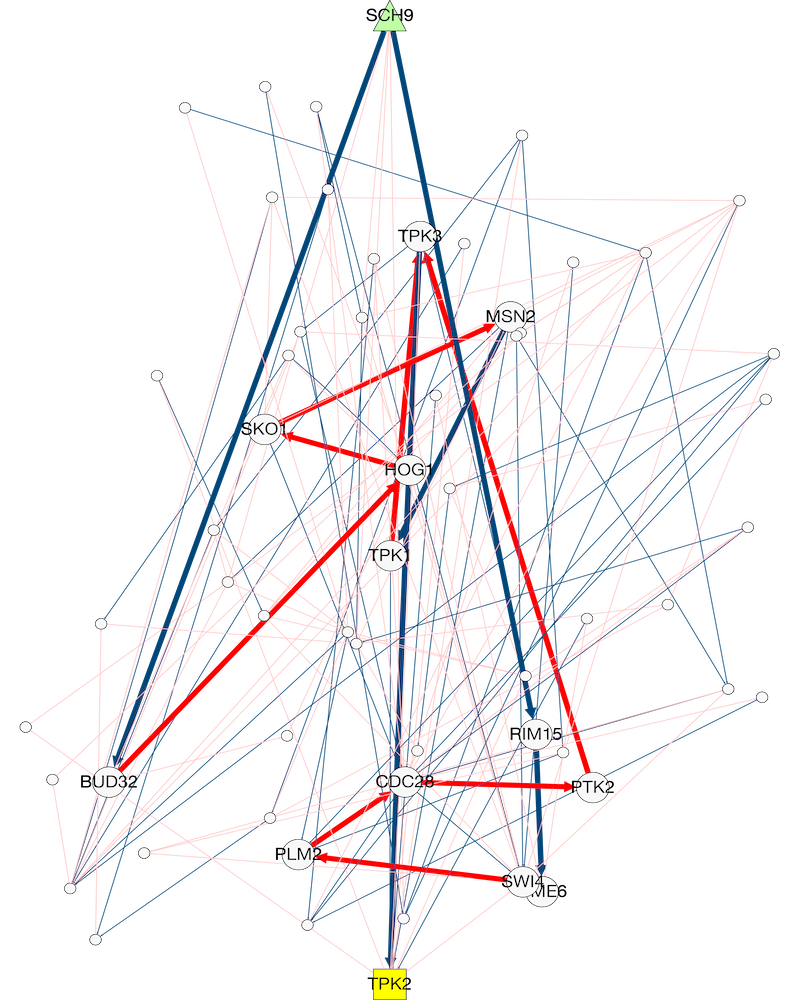}}
  \subcaptionbox{After crowd finished. \label{fig:flud-screenshots-G3-crowd-sa100-crowd-final}}[.24\textwidth][c]{%
    \includegraphics[width=0.24\textwidth,keepaspectratio,frame]{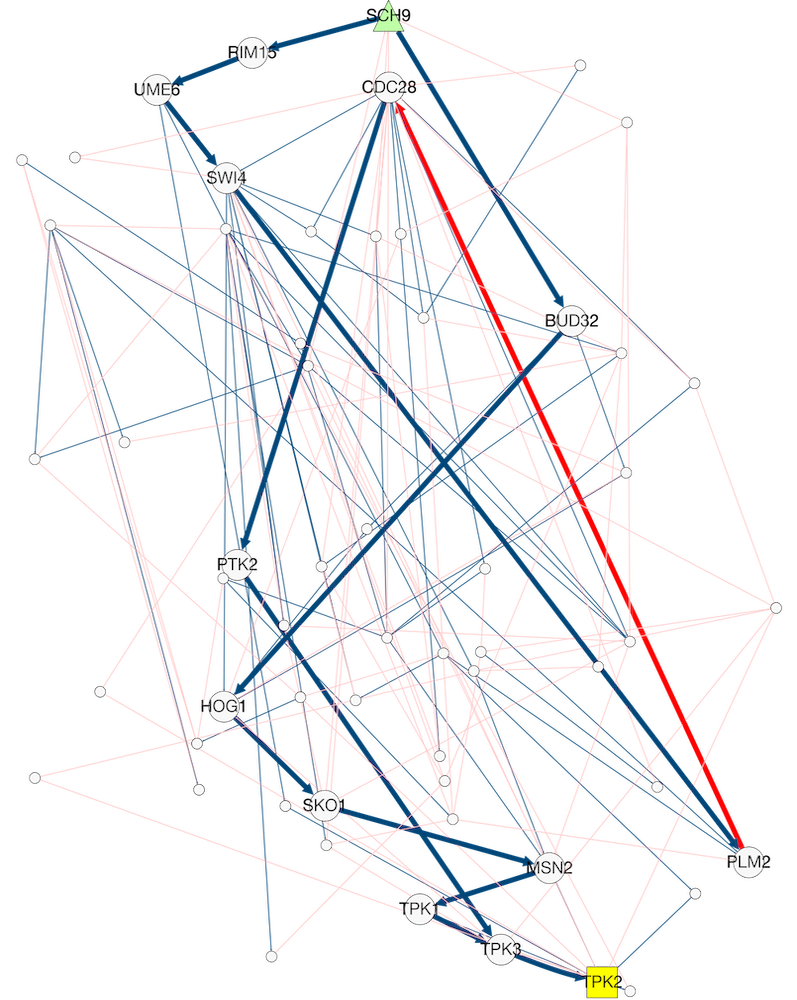}}
  \subcaptionbox{During SA100 process. \label{fig:flud-screenshots-G3-crowd-sa100-sa-intermediate}}[.24\textwidth][c]{%
    \includegraphics[width=0.24\textwidth,keepaspectratio,frame]{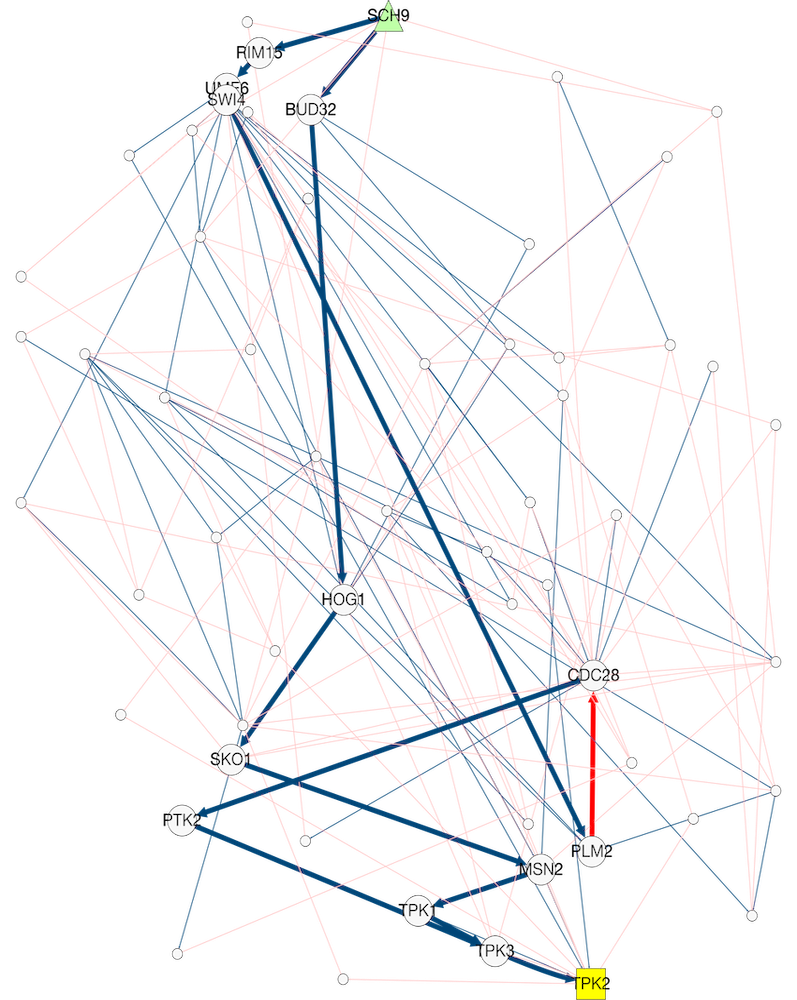}}
  \subcaptionbox{After SA100 finished. \label{fig:flud-screenshots-G3-crowd-sa100-sa-final}}[.24\textwidth][c]{%
    \includegraphics[width=0.24\textwidth,keepaspectratio,frame]{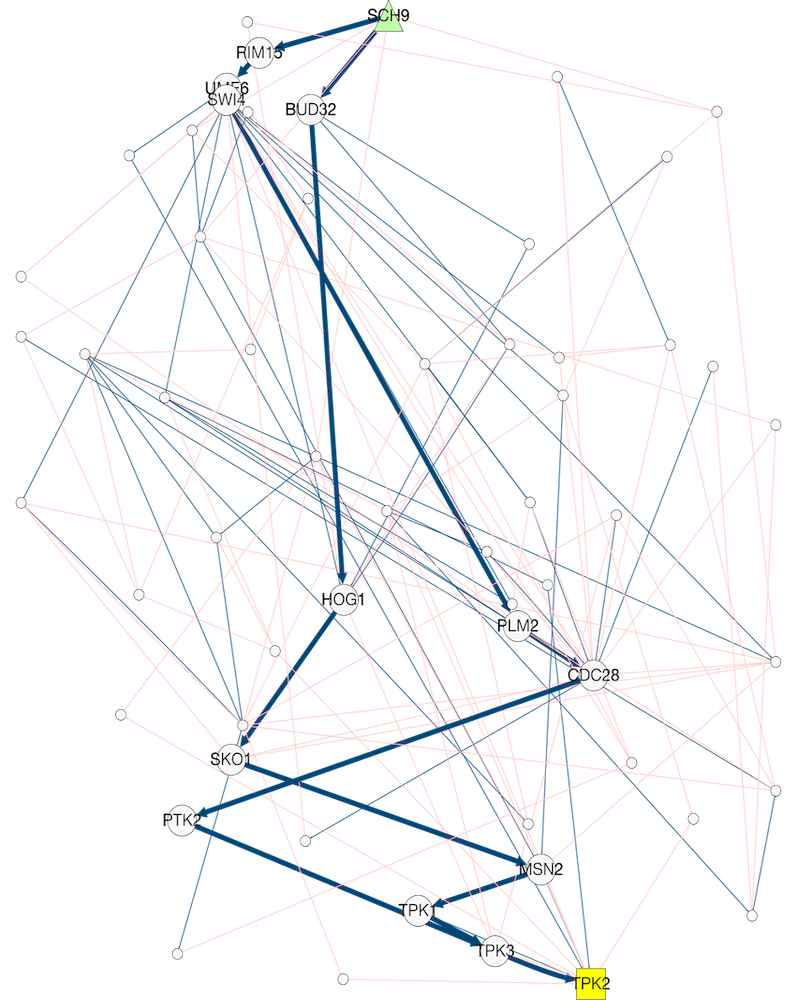}}
    
  \caption{G3 network layouts at four different stages in an example game, illustrating how crowd workers and simulated annealing build on each other's progress with the \emph{Crowd-SA100} approach. a) Layout before crowd worker started playing the game. b) Layout after the crowd worker finished playing the game. c) Layout during the \emph{SA100} annealing process.  d) Layout after the \emph{SA100} annealing process ends. For illustration purposes, the node and edge widths in these layouts have been modified to highlight two representative paths from source node to target node.}
  \label{fig:flud-screenshots-G3-crowd-sa100}
\end{figure}

We also found that simulated annealing's contributions in hybrid approaches increased in comparison to \emph{SA} for graph G3.
This observation is supported by Figure~\ref{fig:crowd-sa-contribution-total-score}, which shows that the simulated annealing in \emph{Crowd-SA100} achieved 1,875\% more average improvement in total score than in \emph{SA}.
To highlight the increase in simulated annealing's performance in \emph{Crowd-SA100} in comparison to \emph{Crowd}, we present an example which shows the progression of the layout as crowd workers and \emph{SA100} collaborate back to back during one of the games (Figure~\ref{fig:flud-screenshots-G3-crowd-sa100}). In this figure, we have highlighted two paths from source to target, for illustration purposes. Figure~\ref{fig:flud-screenshots-G3-crowd-sa100}a shows the layout at the beginning before crowd workers and simulated annealing start optimizing it. There are many upward pointing edges that need corrections to make the highlighted paths downward pointing. During the crowd task, the worker corrected all of these edges except one (Figure~\ref{fig:flud-screenshots-G3-crowd-sa100}b). However, one of the highlighted paths was still left with an upward pointing edge. Next, \emph{SA100} took over and fixed the other path via two major movements in the layout (Figure~\ref{fig:flud-screenshots-G3-crowd-sa100}c--d). We argue that \emph{SA100} was able to optimize DP in this scenario (Figure~\ref{fig:flud-screenshots-G3-crowd-sa100}b) because the number of moves required to create a downward pointing path was small in comparison to the starting layout (Figure~\ref{fig:flud-screenshots-G3-crowd-sa100}a).

\begin{figure}[H]
    \begin{tikzpicture}
        
    \node[inner sep=0pt] at (0, 0.8)
        {\includegraphics[width=\textwidth, keepaspectratio, clip, trim=0in 0in 0in 2.2mm]{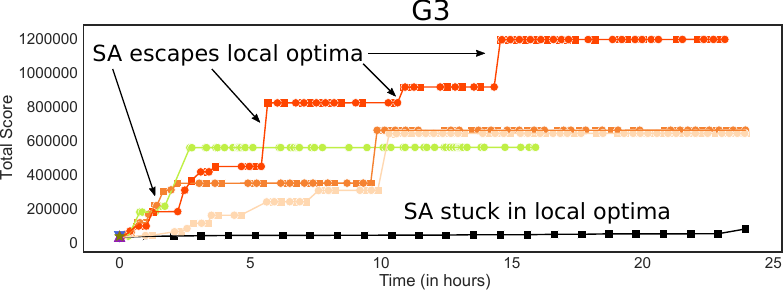}};
        
    \node[inner sep=0pt] at (0.5, -2.5)
        {\includegraphics[width=0.7\textwidth]{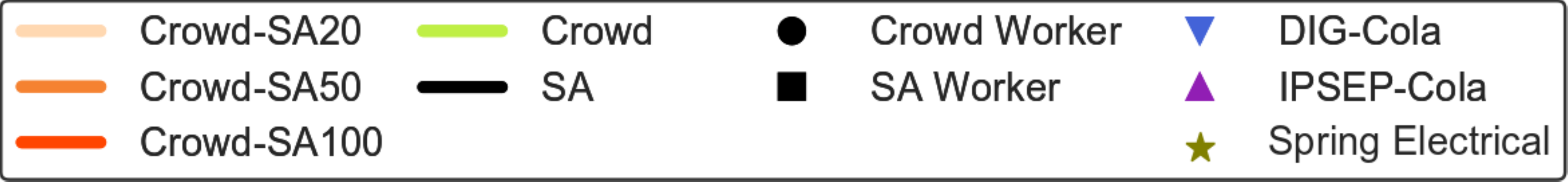}};
        
    \node[inner sep=0pt] at (0, 3.4) {\textbf{G3}};
    
    \end{tikzpicture}
    \caption{Scores achieved by each approach over time for network G3 for a representative sequence. The $x$-axis represents the time taken in hours to reach a particular score and the $y$-axis represents the score. The circle and square markers correspond to crowd workers and simulated annealing respectively.}
    \label{fig:G3-representative-game-sequences}
\end{figure}

We attribute such hybrid collaborations as the main reason behind simulated annealing's ability to escape local optima (see the nearly vertical lines ending in squares in Figure~\ref{fig:G3-representative-game-sequences}) in the hybrid approaches, unlike in the case of SA. Overall, these results indicate the importance of contributions from the crowd workers in the hybrid approach, since \emph{SA100} is similar to the early annealing phase in SA (Figure~\ref{fig:illustration-simulated-annealing-schedules}).

\subsection{How did the players play the Flud game?}
\label{subsec:RQ4-clue-analysis-results}

\begin{figure}[H]
 \centering
  \includegraphics[width=\textwidth]{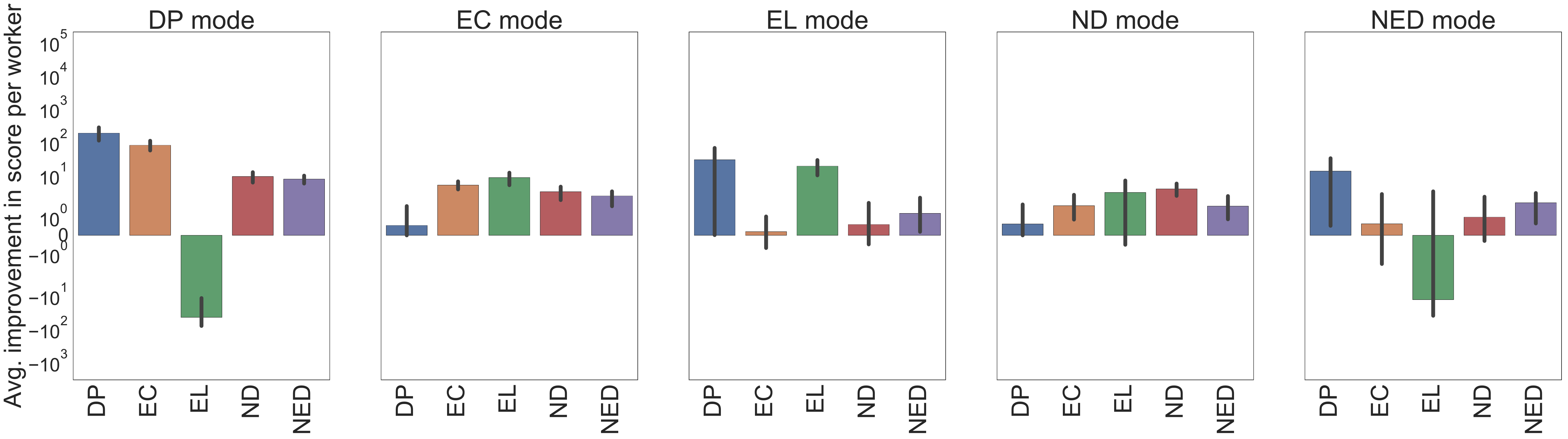}
  \caption{Distribution of average improvement in per-criterion scores achieved by crowd workers in each mode.}
  \label{fig:gain-criterion-score-in-criterion-specific-mode-bar-plot}
\end{figure}

\begin{figure}[htbp]
    \begin{tikzpicture}
    \node[inner sep=0pt] at (0,1)
        {\includegraphics[width=0.55\textwidth]{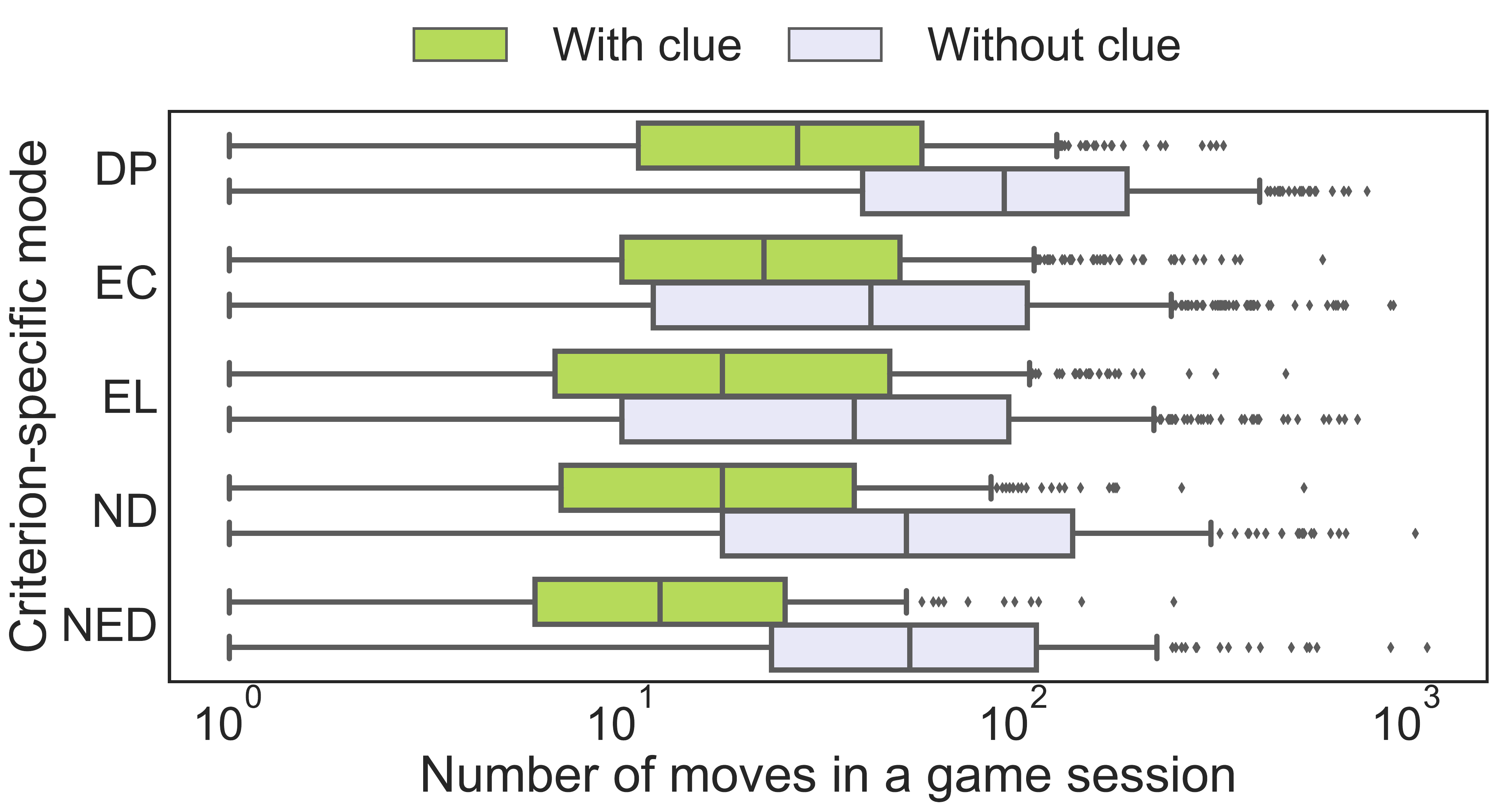}};
    
    \node[inner sep=0pt] at (0,-6)
        {\includegraphics[width=0.78\textwidth]{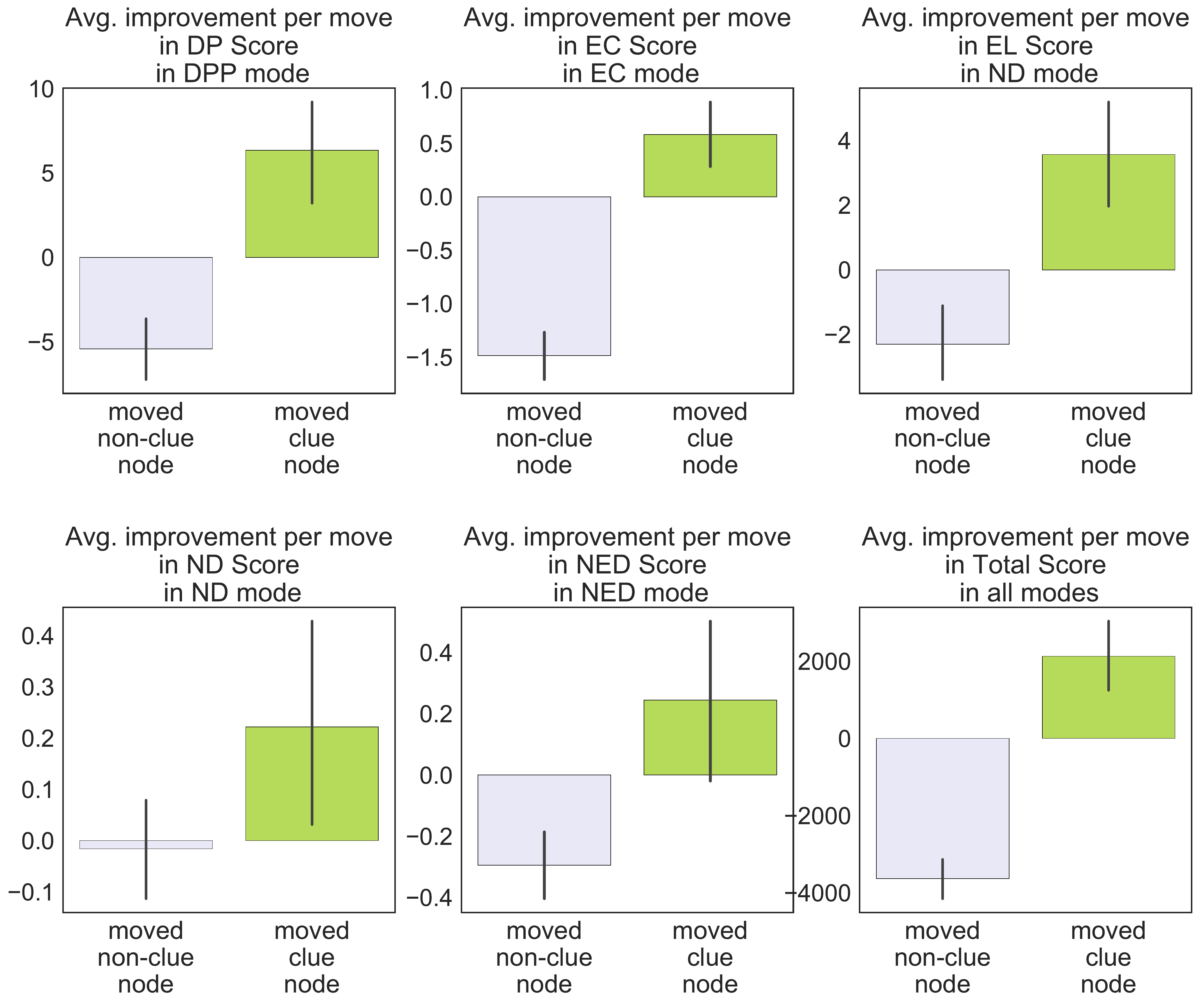}};

        
    \node[inner sep=0pt] at (-6.5, 2.5) {\textbf{(A)}};
    
    \node[inner sep=0pt] at (-6.5, -1.8) {\textbf{(B)}};
    
    \end{tikzpicture}
    \caption{ A) Distribution of number of times crowd workers moved nodes with or without a clue in a criterion specific mode. B) Distribution of average improvement in per-criterion score per move achieved by crowd workers with or without using the criterion-specific clue.}
    \label{fig:gameplay-statistics}
\end{figure}

We analyzed the crowd workers' gameplay, their interaction with the game elements in the Flud interface, and its impact on their performance (Figures~\ref{fig:gameplay-statistics} and \ref{fig:gain-criterion-score-in-criterion-specific-mode-bar-plot}) using the data from Experiment 2. Figure~\ref{fig:gameplay-statistics}A shows the distribution of the number of moves made by the crowd workers in each criterion-specific mode. We found that the crowd workers not only made the highest number of moves (median=126) in the DP mode, but also used the downward pointing path clue (median=28) more often in comparison to other criterion-specific clues (median=19). We attribute this behavior to the higher increase in score 
when the crowd workers are assigned DP mode (mean improvement in total score per session $\approx 73,629$) and use the downward pointing path clue (mean improvement in DP score per move $\approx 6.3$). This observation is supported by Figure~\ref{fig:gameplay-statistics}B and Figure~\ref{fig:gain-criterion-score-in-criterion-specific-mode-bar-plot}. In contrast to the DP criterion, we found that the node-edge separation criterion clue, which gave the lowest average improvement per move  (Figure~\ref{fig:gameplay-statistics}B), was used the least across all criteria (Figure~\ref{fig:gameplay-statistics}A).

Next, in order to understand if the clues helped the players, we compared the average improvement per move for each criterion (Figure~\ref{fig:gameplay-statistics}B). Our results show that when the crowd workers moved a node not involved in a clue, they lost points and decreased the criterion-specific score. In contrast, when players changed the position of a node that was highlighted in a clue, the criterion-specific score increased, on average. These results indicate that the criterion-specific clues helped the crowd workers to improve the criterion-specific scores.

Then, we analyzed the improvement in per-criterion scores made by crowd workers in each criterion-specific mode as shown in Figure~\ref{fig:gain-criterion-score-in-criterion-specific-mode-bar-plot}. We found that the crowd workers improved the criterion-specific score while playing the game in a criterion-specific mode. 



\section{Discussion}



In this paper, we presented Flud, an online game with a purpose that allows humans with no expertise to design biologically meaningful graph layouts with the help of algorithmically generated suggestions. Below, we discuss some key takeaways from our evaluation, focusing on overall performance differences, reflections, and design implications.

\subsection{Crowd outperforms algorithms}
We found that crowdsourcing the layout design task on Flud can generate more meaningful layouts than state-of-the-art algorithms for laying out biological networks. 
Specifically, our results indicated that crowd workers, in general, are considerably better than the state-of-the-art algorithmic approaches at the downward pointing path task, i.e., laying out downward pointing paths from source nodes to target nodes in a directed network. We believe that there are two main reasons for the crowds' superior performance. 

First, biological networks have feedback loops, and the crowd workers are able to carefully position the nodes in a feedback loop with respect to each other in order to optimize the number of downward paths. In contrast, an optimization method such as simulated annealing may get stuck in local optima in the presence of feedback loops or cycles, as seen in the network G3. Second, algorithmically generated suggestions (clues) played an essential role in the players' performance. These suggestions streamlined
the broader layout design goal by asking the players to carefully solve a smaller puzzle (or micro-task), e.g., to move the suggested elements in a certain way in the given network so as to increase the total score. In other words, we were able to reduce the overall challenge level by algorithmically breaking down the layout design task into smaller puzzles that benefit from human judgment and are solvable by untrained crowd workers.

In contrast, algorithmic approaches such as \Digcola and \Ipsep focus on optimizing the number of downward pointing edges instead of paths. We stress that merely counting the number of downward pointing edges~\cite{purchase-helen-metrics-2002} is not sufficient to capture the flow of information. For example, a downward pointing path from a receptor (a protein molecule that receives signals from outside a cell) through intermediary proteins to a transcription factor (a protein that controls the cell's response to the signal) might be crucial to understand the underlying mechanism of cell signaling in a biological network. 

While the \emph{Crowd} approach was able to outperform the automated methods, the hybrid \emph{Crowd-SA100} approach offered faster rate of improvement of the score. Moreover, our results indicated that there is a symbiotic relationship between crowd workers and simulated annealing in the hybrid approaches. On the one hand, we believe that the crowd workers improve the layouts such that it is in a state where simulated annealing can make a non-trivial modification to the layout that increases the score. On the other hand, simulated annealing relieves crowd workers of redundant and trivial tasks by improving the layout whenever possible between crowd worker sessions.

\subsection{Reflecting on design decisions}
We now discuss some of the decisions we made while planning the scoring system, experiments, and incentives.
\subsubsection{Scoring system}
One key decision we made was to normalize the per-criterion scores used to compare the quality of different layouts. While the downward-pointing paths criterion is new, quantitative metrics for rest of layout criteria have been defined in prior work ~\cite{purchase-helen-metrics-2002, purchase-carrington-graph-aesthetics-empirical-2001}. However, many of these metrics are on different scales, which makes it challenging to compare layouts of different network sizes.  Additionally, we cannot use the same priorities (weights) to compute the weighted layout score for different network sizes. Therefore, we normalized each per-criterion score 
to lie between 0 and 1, with higher scores being more desirable. This strategy allowed us to use the same priorities for networks G1--G3 in our experiments. 

Minimizing the number of edge crossings is one of the most popular aesthetic criteria. It has been empirically validated as an important measure for good network drawing~\cite{purchase-carrington-graph-aesthetics-empirical-2001,taylor-applying-2005,tamassia-batini-automatic-drawings-1988}. As an alternative to $\mathit{EC}$ score, we considered \emph{crosslessness}~\cite{purchase-helen-metrics-2002,kwon-would-layout-look-like-tvcg-2018}. 
We discarded it as an option since its values were very close to $EC(G)$ for all the networks we used in our experiments. Moreover, we did not see any benefit of using crosslessness over $EC(G)$ to give feedback to the game players. Hence, we decided to use the $EC(G)$ metric.

Another design decision we made was to choose a very high priority (400) for the downward pointing paths criterion score (DP) in our experiments. We decided to assign the highest priority to DP because it is our sole domain-oriented criterion, and we wanted it to be preferred over the aesthetic layout criteria in case of conflicts. Moreover, due to our poor approximation of the maximum possible number of downward pointing paths in networks with cycles, the DP scores were prone to be very low for such networks. If we had selected uniform priorities, due to these low DP scores, algorithms such as \SA would choose to optimize for aesthetic layout criteria in case of a conflict, despite even trying the priority of four (greater than the priority of rest of the criteria) for the DP criterion. 
Therefore, we decided to select a very high priority of 400 so that the weighted contribution of downward pointing paths to the overall layout score is generally higher than other per-criterion scores. 




\subsubsection{Worker incentives}
In this work, we empirically showed that paid crowd workers recruited from Amazon Mechanical Turk can help researchers by successfully playing the Flud game and achieving high scores that correspond to improved network layouts. We incentivized these workers to submit high-quality layouts by paying them based on their scores~\cite{ho2015incentivizing}. Paid workers offered useful advantages for this research, allowing us to temporarily sidestep the need to build an online community with a critical mass of game players while rapidly prototyping and iterating on the Flud game mechanics in a scaleable, controlled environment.  However, prior work in crowdsourcing suggests~\cite{rogstadius2011assessment, borromeo2016investigation} that unpaid workers or volunteers submit higher quality work in comparison to paid workers. Moreover, the volunteers may be more motivated to perform the task if it supports a worthy cause (for example, scientific research)~\cite{katsuragawa2019pledgework}.
Given that Flud aims to aid scientific researchers, we expect an intrinsically motivated subset of volunteers would spend more time playing the game and make more improvements in a game session. Moreover, as the players get familiar with the task, we expect high-performing players to develop layout strategies they can reuse across different games (networks), as has been observed in other games like EteRNA and FoldIt. In future work, we are exploring what modifications are necessary to the Flud platform to motivate volunteer gameplay, and consequently, how volunteer player performance compares with paid crowd workers from this study.

Currently, Flud only displays the current top score during a game and does not have a leaderboard that shows the total points earned by the top performers. However, prior work shows that leaderboard can play an important role in games to motivate high performance~\cite{hamari2014does}. In this work, we decided not to use a leaderboard since our participants were crowd workers on MTurk who are primarily motivated by compensation rather than by competitive gameplay. Designing a leaderboard to motivate volunteer gameplay in a context where players are essentially collaborating with each other is a promising future direction.  


\subsection{Design implications}
We now discuss some practical implications for our work.

\subsubsection{Mixed-initiative systems are effective in layout tasks}
Prior work shows that humans are adept at spatial reasoning skills and strategic thinking~\cite{williams2016toward}.  These qualities allow users to successfully explore a search space by outperforming algorithms at tasks that require visual perception. In our work, we found that humans are good at a task that requires observation and spatial arrangement to maintain a certain direction of network flow. Corroborating prior work, we found that combining human intelligence with automated approaches can achieve results better than what either could produce alone. One implication of this finding is that systems can leverage mixed-initiative settings to solve for other types of layout tasks that require careful relative arrangement of entities in a two-dimensional space. Examples of such tasks, which manifest in many creative and analytical disciplines, include drawing integrated circuits, mapping social networks, and designing interior spaces in houses and buildings.


%

\subsubsection{Enable preference elicitation to balance multiple criteria}
Balancing multiple criteria in a design task can be challenging for novices. Prior work found that novices are limited by their incomplete knowledge and therefore need preference elicitation support from the system~\cite{an2011mixed}. For example, it may be difficult for a non-expert to tell whether one criterion is more important (and to what extent) than another criterion. It may be even more difficult for a user to understand the consequences of choosing to focus on one constraint before another.  Therefore, preference elicitation support is required to help the non-expert user decide on a certain course of action. Flud overcomes this challenge by transparently showing the per-criterion scores, their priorities, and the relative change (increase/decrease) after each move. These features allows the players to take informative decisions while trying to balance multiple criteria.

\subsubsection{Computationally assign modes to break players out of unproductive play strategies} Prior work shows that in human computation games, players can gravitate towards gameplay strategies that are not advised under certain situations~\cite{williams2016toward}. We saw a similar behavior when we allowed the players to choose the mode instead of fixing a mode at the start of the game. We found that players gravitate towards distance based modes such as edge length and node distribution, ignoring the higher priority modes like downward pointing paths. We attributed this behavior to players inexpertise in choosing the appropriate criterion to focus on at a given stage in the game. We overcame this challenge by passing the control to the Flud system and letting it decide the modes for the player. In our work, we empirically showed that assigning the modes in the order of their priorities generates better results than assigning them randomly. 

\subsubsection{Automatic suggestions offer guidance when players are uncertain} Despite having superior intelligence and cognitive skills, players can sometimes be overwhelmed by the complexity of the game. Under such circumstances, Flud automatically provides players with a suggestion via the clue feature. These suggestions guide players by focusing their attention on a smaller, more tractable puzzle inside the overall game. This approach not only reduces the challenge level, but also helps the player make progress when they get stuck.

\subsection{Limitations}
\label{sec:limitations}

One of the limitations of our evaluation is that the number of networks is small. Since we sought to evaluate multiple game sequences for each of the proposed approaches, we were limited to a small number of networks. A larger number of networks would allow us to show the applicability of the proposed methods to various types of networks. However, the main goals of this paper were to show that the proposed approaches perform better than the automated methods for networks with large number of cycles and that our results are replicable. Therefore, we used three networks with different numbers of cycles and repeated the game sequence for each approach three times.

Another limitation of our Flud implementation is that its network rendering is currently optimized 
for small networks ($<100$ nodes). 
Prior work shows that people with no expertise in network drawing find it difficult to handle large networks~\cite{dwyer-north-comparison-user-auto-graph-layouts-2009}. Therefore, we believe that optimizing the system to support larger networks is not the way forward. Instead, we propose to split the larger networks and ask the Flud players to work on small subnetworks. 
Yuan~\etal \cite{yuan-xin-intelligent-many-users-itvcg-2012} proposed a strategy to combine user-generated layouts into a single layout for a large network. Adapting their method to support the downward pointing path criterion for large networks on Flud would be an important future goal.

\section{Conclusion}
In this work, we explored the potential of crowd-algorithm collaboration for visualizing biological networks. We describe how we gamified the visualization task and made it more accessible to humans, even if they have no biological or computer science expertise. Our results show that such a collaboration between humans and algorithms leads to higher scoring layouts than either from humans or algorithms alone.  Our contributions include (i) a novel mixed-initiative game with a purpose that combines crowdsourcing with computational engines to create high-quality visualizations of biological networks and (ii) experiments that provide empirical evidence of the benefits of mixed-initiative layout schemes compared to algorithmic baselines.

\begin{acks}
We thank Lee Lisle, Parker Irving, and Jeffrey Law. This research was supported by NIH grant 1UH2CA203768-01.

\end{acks}

 




\clearpage

\renewcommand{\thefigure}{A\arabic{figure}}
\renewcommand{\thetable}{A\arabic{table}}
\renewcommand{\thesection}{\arabic{section}}

\setcounter{figure}{0}
\setcounter{table}{0}
\setcounter{section}{0}

\section*{Appendix}

\crefalias{section}{suppsection}
\crefalias{figure}{suppfigure}
\crefalias{table}{supptable}

\section{How we compute per-criterion scores?} 
\label{sec:layout-criteria-computation-details}
To prevent pathological layouts that move nodes indiscriminately far apart from each other, we set the per-criterion scores to zero if any node is moved outside a bounding box of fixed page width $w$ and height $h$. We chose $w=5000$ and $h=6000$ in our implementation. The aspect ratio of our bounding box approximately matches that of a page in a scientific journal, i.e., $0.8$. 


\begin{enumerate}
\item \emph{Downward pointing paths}: We count the number of downward-pointing paths in a layout using a dynamic program that runs in time linear in the number of edges in the network. Note that this algorithm does not assign node positions in order to maximize the number of downward-pointing paths since this task will be undertaken by Flud players. The subgraph composed only of downward-pointing edges is acyclic. Hence, for any node $v$ in the network, we can compute $\pi(v)$, the number of downward-pointing paths that start at $v$ using the following recurrence:
$$\pi(v) = \sum\limits_{\substack{(v,u) \text{ is} \\ downward \\ pointing}} \pi(u)$$
Here, the sum is taken only over outgoing neighbors of $v$ that have smaller $y$-coordinate than $u$, i.e., if the edge $(v,u)$ is pointing downward. The base case is a node $v$ that has no downward-pointing edges leaving it: $\pi(v) = 0$ in this case. We can compute the total number $\pi(G)$ of downward-pointing paths in the graph by summing $\pi(v)$ over all nodes $v$ that have no downward-pointing edges entering them. Note the maximum possible value of $\pi(G)$ is the number of paths (all directed paths, not just downward-pointing ones) in $G$. We count this number~$\rho(G)$ using a depth-first search based algorithm. Although this algorithm has worst-case running time that is exponential in the graph size, it was very efficient in our experiments. 
Finally, we compute a normalized \emph{downward pointing} score as follows: 
$$\mathit{DP}(G) = \frac{\pi(G)}{\rho(G)}.$$


\item \emph{Non-crossing edge pairs}: We count the number~$\chi(G)$ of edge pairs that do not cross by checking for each pair of edges whether they intersect or not. Since the maximum number of edge crossings possible in $G$ is $m(m-1)/2$, we compute the \emph{non-crossing edge pairs score} as  
$$ \mathit{EC}(G) = \frac{2\chi(G)}{m(m-1)}.$$
\item \emph{Edge length}: A trivial but undesirable solution that satisfies this criterion is to place all nodes at the same location. Note that such a layout may have a high score despite the node distribution constraint (described next) because of the per-criterion priorities. Therefore, we require every edge to have a minimum fixed length (300 pixels, in our implementation). We now define the cost $c(e)$ of an edge to be equal to its length $l(e)$ if $l(e) \geq 300$ or equal to a large number (say, $10,000$), otherwise. We normalize the cost of each edge by the largest possible edge length (a diagonal of the screen) and compute the \emph{edge length score} of a layout as
 $$\mathit{EL}(G) = \max \Big\{0, 1 - \frac{1}{m}\sum_{e \in E} \frac{c(e)}{\sqrt{w^2 + h^2}} \Big\}$$ 
where $w$ and $h$ represent the width and height of the bounding box, respectively. 


 \item \emph{Node edge separation}: To evaluate the fidelity of a layout to this aesthetic criterion, we compute the distance $d(v,e)$ between every node $v$ and every edge $e$ in $G$ as follows: we check if the projection of $v$ into the line containing $e$ falls within $e$. If it does, $d(v,e)$ is the distance between $v$ and this projection. Otherwise, $d(v,e)$ is the distance from $v$ to the endpoint of $e$ that is closer to $v$. With these values in hand, we define the \emph{node edge separation score} as 
$$\mathit{NED}(G) = \frac{1}{n}  \mathlarger{\sum}\limits_{u \in V} \frac{\min\limits_{\substack{e=\{t,v\} \in E \\ u \neq t \neq v}} d(u,e)}{\sqrt{w^2 + h^2}}$$
where $w$ and $h$ represent the width and height of the bounding box, respectively. 
\end{enumerate}

\section{Fine-tuning procedure}
At the end of the sequence, we fine-tuned the best layout so far to generate the final output layout by using it as the input to simulated annealing with a small value of the initial temperature ($T_0=10$). Due to the low initial temperature, the random moves are ``local,'' i.e., a node can move only to nearby position. Additionally, we did not accept moves that decrease the score. We expected fine tuning to improve the distance-based components of the layout scores without dramatically affecting the components for downward pointing paths and edge crossings. In their work on incremental improvement of layouts of planar networks, Harel and Sardas \cite{harel-sardas-incremental-planar-1995}  concluded that when applied to a preprocessed network, fine-tuning can yield significant improvements over simulated annealing without any preprocessing. This work serves as an inspiration for us to use fine-tuning as a way to make local moves that improve a player's layout. 

\section{Supplementary Figures}

\begin{figure}[H]
 \centering
  \includegraphics[width=\textwidth]{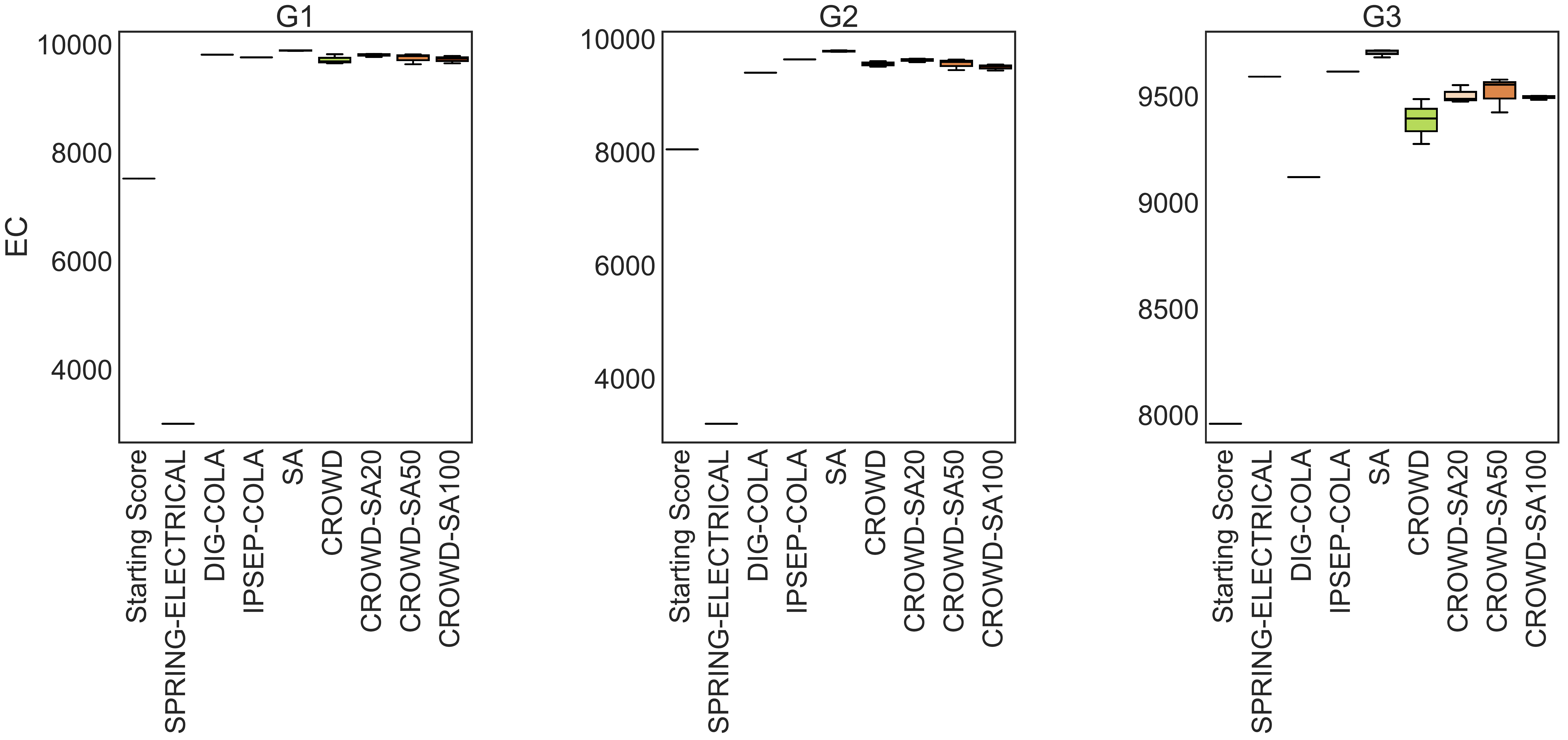}
  \caption{Distributions of the EC scores of layouts created by different approaches. CROWD-SAX corresponds to the hybrid approach where crowd workers and SAX (say SA20) iteratively improve upon one another's results.}
  \label{fig:ec-performance-analysis-box-plot}
\end{figure}

\begin{figure}[H]
 \centering
  \includegraphics[width=\textwidth]{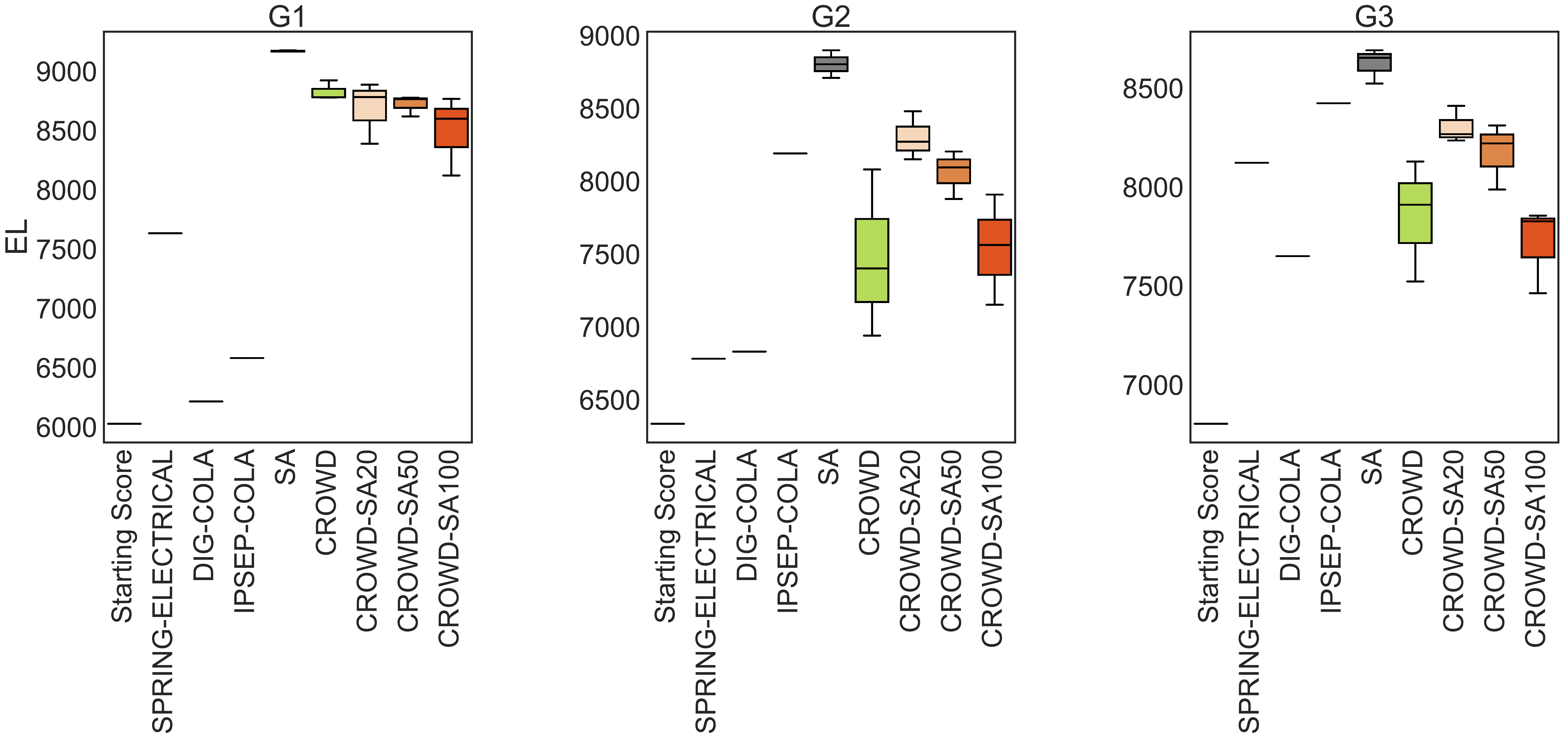}
  \caption{Distributions of the EL scores of layouts created by different approaches. CROWD-SAX corresponds to the hybrid approach where crowd workers and SAX (say SA20) iteratively improve upon one another's results.}
  \label{fig:el-performance-analysis-box-plot}
\end{figure}

\begin{figure}[H]
 \centering
  \includegraphics[width=\textwidth]{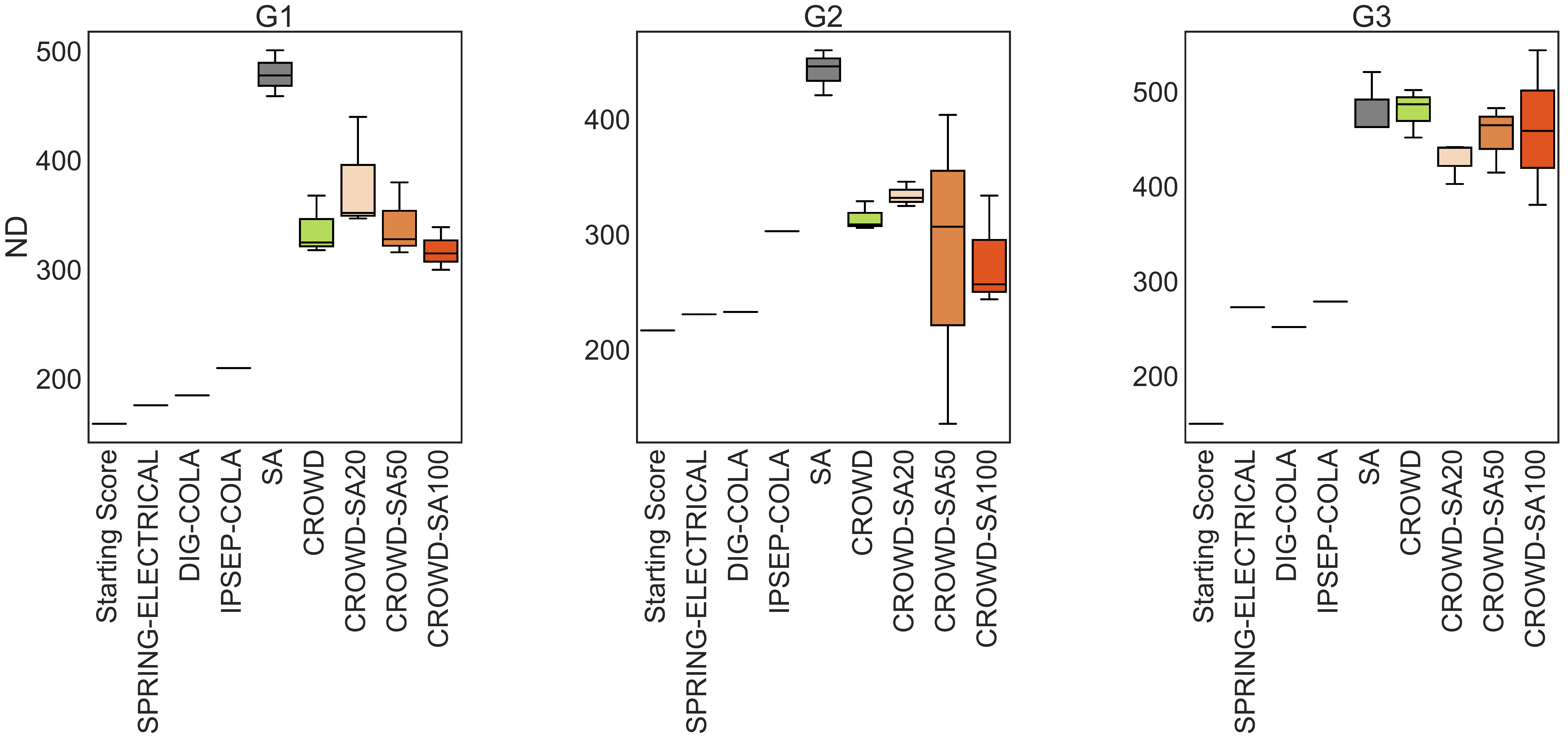}
  \caption{Distributions of the ND scores of layouts created by different approaches. CROWD-SAX corresponds to the hybrid approach where crowd workers and SAX (say SA20) iteratively improve upon one another's results.}
  \label{fig:nd-performance-analysis-box-plot}
\end{figure}

\begin{figure}[H]
 \centering
  \includegraphics[width=\textwidth]{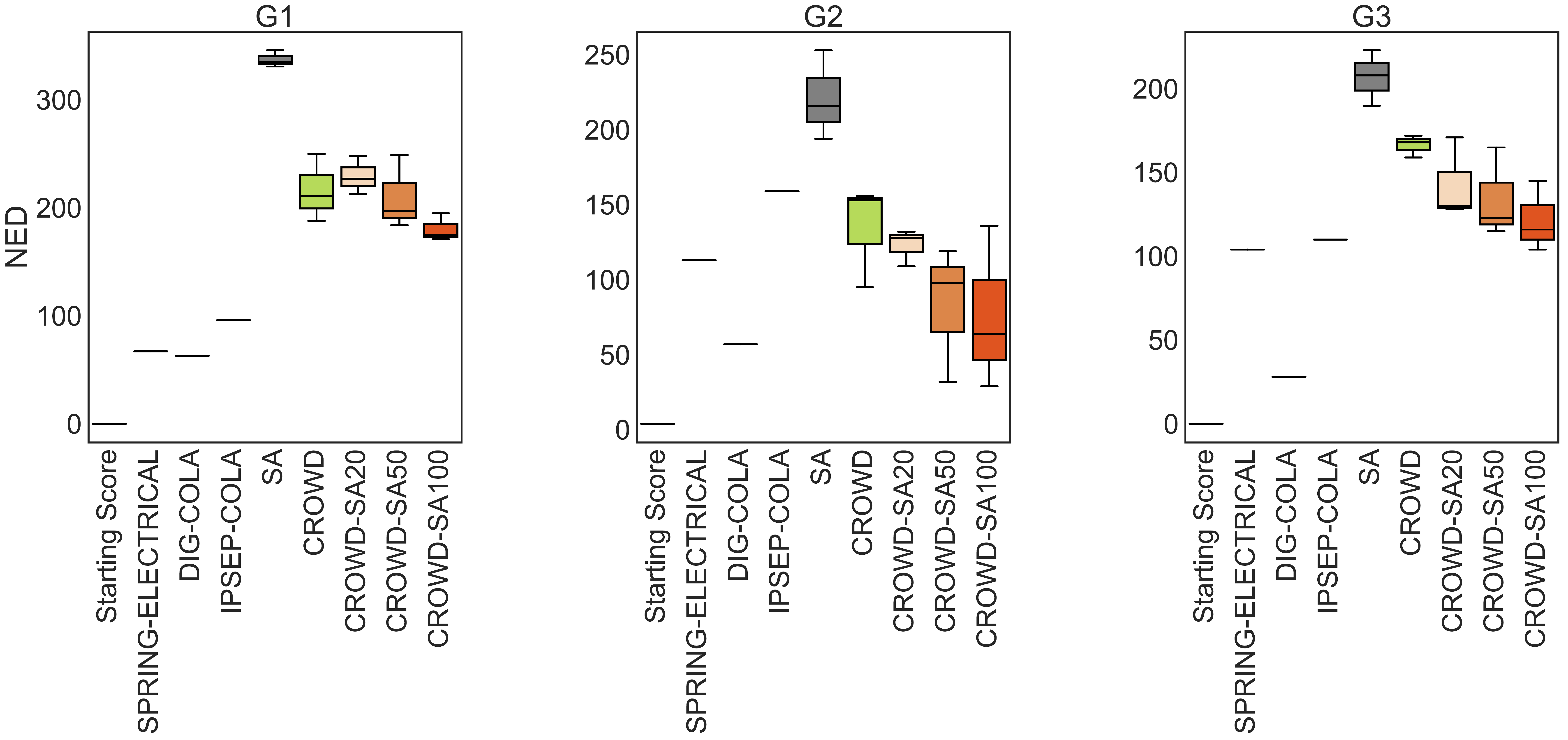}
  \caption{Distributions of the NED scores of layouts created by different approaches. CROWD-SAX corresponds to the hybrid approach where crowd workers and SAX (say SA20) iteratively improve upon one another's results.}
  \label{fig:ned-performance-analysis-box-plot}
\end{figure}

\begin{figure}[H]
 \centering
  \includegraphics[width=\textwidth, keepaspectratio, clip, trim=0in 9in 5.1in 0in]{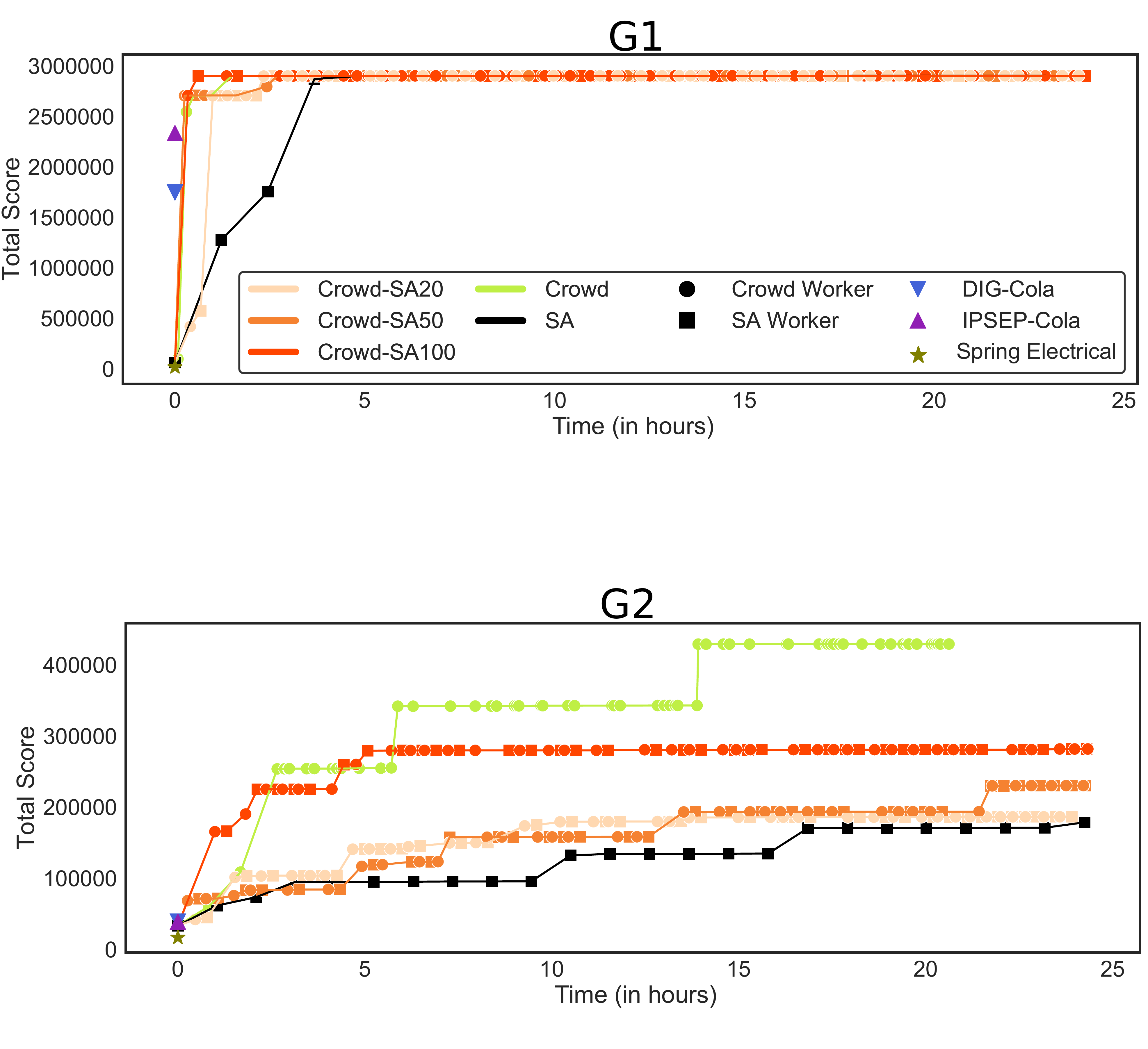}
  \caption{
  Plots showing the rate at which each approach improved the overall layout score. Each row corresponds to a network and shows the time taken by different approaches to increase the total layout score. The $x$-axis represents the time taken in hours to reach a particular score and the $y$-axis represents the score. 
  The circle and square markers correspond to crowd workers and simulated annealing respectively. 
  }
  \label{fig:time-analysis}
 \end{figure}
 
\clearpage

\begin{figure}[H]
    \begin{tikzpicture}
    \node[inner sep=0pt] at (0, 0.0)
        {\includegraphics[width=\textwidth]{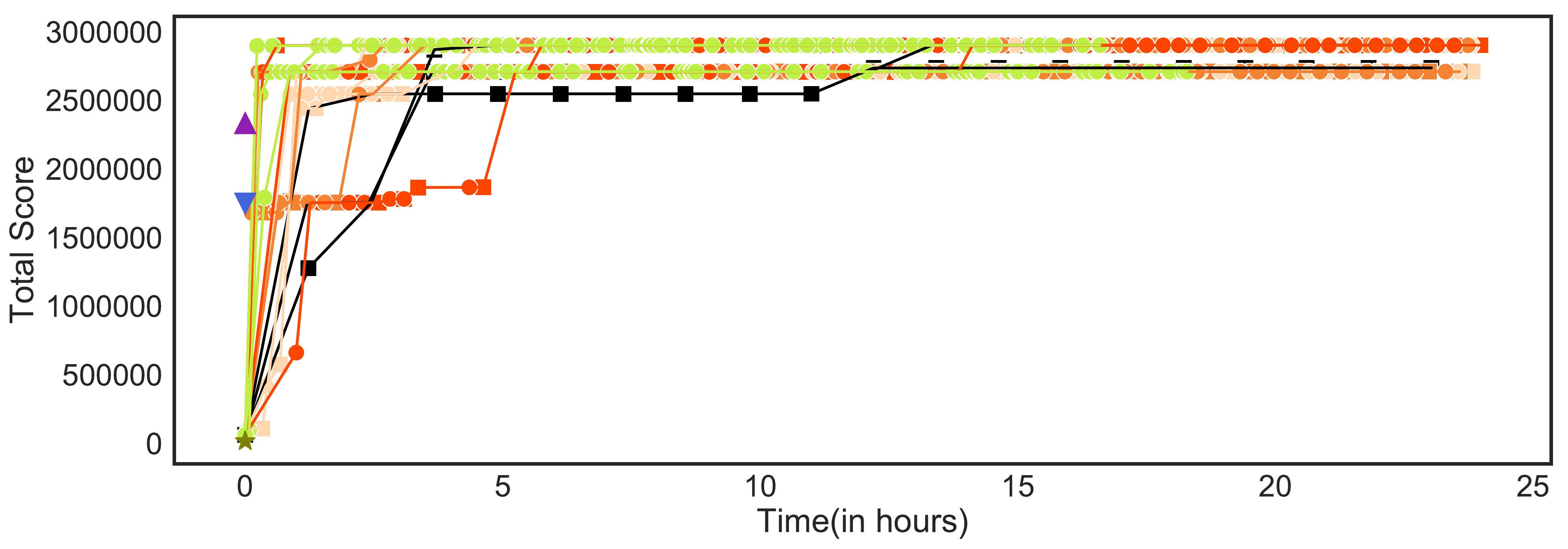}};
        
    \node[inner sep=0pt] at (1.8, -0.9)
        {\includegraphics[width=0.7\textwidth]{time-analysis-legend.pdf}};
        
    \node[inner sep=0pt] at (0, -5)
        {\includegraphics[width=\textwidth]{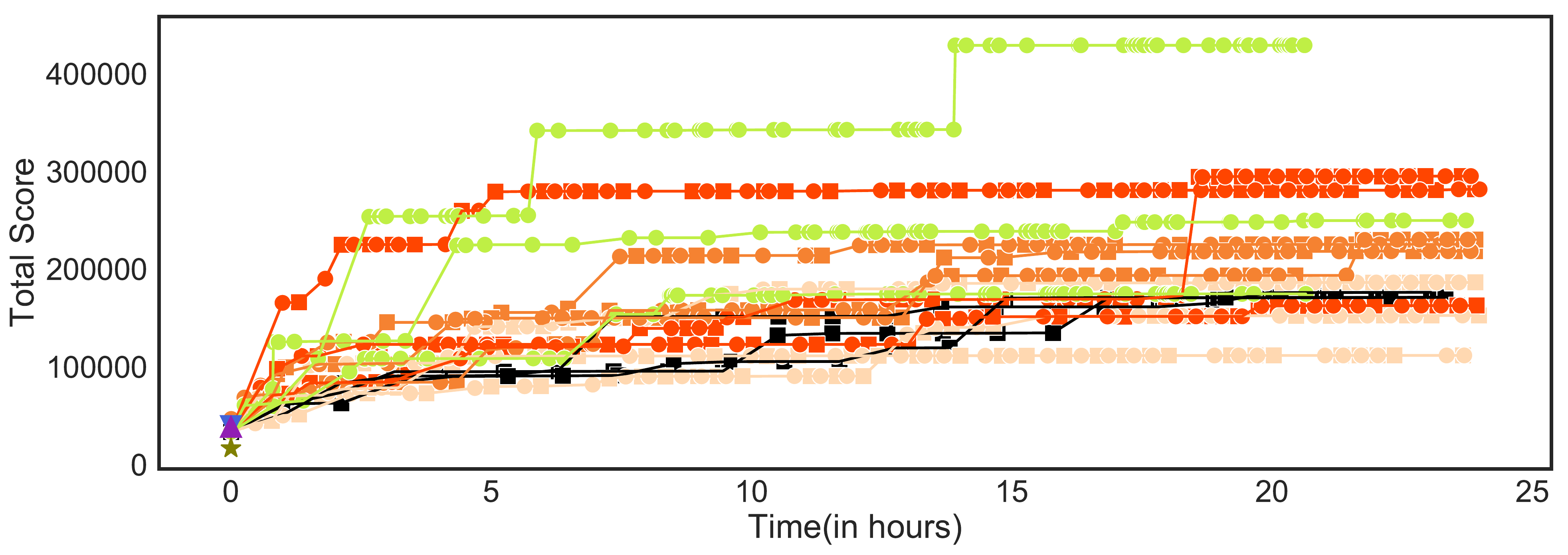}};
        
    \node[inner sep=0pt] at (0, -10.1)
        {\includegraphics[width=\textwidth]{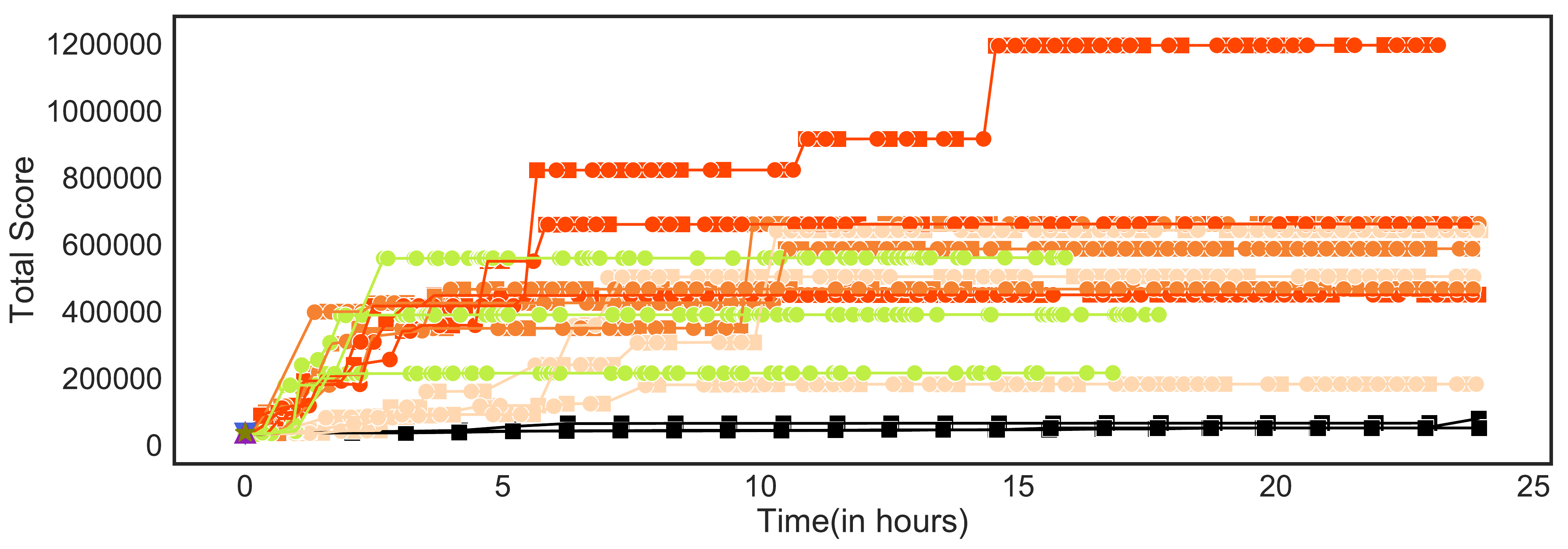}};
        
    \node[inner sep=0pt] at (0.5, 2.5) {\textbf{G1}};
    \node[inner sep=0pt] at (0.5, -2.5) {\textbf{G2}};
    \node[inner sep=0pt] at (0.5, -7.6) {\textbf{G3}};
    
    \end{tikzpicture}
    \caption{   
    Each row corresponds to a network. Each row shows the time taken by three game sequences of different approaches to increase the total layout score. The $x$-axis represents the time taken in hours to reach a particular score and the $y$-axis represents the score. The circle and square markers correspond to crowd workers and simulated annealing respectively. 
    }
    \label{fig:all-game-sequences}
\end{figure}

\begin{figure}[H]

   \subcaptionbox{Crowd approach. \label{fig:flud-best-layout-G3}}[.38\linewidth][c]{%
    \includegraphics[width=0.38\linewidth,keepaspectratio]{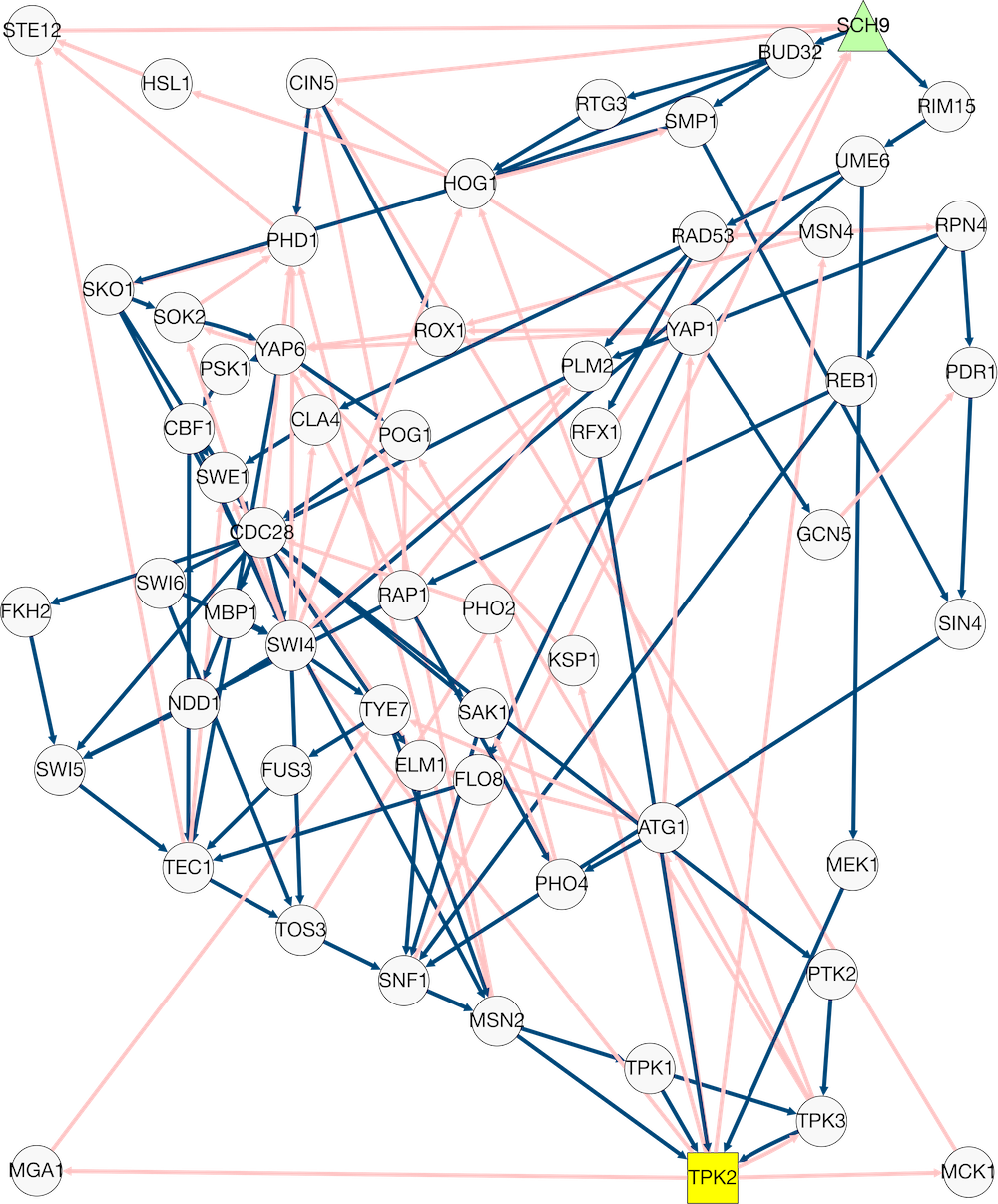}}
    \quad
    \subcaptionbox{Dig-Cola. 
    \label{fig:digcola-best-layout-G3}}[.38\linewidth][c]{%
    \includegraphics[width=0.38\linewidth,keepaspectratio]{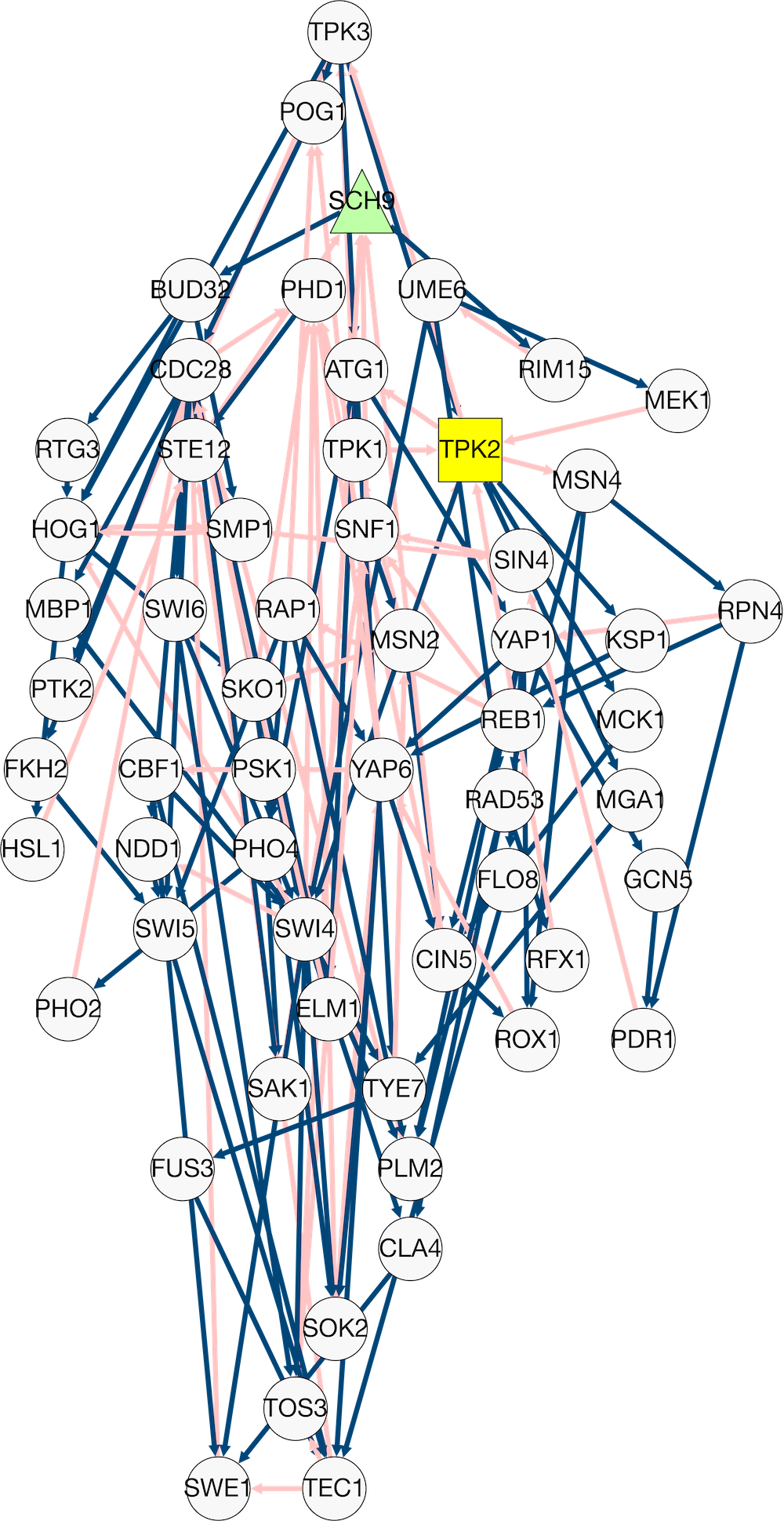}}
    \quad
  \subcaptionbox{SA. 
   \label{fig:sa-best-layout-G3}}[.38\linewidth][c]{%
    \includegraphics[width=0.38\linewidth,keepaspectratio]{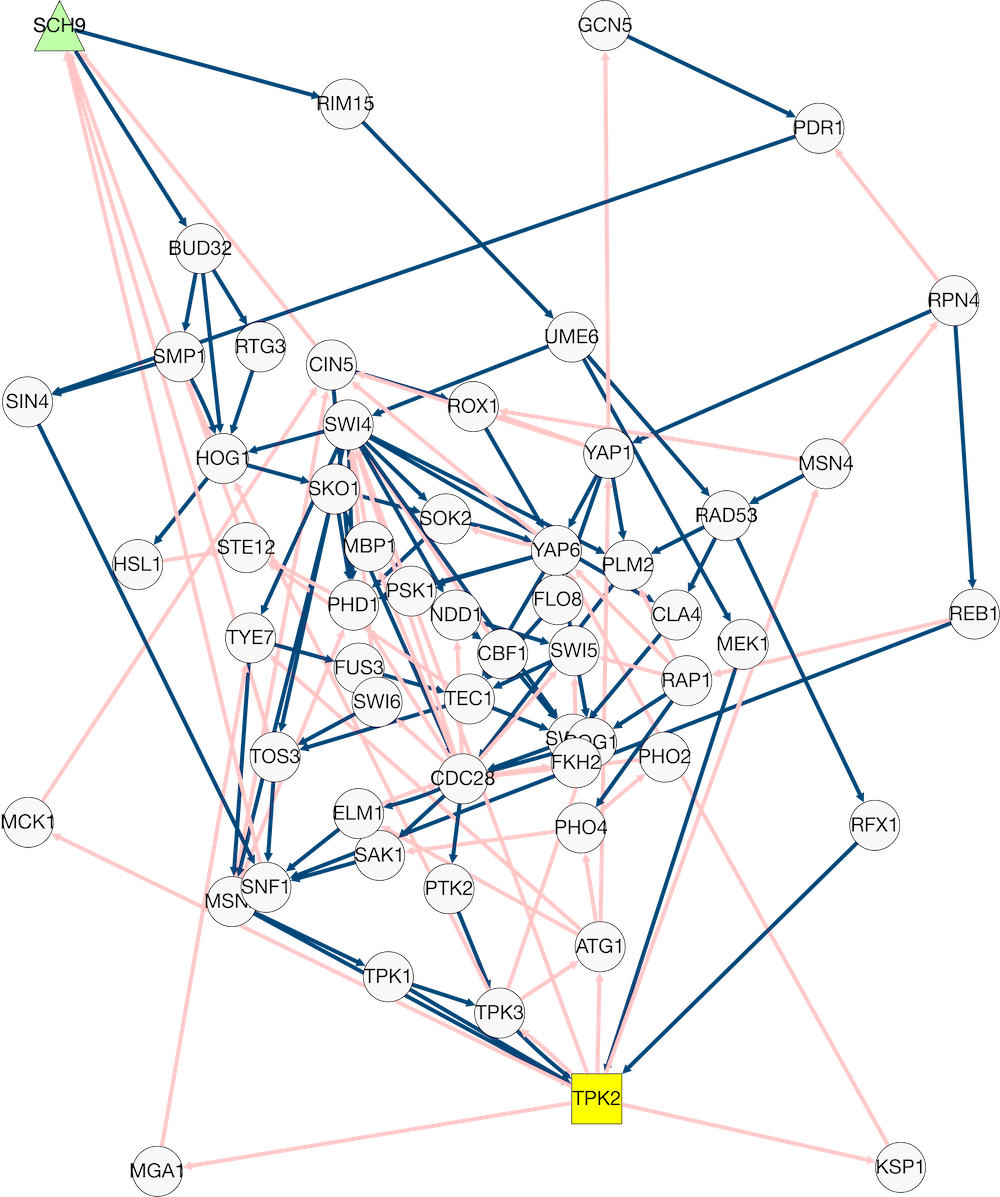}}
    \quad
   \subcaptionbox{Spring Electrical Model. \label{fig:spring-best-layout-G3}}[.38\linewidth][c]{%
    \includegraphics[width=0.38\linewidth,keepaspectratio]{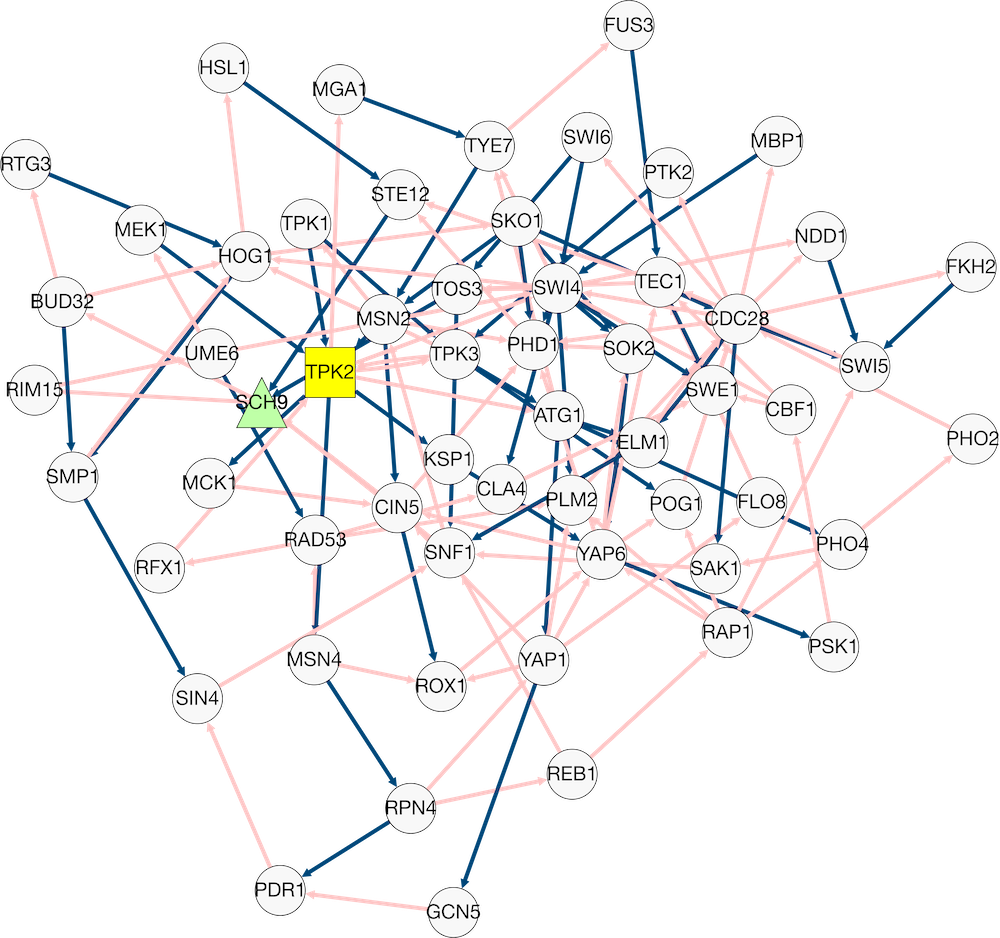}}
  \caption{Best layouts generated by Crowd and baseline methods for network G3. Green and yellow colored nodes represent the source and target nodes, respectively. The upward pointing edges are shown in red whereas downward pointing edges are shown in blue color.  
  }
  \label{fig:best-layouts-G3}
\end{figure}

\clearpage

\end{document}